\documentclass{jfm}

\usepackage{color}
\usepackage{epstopdf}
\begin{document}

\shorttitle{Transition scenario in axisymmetrical hypersonic compression ramp flow} 
\shortauthor{M. Lugrin, S. Beneddine, C. Leclercq, E. Garnier and R. Bur} 

\title{Transition scenario in hypersonic axisymmetrical compression ramp flow}

\author   
 {
 Mathieu Lugrin \aff{1}
  \corresp{\email{mathieu.lugrin@onera.fr}},
  Samir Beneddine \aff{1},
  Colin Leclercq \aff{1},
  Eric Garnier \aff{1}
  \and 
  Reynald Bur \aff{1}
  }

\affiliation
{
\aff{1}
DAAA, ONERA, Paris Saclay University, F-92190 Meudon - France
}

\maketitle

\begin{abstract}
 A high-fidelity simulation of the shock/transitional boundary layer interaction caused by a 15-degrees axisymmetrical compression ramp is performed at a freestream Mach number of 5 and a transitional Reynolds number. The inlet of the computational domain is perturbed with a white noise in order to excite convective instabilities. Coherent structures are extracted using Spectral Proper Orthogonal Decomposition (SPOD), which gives a mathematically optimal decomposition of spatio-temporally correlated structures within the flow. The mean flow is used to perform a resolvent analysis in order to study non-normal linear amplification mechanisms. The comparison between the resolvent analysis and the SPOD results provides insight on both the linear and non-linear mechanisms at play in the flow. To carry out the analysis, the flow is separated into three main regions of interest: the attached boundary layer, the mixing layer and the reattachment region. The observed transition process is dependent on the linear amplification of oblique modes in the boundary layer over a broad range of frequencies. These modes interact nonlinearly to create elongated streamwise structures which are then amplified by a linear mechanism in the rest of the domain until they break down in the reattachment region. The early nonlinear interaction is found to be essential for the transition process.
\end{abstract}
\section{Introduction}\label{sec:intro}
Shock wave-boundary layer interaction (SBLI) is a classical problem of hypersonic flight since shocks appear in the vicinity of any geometrical discontinuity, such as control surfaces. There are two canonical cases for the study of SBLI at high supersonic/hypersonic speed: impinging oblique shock-boundary layer interaction (OSBLI) and SBLI caused by compression ramps.
In both cases, the adverse pressure gradient imposed by the shock will, if it is strong enough, cause the separation of the boundary layer (BL) and thus create a separation bubble. 
The shock-bubble system brings one of the main limitations of SBLI on high-velocity flight: they tend to initiate low-frequency large-scale motion in the flow, causing, among other things,  unsteady thermal loading. \citet{Clemens_Review_upstream_downstream_mecanism} presented and interpreted results from recent studies on that subject. This low-frequency dynamics is an important feature of SBLI, and may be linked to an unstable global mode of the recirculation bubble. Most of the studies on the subject focused on turbulent boundary layers, for instance with the direct numerical studies of \citet{DNS_adams_2000, wu2007direct} or \citet{priebe2012low}. 
However, taking into account the transitional process is essential when designing hypersonic vehicles. \citet{BURTransitionalSWBLI} studied the impact of transition through a SBLI on the European pre-X demonstrator and showed how crucial it is to get a better understanding of the transitional process for this type of flows. For instance, the wall heat-flux peak, another main limiting factor of hypersonic flight, could be more than 20 to 30\% higher than the turbulent one in the transition region.
Along this line, \citet{benaytransitionalSWBLI} experimentally studied a canonical case of SBLI and documented the impact of transition on the topology of the flow and the heat fluxes on the model. However, their study did not bring any information on the transition dynamics.
More recently, the interest for the transition process seems to be growing, \citet{hildebrand2018simulation} conducted a DNS and a global stability analysis on a OSBLI at a transitional Reynolds number, but focused mainly on the globally unstable mode of the separated region rather than on convective instabilities developing along the geometry. Yet, \citet{arnal1989laminar} showed that the transition process in hypersonic flow is highly dependent on receptivity, making the amplification of free-stream disturbances via the non-normality of the linearized Navier--Stokes operator  \citep{schmid2007nonmodal} a better candidate than the linear growth of an unstable global mode. 
While some boundary layer instabilities such as \citet{MACK} second mode are already known to locally dominate the hypersonic flat plate boundary layer, many studies   \citep{fasel1993direct,chang1994oblique,laible2009numerical,mayer2011direct,franko_lele_2013,leleAdverse,fasel2015numerical} show that it is not the only possible cause of transition: oblique breakdown, which is linked to the streaks created by the non-linear interaction of first oblique modes is also a possible candidate. This mechanism was first discovered by \citet{thumm1991numerische} (see also \citet{fasel1991direct,fasel1993direct}) for a supersonic (Mach 1.6) boundary layer using DNS. It was shown that
the nonlinear interaction of a pair of oblique waves with opposite spanwise wavenumbers generates steady
streamwise structures with twice the spanwise wavenumber which grow rapidly in the streamwise direction. \citet{schmid1992new} then confirmed for a plane
channel flow that this mechanisms may also be relevant for incompressible flows.  In this context, it is not possible to identify \emph{a priori} a single dominant transition mechanism for a Mach 5 SBLI. 

Another open debate is the origin of the steady longitudinal structures that appear in hypersonic compression ramp flows, and which often seem to be crucial in the transition process. Some studies suggest that they are due to centrifugal effects and are thus G\"{o}rtler vortices. \citet{NAVARROMARTINEZ2005225} performed a DNS of a hypersonic compression ramp and proposed that the development of steady eddies was linked to centrifugal effects. They also indicated that these vortices were responsible for a spanwise inhomogeneity and an increase of the peak heat flux at reattachment of the order of $20\%$. That kind of heat or friction streak has been observed in many experiments \citep{benaytransitionalSWBLI, GORTLER_oil_CJ_cowl_murray_hillier_williams_2013}. Using optical measurement techniques,
\citet{zhuang2018gortler} also showed the presence of elongated vortical structures in an OSBLI case, which they associated with G\"{o}rtler vortices.
However, the mechanism proposed by \citet{Gortler_original} is not the only one that can lead to the amplification of steady vortices.   { For instance, \citet{dwivedi2018reattachment} showed that a baroclinic mechanism could also lead to the growth of such structures.Another possible mechanism would be the 'lift-up' effect such as pointed by the work of \citet{bugeat20193d} (albeit for attached boundary layer only).
It is unclear if these vortices are directly due to any of these mechanisms or if} the already discussed non-linear interaction linked to oblique breakdown plays a role.

The work presented in this paper aims at describing the transition scenario in a hypersonic flow along an axisymmetrical compression ramp.
To do so, a   {Quasi Direct Numerical Simulation (QDNS, such as defined by \citet{spalart2000strategies})} is carried out. A white-noise perturbation is introduced in the inlet of the computational domain,   {in order to excite convective instabilities in the flow}. The unsteady data are then analysed using Spectral Proper Orthogonal Decomposition (SPOD) to extract coherent unsteady features. To get a better physical understanding of the flow, a non-normal linear stability analysis (a resolvent analysis) is conducted on the mean flow associated with the QDNS and compared to the SPOD results.

\begin{figure}
    \centering
    \includegraphics[width=\linewidth]{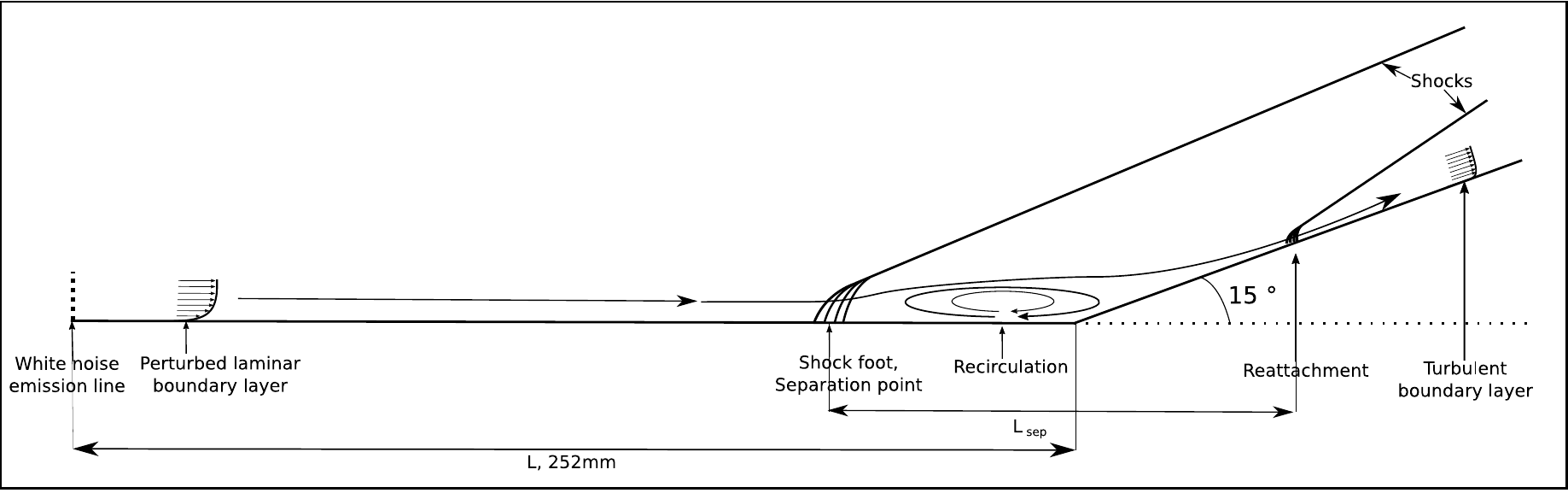}
    \caption{Schematic of a compression ramp, showing the topology of the flow with first the attached BL, then the SBLI, followed by the separated region caused by the adverse pressure gradient and finally the reattachment. }
    \label{fig:schematic}
\end{figure}

The geometry and flow parameters are based on an experimental and numerical database from ONERA that has been studied by \citet{benaytransitionalSWBLI} and \citet{BURTransitionalSWBLI} among others. The geometry under study is a hollow cylinder-flare. Some key features of the configuration and the flow are presented in figure \ref{fig:schematic}. The model is a cylinder of diameter $D=131 mm$ and length $L=252 mm$, followed by a $15$-degree flare. The total length of the geometry is $350 mm$. Free-stream conditions are presented in table \ref{table:freestream}   {and are based on the $Re_L=1.9 \times 10^6$ case studied by \citet{benaytransitionalSWBLI}}, the free-stream Mach number is set to 5. These specific flow conditions have been chosen as they led to a transition in the interaction region during the experiments at the ONERA R2Ch blowdown facility conducted by \citet{benaytransitionalSWBLI}. The Reynolds number based on the momentum thickness computed at the separation point is equal to $\Rey_\theta=724$. 

The article is organised as follows: section \S \ref{sec:numsim} presents the QDNS setup and provides both theoretical and practical details about the tools used for post-processing and analysing the unsteady data (SPOD and global energy computation). Section \S \ref{sec:RESOLVENT} then introduces the resolvent analysis theory and presents the numerical strategy used in the article to carry out tridimensional resolvent analyses. The following section \S \ref{sec:results} focuses on the results of these analyses. In particular, the three regions of interest (attached boundary layer, mixing layer, and reattachment region) are studied in different subsections, following a methodology explained at the beginning of the section. Finally, before concluding in \S \ref{sec:concl}, the results are summarised in \S \ref{sec:scenario}, where we also provide an overall view of the proposed scenario for the transition process.

\section{Numerical simulations}\label{sec:numsim}

\subsection{Quasi-direct numerical simulation setup}
\label{sec:setup}
\begin{table}
 \begin{center}
  \begin{tabular}{cc}
      $T_{\infty}$  & 86.6 K \\
      $M_\infty$  & 5 \\
      $P_\infty$  & 1228 Pa\\
      $U_\infty$ & 933 m/s \\
      \hline
      $T_{wall}$ & 290 K\\
      \hline
      $\Rey_\theta$ & 724 \\
      $\delta$ &  1.975 mm\\ 
      $\frac{\delta}{R}$ & 0.03 \\
      $L_{\mbox{sep}}$ &  129 mm \\
  \end{tabular}
  \caption{Free-stream conditions and characteristic values for the simulation, the Reynolds numbers based on momentum thickness and displacement thickness are computed upstream of the separation point.}{\label{table:freestream}}
 \end{center}
\end{table}
A QDNS of the 3D unsteady flow has been performed using the high-performance finite volumes multi-block structured FAST (Flexible Aerodynamic Solver Technology) compressible Navier--Stokes solver from ONERA \citep{peron2017immersed}. 
The temperature of the wall is imposed at 290K to reproduce the experimental conditions of \citet{benaytransitionalSWBLI,BURTransitionalSWBLI}.   {Standard supersonic inflow, outflow and farfield conditions are used for the other boundaries. These are characteristics-based boundary conditions that avoid numerical reflections.}
The computational domain spans over~$60$ degrees in the azimuthal direction with periodic boundary conditions on the sides. This numerical periodicity, which is necessary for the simulation to be affordable, constrains the azimuthal wavenumbers that may exist within the simulated fields, which can only be multiples of~6. To excite convective instabilities, noise is injected at the upstream inlet of the domain   {(see \S\ref{sec:inletpert} for details)}.
\begin{table}
 \begin{center}
  \begin{tabular}{cc}
      \textbf{Grid size}  &  \\
      $n_x$  & 1409 \\
      $n_r$   & 204\\
      $n_\theta$ & 600\\
  \end{tabular}
  \caption{Grid for the QDNS.}{\label{table:numerical}}
 \end{center}
\end{table}

The domain is discretised using a structured axisymmetric mesh whose main parameters are presented in table \ref{table:numerical}. 
{  
The mesh sizing (presented in appendix \ref{sec:grid}) is such that the flow upstream of the reattachment point is fully resolved with respect to DNS standards.
On the flare, where the flow becomes turbulent and the wall-shear-stress is maximum, the sizing of the mesh becomes slightly under-resolved, and corresponds to a highly resolved LES of SBLI \citep{garnier2002large,teramoto2005large,bonne2019analysis} rather than a DNS. Therefore, the computation corresponds to a QDNS such as described by \citet{spalart2000strategies}, since the resolution is in between the typical LES and DNS resolution \citep{garnier2009large,georgiadis2010large}.

A mesh corresponding to a strict DNS in the reattachment region of the flow would require about ten times more grid points, which would drastically increase the computational cost and make the SPOD analysis impossible. However, the dynamics of the reattachment region is not the primary focus of this paper. The goal instead, is to capture the mechanisms of transition, which are driven by coherent structures developing upstream of the reattachment indeed. Therefore, we only need DNS resolution upstream of the transition point and LES resolution downstream, provided the feedback from the downstream region is negligible. This hypothesis was checked by performing a full-fledged DNS over a long enough period of time to converge the mean flow and transition point (but too short for SPOD analysis). Results shown in appendix \ref{sec:grid} indicate that our quasi-DNS trade-off yields an accurate description of both and may therefore be considered appropriate for studying transition, at a fraction of the cost. This conclusion is in line with previous studies \citep{teramoto2005large} which already showed that LES may be a satisfactory tool for the study of transition in such flows.
}

  {
Viscous fluxes are computed using a second-order centered scheme, and convective fluxes are computed using the second-order upwind AUSM(P) scheme proposed by \citet{AUSMONERA} with a third-order MUSCL reconstruction. The use of an upwind scheme is important in the under-resolved zone of the computation as it maintains the smoothness of the solution by offsetting the energy cascade \citep{spalart2000strategies} as is commonly done for Monotonically Integrated Large Eddy Simulation (MILES). This version of the AUSMP(P) was already successfully used by \citet{bonne2019analysis} in their MILES of an OSBLI case.}
The time integration is performed via an explicit third order 3-steps Runge-Kutta scheme. The time step is set to $10^{-8} s$ to ensure a CFL number lower than 0.5 in the whole domain.

\begin{figure}
    \centering
    \includegraphics[trim= 0.5cm 2.5cm 0.5cm 2.5cm,width=1\linewidth,clip=true]{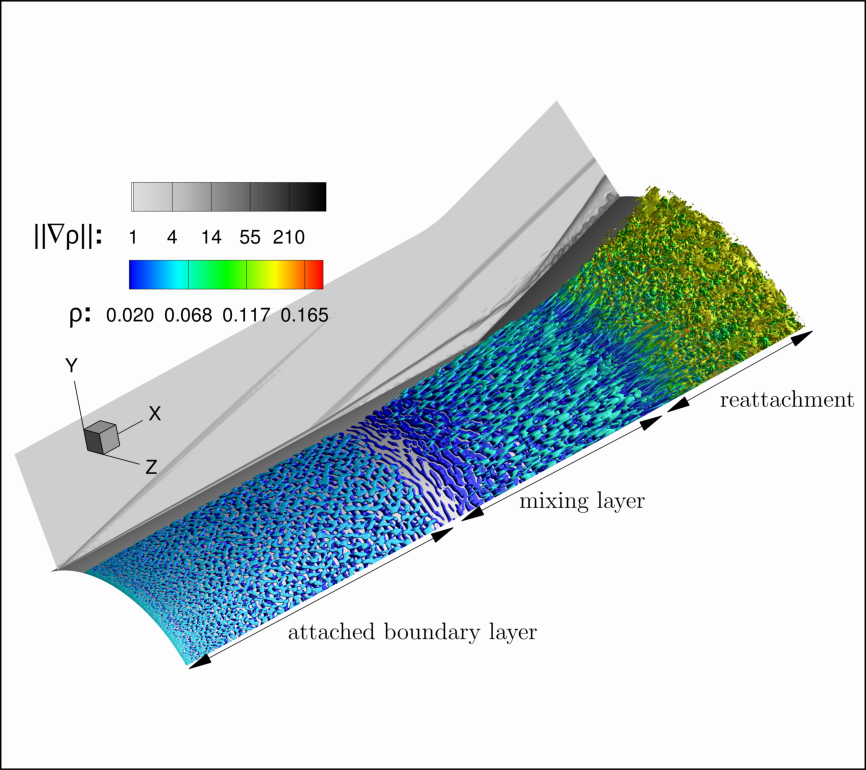}
    \caption{Isosurface of Q criterion ($Q=9\times10^{-6}\frac{U^2}{\delta^2}$) coloured by density and numerical Schlieren visualisation for an instantaneous snapshot of the QDNS.}
    \label{fig:QCRIT_INTRO}
\end{figure}

An isosurface of the Q criterion ($Q=9\times10^{-6}{U^2}/{\delta^2}$) coupled to a numerical Schlieren visualisation of one snapshot from the DNS is presented in figure \ref{fig:QCRIT_INTRO}. It shows the three main regions of interest of the study: the attached boundary layer upstream from the interaction, then the mixing layer between the separation shock and the reattachment point, and finally the reattachment region.\\

\subsection{Inlet perturbation}
\label{sec:inletpert}
As mentioned in \S\ref{sec:setup}, noise is added at the inlet of the domain to excite all possible convective instabilities   
In several papers \citep{mayer2011direct,franko_lele_2013,leleAdverse}, the inlet disturbance is chosen in order to excite a particular instability mechanism within the boundary layer. In the present work, it was chosen not to decide a priori which mechanism was going to be dominant and to let all of them compete in the simulation. Consequently, a generic spatio-temporally white perturbation has been injected at the inlet, which is able to excite vortical, acoustic, and entropic modes. This is reminiscent of the work of \citet{hader2018towards}, who injected broadband pressure fluctuations into their numerical simulation to study natural transition mechanisms in hypersonic boundary layers. 
Note that other choices of generic disturbances may have been considered. The particular receptivity of the chosen noise is studied in the article.  

It is worth mentioning that the amplitude of the inlet noise is not a free-stream turbulence level and cannot be linked directly to the turbulence level of the R2ch blowdown facility. The article does not aim at reproducing the actual free-stream noise of the hypersonic wind tunnel, which is composed of various complex fluctuations \citep{schneider2008development}   with noise radiating from the nozzle and shear layer plus possible perturbations coming from the upstream parts of the blowdown tunnel. Instead, it aims at studying a flow configuration with a generic inlet disturbance, which excites a variety of modes that would develop, compete, and interact together.

However, injecting true spatio-temporally white noise raises numerical difficulties as spatial schemes are not designed to work with very short wavelength oscillations (of the order of a few cells). Because of that, high-amplitude white-noise injection requires filtering of the very small wavelength oscillations to avoid numerical instabilities.
The present section discusses the effect of the noise amplitude on the flow (including high amplitude noise). Therefore, it required such filtering, which is performed by a convolution of the disturbance signal by a Gaussian kernel that spans over seven cells in every direction. 

\begin{figure}
    \centering
    \includegraphics[trim =  0.0cm 0.0cm 0.0cm 0cm,width=0.6\linewidth,clip=true]{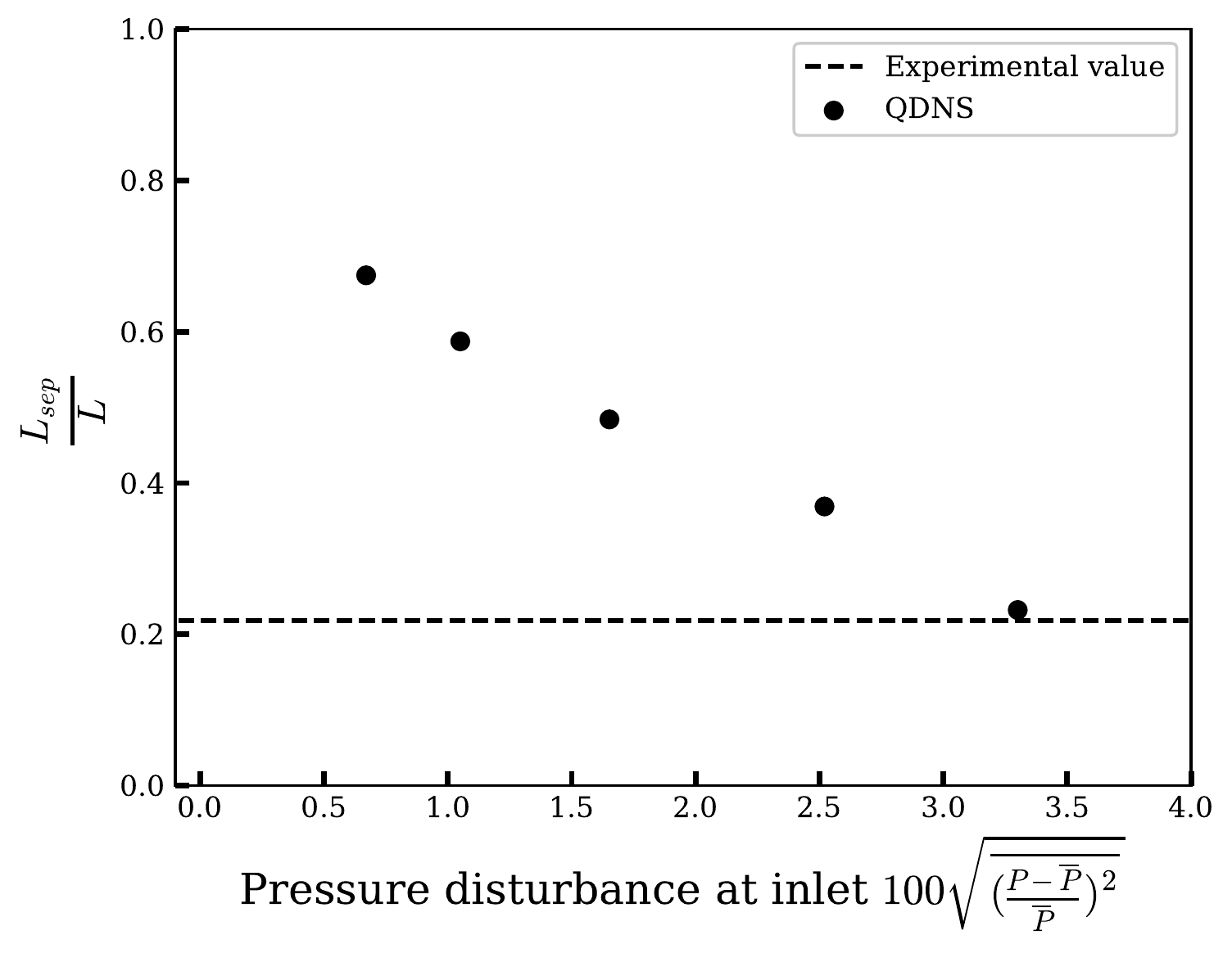}
    \caption{Size of the separated region of the mean flow with different levels of  filtered noise and experimental value corresponding to the same free-stream conditions from \citet{benaytransitionalSWBLI}, showing the impact of the disturbance on the topology of the mean flow.}
    \label{fig:freestream_noise}
\end{figure}

\begin{figure}
    \centering
    \includegraphics[width=0.70\linewidth]{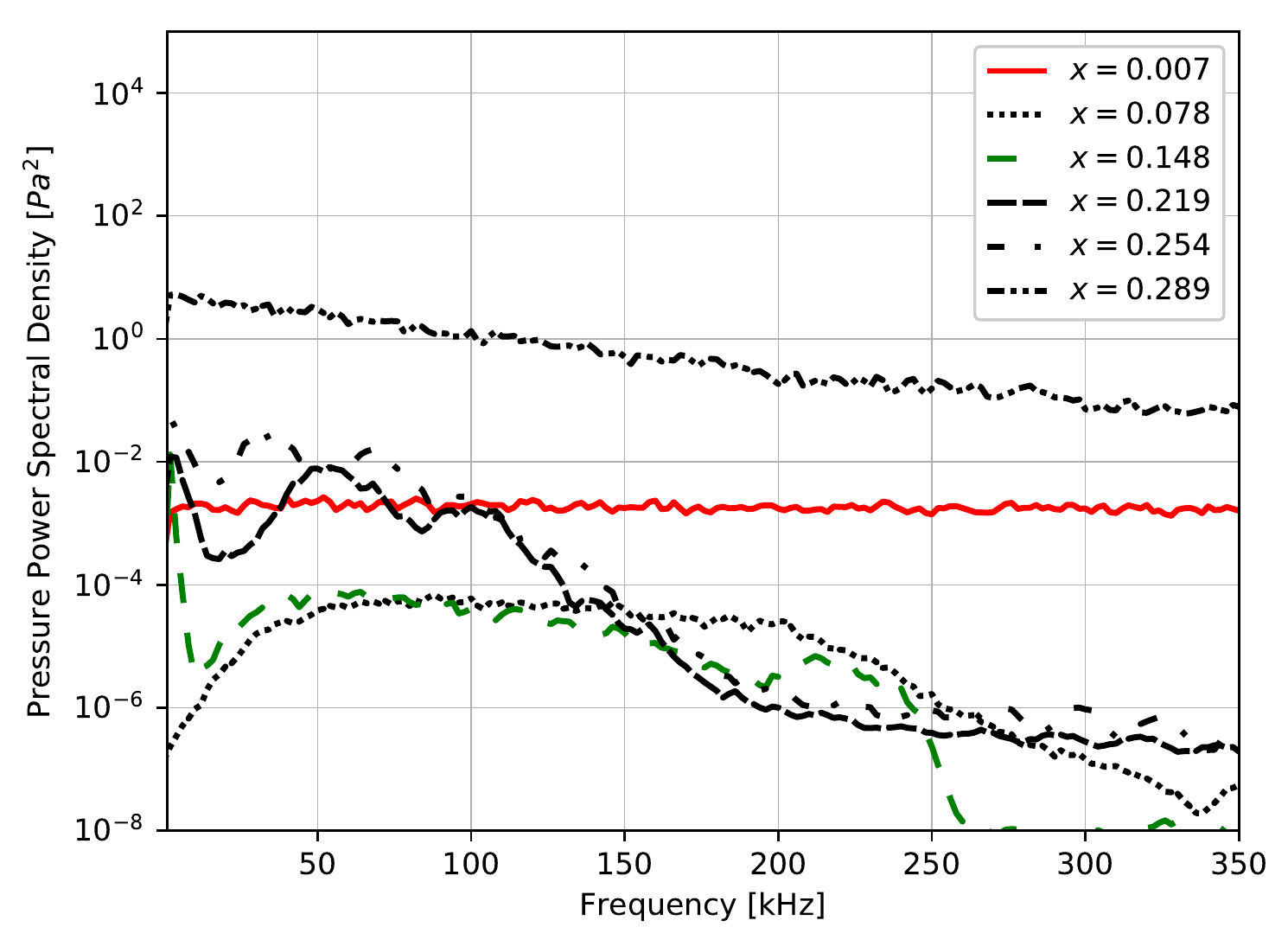}
    \caption{Power spectral density of wall pressure fluctuations at different longitudinal locations of the QDNS.}
    \label{fig:PSD_capt}
\end{figure}

Five QDNSs have been performed, each with a different level of filtered inlet noise (the noise levels are presented in figure \ref{fig:freestream_noise}), yielding five mean flows computed by averaging in time and along the azimuthal direction the simulation results. From these mean flows, a bubble length $L_{\mbox{sep}}$ can be computed, which gives the results presented in figure \ref{fig:freestream_noise}. 
The level of noise impacts the transition location and therefore influences $L_{\mbox{sep}}$ since both the separation and reattachment dynamics strongly depend on the laminar/turbulent nature of the flow.
More importantly, these results show that for the appropriate level of inlet perturbation, the QDNS yields results in agreement with the experimental data from \citet{benaytransitionalSWBLI}, which validates the present computational parameters. But it also reveals how sensitive the flow is to external noise, which raises the question of the level of perturbation to choose for the present study. We choose not to reproduce the experimental conditions from \citet{benaytransitionalSWBLI}.
This is mainly because the available experimental results   {only contain time-averaged data and do not bring any unsteady information on the dynamic of the flow that could be used for comparison.}
The chosen inlet perturbation involves a lower level of noise, which corresponds to an RMS pressure amplitude of 1.5\% of the free-stream value at the inlet and yields $L_{\mbox{sep}}\approx0.5L$ (see figure \ref{fig:freestream_noise}).
With such a low level of disturbance, one avoids the numerical stability issues mentioned above, such that the spatial filtering of the inlet perturbation becomes unnecessary. Therefore, to be in the most generic case, all the following results are based on an unfiltered white noise whose amplitude yields the same recirculation length $L_{\mbox{sep}}\approx0.5L$. Quantitative characterisation of the white noise actually injected is given in figure \ref{fig:PSD_capt}: the red curve ($X=0.028L$) displays the temporal spectrum of the wall pressure fluctuations a few millimeters downstream from the inlet, showing that the PSD is flat as expected. With that precaution, which avoids unnecessary numerical treatment, whatever instabilities growing in the simulations are most likely due to a physical process only.   {Technical details about the injected noise are presented in appendix \ref{sec:noise}.}

{}

\begin{figure}
    \centering
    \includegraphics[trim =  0.0cm 0.0cm 0.0cm 0cm,width=0.6\linewidth,clip=true]{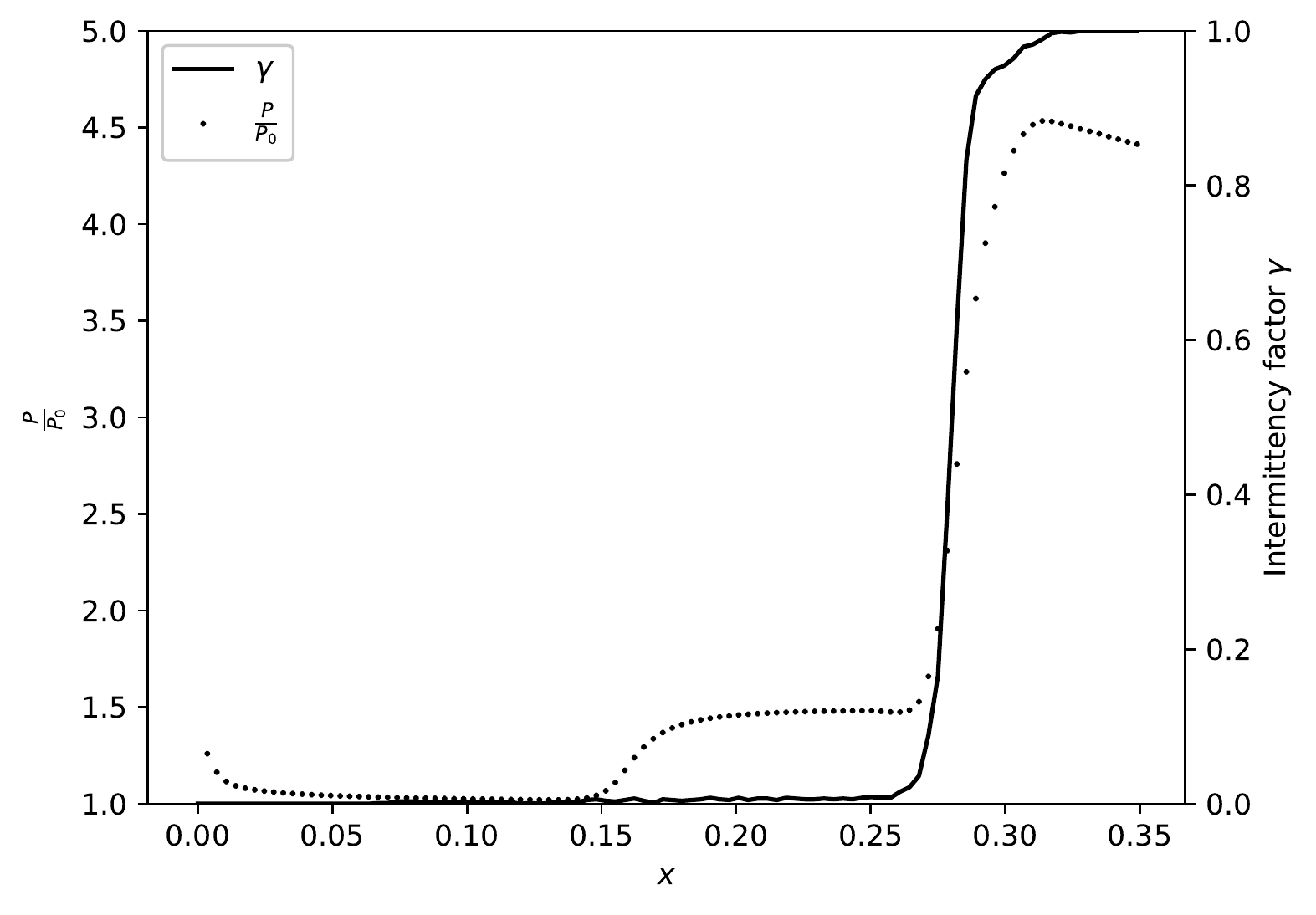}
    \caption{  {Intermittency factor and wall pressure distribution along the geometry showing that the transition is occurring near the reattachment point. }}
    \label{fig:intemrittency}
\end{figure}

The impact of the noise on the topology of the flow highlights the importance of boundary layer instabilities, which play a critical role in the transition by selecting and amplifying white noise to trigger secondary instabilities or non-linear interactions later.
In the case presented here, the transition point is located at the reattachment: the incoming boundary layer upstream of the interaction is laminar (incompressible shape factor around $2.7$). At reattachment, the incompressible shape factor is close to $1.5$, characteristic of a turbulent boundary layer. Figure \ref{fig:QCRIT_INTRO} qualitatively shows how the transition process is increasing in intensity as it goes through each studied zone, eventually creating turbulent structures on the flare. 

{  To confirm the assumption that the flow is transitioning at reattachment, figure \ref{fig:intemrittency} presents both the intermittency factor and the wall pressure distribution along the geometry.
The local intermittency value $\gamma(x)$ represents the probability of being in a turbulent spot at a given time. It is commonly used to describe the transition process (see for instance \citet{sandham2014transitional}).
An intermittency factor of 0 thus means that the boundary layer is fully laminar, with no turbulent spot, while a factor of 1 means that the flow is fully turbulent.
Everything between 0 and 1 is considered transitional.
In the present case, the intermittency factor is computed from spectrograms of wall pressure fluctuations along the geometry, following an idea of \citet{arnal}.
First, a range of "laminar" perturbation frequencies is defined. The presence of a turbulent spot is assumed if fluctuations are detected outside of this range (at higher frequencies). In the present case, it was decided to define the laminar range from 0Hz up to 600kHz. These values have been chosen such that the upper limit is more than twice the highest frequency of the common hypersonic boundary layer instabilities (results have shown that the shape of gamma is not impacted by a change of this threshold toward upper frequencies).
The results presented in figure \ref{fig:intemrittency} shows that the intermittency is strictly 0 in the whole attached boundary layer and really close to 0 for most of the separated region (which is caracterised by the pressure plateau at ${P}/{P_0}=1.6$).
In the final part of the mixing layer the intermittency first slightly increases and then brutally reaches 1 at the reattachment (which is characterised by a steep increase in pressure).}

  {For all the different cases considered in this section, associated with different levels of noise, the transition location obviously changes. But so does the reattachment point, such that eventually, the transition to turbulence always occurs close to the reattachment,  and the transition scenario was always the one presented here}.

  {The relation between the recirculation bubble topology and the upstream perturbations has also been documented by \citet{marxen2010mean} for incompressible separation. They showed that the transition caused by upstream perturbations leads to the shrinkage of the bubble from both sides.
The fact that the flow topology is highly dependent on the level of free-stream noise is one of the primary motivations to use the mean flow instead of a base flow for the stability analysis, as it was already advised by \citet{marxen2010mean}.}

\subsection{Spectral Proper Orthogonal Decomposition}
\label{sec:SPOD}
\begin{table}
 \begin{center}
  \begin{tabular}{cc}
      \textbf{Grid size and resolution:}  &  \\
      $n_x$  & 353 \\
      $n_r$   & 102 \\
      $n_\theta$ & 257\\
      $\theta$ resolution & $\approx0.23^\circ$\\
      $m$ resolution & 6\\
      \\
      \textbf{Temporal sampling:} & \\
      Sampling rate & 200 kHz \\
      Number of samples & 1664 \\
      Number of samples in each realisation & 128 \\
      Frequency resolution & 1562.5 Hz\\
      $N_r$ & 26 \\
      \end{tabular}
  \caption{Numerical parameters for the SPOD.}{\label{table:numericalSPOD}}
 \end{center}

\end{table}
Convective amplification mechanisms are known to generate coherent structures \citep{beneddine2016conditions}, which may be studied through a Spectral Proper Orthogonal Decomposition (SPOD).   {This variant of the classical Proper Orthogonal Decomposition (POD) was first introduced by \citet{lumley1970stochastic}} and has been widely used by the turbulence community since then (see for instance \citet{gudmundsson2011instability}).   {It has been recently studied from a mathematical point of view by \citet{towne2018spectral}, who showed that it is by construction the optimal decomposition to identify spatio-temporally correlated structures within statistically stationary flow.}

To perform the decomposition, one has to first sample snapshots from the simulation, then gather them in $N_{r}$ (possibly overlapping) realisations of the flow. 
Each realisation contains a temporal sequence of snapshots vectors $\left(\mathbf{s}_{t_0},\mathbf{s}_{t_0+\Delta t},\dots\right)$, where the components of  $\mathbf{s}_{t}$ are the values of the 3D flow field at time $t$. 
A Direct Fourier Transform (DFT) is then applied both in the temporal and azimuthal direction, giving Fourier mode vectors $\mathsfbi{\hat{S}}^{k}\!(\omega,m)$, where $k$ is the realisation number, $\omega$ the angular frequency and $m$ the azimuthal wavenumber of the mode. Due to the spectral transformation in the azimuthal direction, the vectors $\mathsfbi{\hat{S}}^{k}\!(\omega,m)$ correspond to bi-dimensional fields: they contain complex values associated with each flow variables at each pair $(x,r)$ from the mesh.
For a given pair ($\omega$,$m$) of interest, the Fourier modes  of all realisations are then stacked in a matrix $\hat{\mathsfbi{X}}_{\omega,m}$, which reads
\begin{equation}
    \hat{\mathsfbi{X}}_{\omega,m} =\left[ \mathsfbi{\hat{S}}^{0}\!(\omega,m), \  \mathsfbi{\hat{S}}^{1}\!(\omega,m), \  \cdots, \  \mathsfbi{\hat{S}}^{N_r-1}\!(\omega,m) \right].
\end{equation}
This matrix is then processed similarly to a snapshot matrix in a classical space-only POD decomposition:
the $i$-th SPOD mode $\mathbf{\Phi_i}^{(\omega,m)}$ can be computed from the $i$-th left singular vector of $\hat{\mathsfbi{X}}_{\omega,m}$, which may be computed by solving the eigenproblem associated with the cross spectral density matrix  
\begin{equation}
    \hat{\mathsfbi{X}}_{\omega,m} \hat{\mathsfbi{X}}_{\omega,m}^\star \mathsfbi{Q_e} \ \mathbf{\Psi_i}^{(\omega,m)} = \lambda_i \mathbf{\Psi_i}^{(\omega,m)},
    \label{eq:eigSPOD}
\end{equation}
  {with $\mathsfbi{Q_e}$ the inner product associated with the  energy norm defined by \citet{chu1965energy} which is presented in the appendix \ref{inner}.} This norm is commonly used in order to describe fluctuation energy in compressible flow \citep{hanifi1996transient,JOSEPHGEORGE20115280,bugeat20193d} and is more adapted than a simple kinetic energy norm often used for (quasi-)incompressible flows. The SPOD modes are ordered with respect to their contribution to the global dynamics, \emph{i.e.} $\lambda_0 > \lambda_1 > \lambda_2>\dots$, and for a given pair $(\omega,m)$, the relative contribution of the $i$-th SPOD mode is measured by the ratio $r_i = \lambda_i / \sum_k \lambda_k$. In the following, we will focus in particular on the leading SPOD mode, and $r_0$ will be systematically specified to quantify how dominant it is compared to the remaining ones.

  {In practice, the eigenmodes are computed by using the snapshots method of \citet{berkooz1993proper} which is a less costly but equivalent decomposition based on $ \hat{\mathsfbi{X}}_{\omega,m}^\star  \mathsfbi{Q_e} \hat{\mathsfbi{X}}_{\omega,m}$ rather than (\ref{eq:eigSPOD}). This provides the right singular vectors of $\hat{\mathsfbi{X}}_{\omega,m}$, from which one can easily retrieve the SPOD modes }(see for instance \citet{towne2018spectral} for details).
The snapshot vector size is also reduced by downsampling the mesh since spatially coherent structures are always significantly bigger than the dissipation scale that needs to be resolved in the QDNS (see table~\ref{table:numericalSPOD}).
The resulting $i$-th singular vector is a discrete 2D field $\mathbf{\Psi_i}^{(\omega,m)}$ corresponding to a slice of the SPOD mode in the azimuthal direction.
This framework is well-adapted for a spectral study of periodic structures that develop on an axisymmetric geometry such as streamwise vortices. 
For visualisation purpose, the structure of SPOD modes will be displayed in the present work by showing isocontours of the real part of $\mathbf{\Psi_i}^{(\omega,m)} e^{i m\theta}$, which will be called SPOD mode in the captions for conciseness.

Note that the spectral resolution in $m$ and $\omega$ of the SPOD is set by the azimuthal span of the computational domain, the temporal length of each realization and the sampling frequency of the snapshots, respectively. These elements, as well as other parameters of the SPOD are specified in table~\ref{table:numericalSPOD}. The sampling frequency is set to 200kHz in order to capture the most energetic physical mechanisms in the QDNS,   {A study using a low-pass filter (not shown here) has been carried out to ensure that this sampling frequency does not yield any noticeable aliasing}. The power spectral densities of pressure fluctuations for probes distributed along the wall presented in figure \ref{fig:PSD_capt} confirm that the transition process comes from a rather low-frequency mechanism and that high-frequency ones (\emph{i.e.} $f \geq 100kHz$) are not important in the present context   {(justifications regarding linear amplification mechanisms will be presented later in \S \ref{sec:BL}))}.

\subsection{Fluctuation energy distribution}
\label{sec:energy}
\begin{figure}
    \centering
    \includegraphics[width=0.55\linewidth]{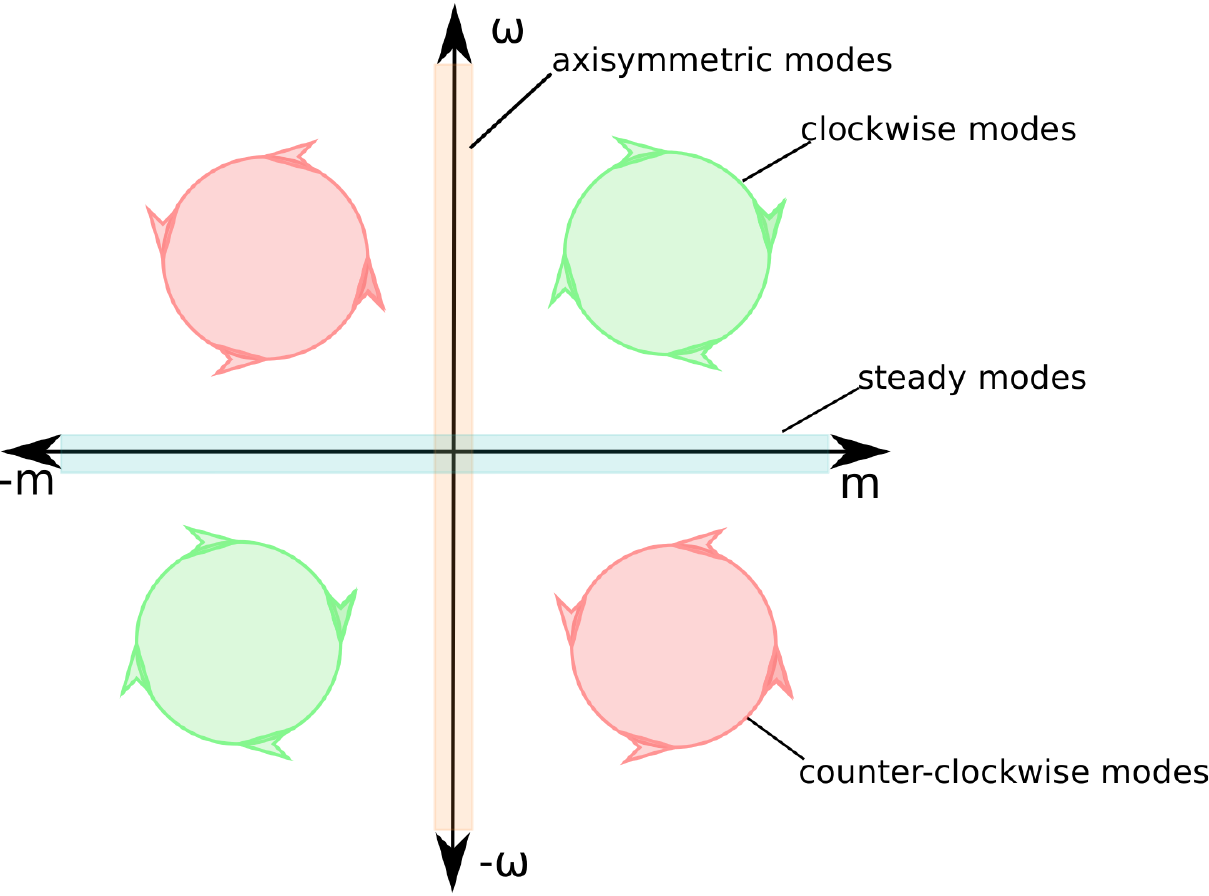}
    \caption{Schematic of the fluctuation energy distribution map representing the corresponding structures for each zone.}
    \label{fig:Energy_map}
\end{figure}

The matrix formulation used for the SPOD in \S\ref{sec:SPOD} is convenient to compute the global energy of the fluctuation in the simulation associated to a pair $(\omega,m)$
\begin{equation}
    E_{Chu}(\omega,m)=\frac{Tr(\hat{\mathsfbi{X}}_{\omega,m}^\star\mathsfbi{Q_{e}} \hat{\mathsfbi{X}}_{\omega,m})}{N_r},
    \label{eq:EnergyBalance}
\end{equation} 
Equation (\ref{eq:EnergyBalance}) may be used to produce energy distribution maps that reveal regions in the ($\omega$-$m$)-domain where fluctuations are particularly energetic. A schematic of such a colormap is presented in figure \ref{fig:Energy_map}, the top right quarter containing the clockwise modes and the bottom right the counter-clockwise modes. Modes along the frequency axis are axisymmetric and modes along the wavenumber axis are steady by construction. As the data from the QDNS are real,   { the Fourier transformed snapshots display Hermitian symmetry:
\begin{equation}
    \hat{X}_{-\omega,-m}=\overline{\hat{X}_{\omega,m}}
\end{equation}
Because of that, the energy map is symmetric around the origin (\textit{i.e.} the top right/left quarter is the same as the bottom left/right one).
Additionally, as the flow is statistically homogeneous in the azimuthal direction, the clockwise and counter-clockwise modes mirror each other as well.}
  {Note that for visualisation purposes, the energy maps are displayed as continuous colormaps. However, the actual values are only defined in discrete pairs ($\omega$-$m$) (that are represented in the background of the maps as dots): $m$ is a multiple of 6 because of the spanwise extent of the domain, and the resolution of the $\omega$ axis is set by the temporal length of the time sequences that are Fourier-transformed (see \S\ref{sec:SPOD} and table \ref{table:numericalSPOD}). }

\section{Mean flow resolvent analysis}
\label{sec:RESOLVENT}
\subsection{Resolvent   {analysis}}
\label{sec:resolv}
Global stability analysis is widely used to study the dynamics of fluid flows. In many cases, studying the spectrum of the linearised Navier--Stokes operator gives important information on unstable global modes to understand the origin of unsteady features of the flow. However, as the linearised Navier--Stokes operator is non-normal (\textit{i.e.} its eigenfunctions are non-orthogonal), initial conditions or external forcings of very low amplitude can trigger high-amplitude fluctuations even when a flow is globally stable. The global resolvent analysis (sometimes called input/output analysis) allows to study the impact of non-normality of the operator on the amplification of such disturbances. Compared to local approaches commonly used to study transition (such as local stability analysis, Parabolised Stability Equation (PSE) analysis, \textit{etc}), no assumption about the parallelism of the flow is required, which makes it perfectly adapted for the study of convective instabilities in the presence of shocks and separation.

{
Several papers have unveiled the links between resolvent and local stability analyses, and it is now well established that resolvent modes match local stability results in zones where the flow is nearly parallel and dominated by some locally unstable modes. As such, resolvent analyses may be viewed as a generalisation of the classical local stability approach, with the difference that it may deal with more complex situations that cannot be factored in by a local stability approach or by PSE (see \citet{article}, \citet{beneddine2016conditions}, \citet{bugeat20193d} for instance).
 
In the context of hypersonic boundary layers, the recent work of Bugeat \citep{bugeat2017stabilite, bugeat20193d} specifically shows how resolvent modes are related to classical Linear Stability Theory (LST) results from the literature about the well-known modes 1 and 2.

Following the work of \citet{brandt2011effect} and \citet{bugeat20193d}, we can separate the non-normal mechanisms presented in this study into two categories, the 'convective type non-normalities' and the 'component-type non-normalities'. The former are linked to the advection of perturbations in the mean flow and these are usually referred to as 'modal' instabilities in the LST framework. The latter are linked to the transport of mean-flow momentum by the perturbations (the 'lift-up' effect for example) that would be referred to as non-normal instabilities in the local framework. Note that the vocabulary from the literature is somewhat ambiguous, as "modal" and "non-modal" terms are used differently in the global stability and the local LST frameworks, sometimes to characterise the same underlying physical mechanism. In the following, we focus on global resolvent analysis and use the global stability point of view to refer to the nature of the modes.
}

Starting from $\boldsymbol{q} =(\rho,\rho \boldsymbol{u}, \rho E)$ the state vector of the flow and $\boldsymbol{\mathcal{N}}$ the compressible Navier-Stokes operator, the temporal evolution of $\boldsymbol{q}$ is governed by an equation of the form
\begin{equation}
    \frac{\partial \boldsymbol{q}}{\partial t}= \boldsymbol{\mathcal{N}}(\boldsymbol{q}) + \boldsymbol{f_0},
    \label{GE}
\end{equation}
with $\boldsymbol{f_0}$ a forcing term corresponding to the injected noise perturbation. By introducing the mean flow $\boldsymbol{\bar{q}_{0}}$  as defined in \S\ref{sec:inletpert} and $\mathsfbi{J}=\left. \frac{\partial\boldsymbol{\mathcal{N}}}{\partial\boldsymbol{q}}\right|_{\boldsymbol{\bar{q}_{0}}}$ the linearisation of $\boldsymbol{{\mathcal{N}}}$ about $\boldsymbol{\bar{q}_0}$ , the fluctuation around the mean flow $\boldsymbol{q'}=\boldsymbol{q}-\boldsymbol{\bar{q}_0}$ is governed by
\begin{equation}
    \frac{\partial \boldsymbol{q'}}{\partial t} =\mathsfbi{J}\boldsymbol{q'} + \boldsymbol{f_0} + \boldsymbol{\mathcal{F}}(\boldsymbol{q'},\boldsymbol{\bar{q}_0}),
    \label{Jq_0}
\end{equation}
with $\boldsymbol{\mathcal{F}}(\boldsymbol{q'},\boldsymbol{\bar{q}_0}) = \boldsymbol{\mathcal{N}}(\boldsymbol{q}) - \mathsfbi{J}\boldsymbol{q'} - \boldsymbol{f_0} $ a term gathering the non-linear part of the Navier--Stokes operator. Following the formalism of \citet{beneddine2016conditions}, one may then define $\boldsymbol{f'} = \boldsymbol{f_0} + \boldsymbol{\mathcal{F}}(\boldsymbol{q'},\boldsymbol{\bar{q}_0})$ as a forcing term containing the non-linear forcing and the injected perturbation such that equation (\ref{Jq_0}) reduces to
\begin{equation}
    \frac{\partial \boldsymbol{q'}}{\partial t} =\mathsfbi{J}\boldsymbol{q'} + \boldsymbol{f'}.
    \label{Jq}
\end{equation}
The Fourier transform of equation (\ref{Jq}) reads
\begin{equation}
    \boldsymbol{\hat{q'}}= \mathcal{R}\boldsymbol{\hat{f'}},
    \label{equ:resolvent}
\end{equation}
with 
$\boldsymbol{\hat{f'}}$ and $\boldsymbol{\hat{q'}}$ the Fourier transform of $\boldsymbol{f'}$ and $\boldsymbol{q'}$, respectively, and
$\mathcal{R}$ the resolvent operator defined as  $\mathcal{R}=(i\omega\mathsfbi{I}-\mathsfbi{\mathsfbi{J}})^{-1}$. 
This compact equation shows that the flow may be seen as an input-output system, where a forcing $\boldsymbol{\hat{f'}}$ generates a response $\boldsymbol{\hat{q'}}$ through the resolvent operator.
Then, a resolvent analysis consists in computing for every frequency $\omega$ of interest an optimal forcing $\boldsymbol{\phi_0}$  which maximises the gain defined as
\begin{equation}
    \mathcal{G}(\omega) = \frac{ \langle\mathcal{R} \boldsymbol{\phi},\mathcal{R}\boldsymbol{\phi} \rangle_e }{ \langle\boldsymbol{\phi},\boldsymbol{\phi}\rangle }, 
    \label{equ:gain}
\end{equation}
where $\langle . , .\rangle_e$ represents the energy of the fluctuation as defined in \S\ref{sec:energy} 
and $\langle . , . \rangle$ the scalar product associated with the $\mathcal{L}_2$ norm
\begin{equation}
\langle \boldsymbol{\phi},\boldsymbol{\phi} \rangle = \boldsymbol{\phi}^* \mathsfbi{Q} \boldsymbol{\phi},
\end{equation}
with $\mathsfbi{Q}$ the weight matrix defined in appendix \ref{inner} . 
The optimal forcing and the associated gain are given by the dominant right singular vector and dominant singular value of $\mathcal{R}$, and they may be computed by solving
\begin{equation}
     \mathcal{R}^* \mathsfbi{Q_e} \mathcal{R} \boldsymbol{\phi_i} = \mu_i^2  \mathsfbi{Q} \boldsymbol{\phi_i}.
     \label{equ:eigprob}
\end{equation}
The highest eigenvalue  $\mu_0^2$ of equation (\ref{equ:eigprob}) is the optimal gain, the corresponding eigenvector $\boldsymbol{\phi_0}$ is the optimal forcing. {  These quantities are functions of the frequency~$\omega$. Additionally as shown in section \ref{sec:ResolvSchmid}, one may perform a Fourier transform of equation (\ref{Jq_0}) in the azimuthal direction such that the gain and optimal forcing are not only  functions of $\omega$, but also functions of the azimuthal wavenumber $m$. } 

Computing lower-magnitude eigenvalues $\mu_{i\geq1}^2$ of (\ref{equ:eigprob}) gives sub-optimal forcings $\boldsymbol{\phi_{i\geq1}}$. After normalization, these forcings yield an orthonormal basis of the forcing space, \textit{i.e.} $\langle \boldsymbol{\phi_i},\boldsymbol{\phi_j} \rangle = \delta_{ij}$. The optimal responses given by $\boldsymbol{\psi_i}=\mathcal{R}\boldsymbol{\phi_i}/||\mathcal{R}\boldsymbol{\phi_i}||_e$ gives a similar basis of the response space, and equation (\ref{equ:resolvent}) may then be decomposed as
\begin{equation}
    \boldsymbol{\hat{q'}} =\boldsymbol{\psi_0} \mu_0  \langle\boldsymbol{\phi_0},\boldsymbol{\hat{f'}}\rangle  + \sum_{i\geq 1} \boldsymbol{\psi_i} \mu_i  \langle\boldsymbol{\phi_i}, \boldsymbol{\hat{f'}}\rangle.
    \label{equ:reponse}
\end{equation}

Physically, when there exists one strong convective instability mechanism within the flow (such as first or second mode instabilities), the optimal gain becomes very high, and the resolvent analysis yields $\mu_0\gg \mu_{i\geq1}$ (see \citet{beneddine2016conditions}).
{  When this occurs, the first term of the right-hand-side of (\ref{equ:reponse}) is expected to be dominant, as long as the noise contained in $\boldsymbol{\hat{f'}}$ does not preferentially excite a suboptimal forcing in a way that shifts the dominance (which was never observed in the present study). 
 Then, $\boldsymbol{\hat{q'}}$ is going to be dominated by the first optimal response $\boldsymbol{\psi_0}$ as a result of this strong linear amplification mechanism. Therefore, the resolvent analysis may explain the appearance of coherent structures, and as such, it is an important tool to confront with SPOD analyses.

However, $\boldsymbol{\hat{f'}}$  may project better onto $\boldsymbol{\phi_0}$ for some given values of $(m,\omega)$. This may be investigated by introducing the coefficient $c_0 = \mu_0^2  |\langle\boldsymbol{\phi_0},\boldsymbol{\hat{f'}}\rangle|^2$, which represents the combination of two mechanisms: the ability of the linear operator to optimally amplify a certain type of structure (through $\mu_0$) and the strength of the excitation of this mechanism by $\boldsymbol{\hat{f'}}$, which contains both the injected noise and the non-linear terms. Situations where $E_{Chu}(\omega,m)$ is high while the dominant amplification mechanism is weak (\textit{i.e.} $\mu_0^2(\omega,m)$ is small) may be explained by receptivity processes that are accounted for by $c_0(\omega,m)$. In general, in the context of strong nonlinear interaction and no dominant linear instability mechanism, $c_0(\omega,m)$ is not expected to match $E_{Chu}(\omega,m)$ since there is no reason for the forcing term $\boldsymbol{\hat{f'}}$ to specifically excite a given linear mechanism. However, if $c_0(\omega,m)$ matches $E_{Chu}(\omega,m)$, then for this particular pair $(\omega,m)$, the forcing term projects well onto $\boldsymbol{\phi_0}$ such that high-energy structures stem from a weak (but strongly excited) linear mechanism. As shown in the following, such a situation where the nonlinearities excite a very specific linear amplification mechanism is central for the transition scenario of the studied flow configuration.

It is also interesting to discriminate the contribution of the injected noise from the contribution of the non-linear terms in the receptivity processes. To do so, one may simply compute $c_r$, which is defined in the same way as $c_0$ but with a scalar product spatially restricted to the stencil of the injection plane of the noise (\textit{i.e.} the support of the forcing term). If $c_r$ is close to $c_0$, the receptivity is linked to the nature of the noise alone. Otherwise, the receptivity of the non-linear terms also comes into play.

In order to preserve the stochastic framework introduced in section \ref{sec:SPOD} and to conform with the SPOD approach, the actual computation of $c_{0}$ in the following is $c_{0} = \mu_0^2  E[|\langle\boldsymbol{\phi_0},\boldsymbol{\hat{f'}}\rangle|^2]$, where $E[.]$ is the expected value estimated from an average of values computed for several realisations (using the same time sequences as for the SPOD analysis described in section \ref{sec:SPOD}), $\boldsymbol{\hat{f'}}$ being computed as $\mathcal{R}^{-1}\boldsymbol{\hat{q'}}$}.

Note that it is possible to localise the resolvent analysis to a given region of the flow by setting all coefficients of $\mathsfbi{Q_e}$ associated with cells outside of this region to zero. The gain is then defined as the maximal energy restrained to this specific zone, and as such, the response is constrained in space (but the forcing is not). This approach is used in the following to study coherent structures in specific domains of the flow.

\subsection{Azimuthal decomposition of the resolvent analysis}
\label{sec:ResolvSchmid}
The computation of the resolvent operator requires the jacobian matrix. Following the procedure described by \citet{bened2017}, $\mathsfbi{J}$ is obtained by a finite-differences linearisation of the discrete equations implemented in FAST. The largest eigenvalues of (\ref{equ:eigprob}) may then be solved using the Arnoldi algorithm coupled with an LU solver for the inversion phase (using ARPACK \citep{lehoucq1998arpack} and MUMPS \citep{amestoy2001fully}). Unfortunately, given the size of the matrices involved, the computational cost of this strategy is not affordable. But since the mean flow is axisymmetric and the mesh is homogeneous in the azimuthal direction, the Jacobian operator may be rearranged in a block diagonal form as proposed by \citet{schmid2017stability} to make the computation significantly cheaper. This cost-reduction method, which has also been used in \citet{paladini2019transonic}, is briefly presented below. 

Since the solver FAST works internally with Cartesian coordinates, one has to first carry out a transformation to cylindrical coordinates to retrieve the axisymmetry of the flow using the following relation
\begin{equation}
\setlength{\arraycolsep}{4pt}
\renewcommand{\arraystretch}{1.5}
\left(
\begin{array}{c}
   \rho \\
    \rho {u_x}\\
  \rho {u_r}\\
   \rho u_\theta \\
   \rho E \\
\end{array}  \right)
=   \left[
\begin{array}{ccccc}
  1  & 0  & 0 & 0  & 0   \\
   0 & 1 & 0 & 0  & 0 \\
  0 & 0 & cos(\theta) & sin(\theta)  &  0 \\
  0 & 0 & -sin(\theta) & cos(\theta)  &  0 \\
   0 & 0 & 0 & 0 &  1 \\
\end{array}  \right]
\left(
\begin{array}{c}
   \rho \\
    \rho {u_x}\\
  \rho {u_y}\\
   \rho u_z \\
   \rho E \\
\end{array}  \right)
\label{rotation}
\end{equation}
Under appropriate indexing of the degrees of freedom, the Jacobian operator can then be rearranged into the block-circulant form
\begin{equation}
\setlength{\arraycolsep}{2pt}
\renewcommand{\arraystretch}{1.3}
\mathsfbi{J} = \left[
\begin{array}{ccccc}
  \mathsfbi{A}_0 & \mathsfbi{A}_1 & ... & \mathsfbi{A}_{n-2}  & \mathsfbi{A}_{n-1}\\
 \mathsfbi{ A}_{n-1} & \mathsfbi{A}_0 & ... & \mathsfbi{A}_{n-3}& \mathsfbi{A}_{n-2}\\
 \mathsfbi{ A}_{n-2} & \mathsfbi{A}_{n-1} & ... & \mathsfbi{A}_{n-4} & \mathsfbi{A}_{n-3}\\
  \vdots & \vdots  & \vdots  &\vdots &\vdots \\
  \mathsfbi{A}_1 & \mathsfbi{A}_2& ... & \mathsfbi{A}_{n-1} & \mathsfbi{A}_{0}\\
\end{array}  \right] ,
\label{schmid}
\end{equation}
where each line of blocks corresponds to a given azimuthal slice of the mesh and the block matrices $\mathsfbi{A}_0$,   {..., $\mathsfbi{A}_{n-1}$} have a size corresponding to such a slice (the size of a 2D problem). The block-circulant nature of the matrix comes from the numerical and physical equivalence of all azimuthal slices of the mean flow, which cannot be distinguished from one another.
As shown by \citet{schmid2017stability}, this block circulant matrix can then be transformed into a block-diagonal matrix
\begin{equation}
\setlength{\arraycolsep}{2pt}
\renewcommand{\arraystretch}{1.3}
\mathsfbi{\tilde{J}} = \left[
\begin{array}{ccccc}
  \mathsfbi{\tilde{A}}_0 &  & &  & \\
  & \mathsfbi{\tilde{A}}_1 &  & &\\
   & & \ddots  & &\ \\
   & & & & \mathsfbi{\tilde{A}}_{n-1}\\
\end{array}  \right], 
\label{schmid_diag}
\end{equation}
with
\begin{equation}
    \mathsfbi{\tilde{A}}_m = \mathsfbi{A}_0 + \rho_m \mathsfbi{A}_1 + \rho_m^2 \mathsfbi{A}_2 + ... + \rho_m^{n-1} \mathsfbi{A}_{n-1},
\end{equation}
and $\rho_m=e^{\frac{i 2 \upi m}{n}}$ corresponding to an $m$-root of unity.\\
Then, the analysis of the global 3D resolvent may be done by performing $n$ smaller resolvent analyses by successively considering for $m=0,\dots,n-1$ the operator 
\begin{equation}
 \tilde{\mathcal{R}}(m,\omega)=(i\mathsfbi{I}\omega-\mathsfbi{\mathsfbi{\tilde{A}}}_m)^{-1}
\end{equation}
For each value of m, the 3D optimal forcing and response, denoted as $\boldsymbol{\phi}_m$ and $\boldsymbol{\psi}_m$, respectively, are the singular vectors of $\mathcal{R}$ and can be computed from those of $\tilde{\mathcal{R}}$ : $\boldsymbol{\tilde{\phi}}_m,\boldsymbol{\tilde{\psi}}_m$ as
\begin{equation}
\setlength{\arraycolsep}{2pt}
\renewcommand{\arraystretch}{1.3}
\boldsymbol{\phi}_m = \left(
\begin{array}{c}
   \boldsymbol{\tilde{\phi}}_m \\
    \rho_m \boldsymbol{\tilde{\phi}}_m\\
  \rho_m^2 \boldsymbol{\tilde{\phi}}_m\ \\
   \vdots \\
   \rho_m^{n-1} \boldsymbol{\tilde{\phi}}_m \\
\end{array}  \right)
,\quad\boldsymbol{\psi}_m = \left(
\begin{array}{c}
   \boldsymbol{\tilde{\psi}}_m \\
    \rho_m \boldsymbol{\tilde{\psi}}_m\\
  \rho_m^2 \boldsymbol{\tilde{\psi}}_m\ \\
   \vdots \\
   \rho_m^{n-1} \boldsymbol{\tilde{\psi}}_m \\
\end{array}  \right).
\label{vector_3D}
\end{equation}
This shows that $m$ is actually the azimuthal wavenumber of the resolvent mode.
With this formulation, the resolvent operator does not only depend on the frequency but also $m$. This leads to an analysis in the $(\omega, m)$-domain that allows direct comparison of the resolvent gain with the energy map (see \S\ref{sec:energy}). For that reason, resolvent analyses are performed for values of $m$ corresponding to multiples of 6 to be consistent with the DNS, and gain values are displayed for a $(\omega, m)$-domain encompassing that of the energy maps (including negative values of $m$ and $\omega$). 
Thus, the interpretation of the map is the same as that presented in  figure \ref{fig:Energy_map} and \S\ref{sec:energy}.\\

\section{Results}\label{sec:results}
\newcommand{\SizeMF}{0.55}
\newcommand{\SizeDDD}{0.65}

\subsection{Low-frequency dynamics of the recirculation bubble}
\label{sec:bulle}
\begin{figure}
    \centering
    \includegraphics[width=1\linewidth]{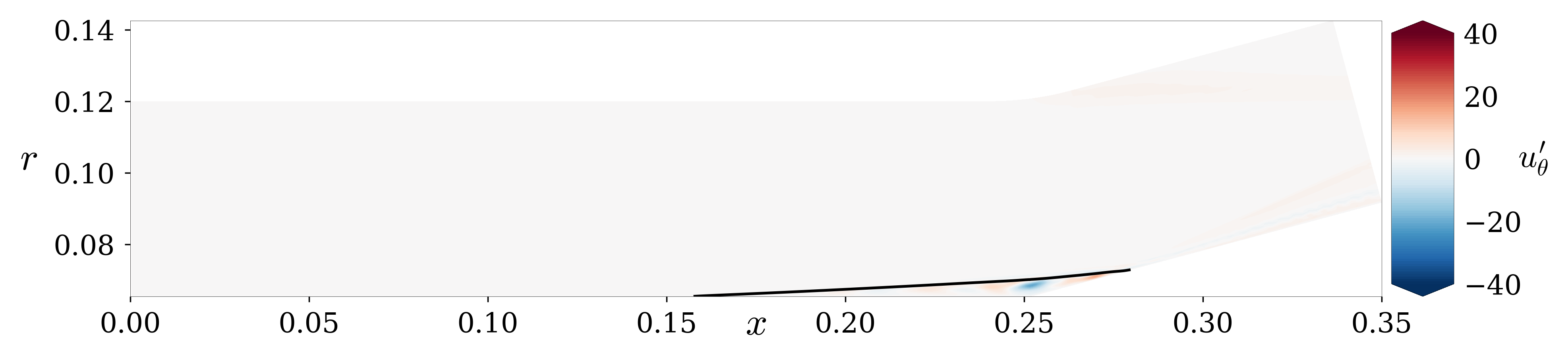}
    \caption{SPOD leading mode ($r_0>60\%$) for the full domain at $m=0$ and $f=1.5kHz$, the black line represents the limit of the recirculation bubble.}
    \label{fig:SPODBulle}
\end{figure}

The SPOD analysis of the QDNS revealed the existence of low-wavenumber modes ($m=0$ and $6$) at very low frequency. Figure \ref{fig:SPODBulle} shows that the structure of these SPOD modes corresponds to a bubble dynamics. Such modes systematically appear in the energy maps presented in the following sections. Unfortunately, due to the time duration of the QDNS, the lowest frequency that can be resolved with the SPOD is 1.5kHz, and there is no guarantee that the actual dominant frequency of these modes is not lower. Similarly, the limited azimuthal span of the simulation restricts the possible values of $m$, such that these modes may be actually related to non-zero wavenumbers below 6.  

{  
Several previous studies in incompressible  \citep{gallaire2007three,marquet2009direct} and compressible \citep{robinet2007bifurcations,hildebrand2018simulation,sidharth2018onset} flows suggest that these structures are quasi-steady modes resulting from a global instability rather than convective amplification mechanisms, which slowly breaks the axisymmetry of the mean flow in the recirculation region. Due to the strong separation of both temporal and spatial scales between these bubble modes and the elongated structures that breakdown to turbulence (visible in figure \ref{fig:QCRIT_INTRO}), it is unlikely that structures responsible for transition stem from the bubble dynamics. Therefore, it is not included in the final transition scenario proposed in this paper. The rest of the article focuses on resolvent analyses about the axisymmetric mean flow obtained from the QDNS, without accounting for the hypothetical loss of axisymmetry that might be observed on a very long QDNS spanning the whole 360 degrees domain. Note that such desymmetrization of the recirculation region has not been reported in the experimental results from \citet{benaytransitionalSWBLI}. 

Nonetheless, on a long timescale, it is possible that these modes modify the mean flow (by breaking its symmetry) in a way that affects the convective instability modes that are studied below. Addressing this question would require to perform resolvent analyses about non-axisymmetrical flow fields associated with a QDNS grid on a full 360 degrees domain. This implies computational resources that outdo by roughly two orders of magnitude those used for the largest global stability analyses existing in the literature (such as that of \citet{timme2018global}). Therefore, it is unfortunately out of reach, and thus, remains an open question for now. }

However, while not directly related to the transition scenario, characterising the bubble global instability is an interesting (and affordable) point to address in future works to better understand the full dynamics of such compression ramp flows. It would require other approaches (global stability analyses rather than resolvent computations) and longer DNS, possibly spanning the entire domain in azimuth to capture the lower wavenumbers that may exist.

\subsection{Methodology for the study of subdomains}
In the next three sections, which correspond to the study of the subdomains of interest defined in \S\ref{sec:setup}, the physical analysis follows the methodology below:
\begin{enumerate}
    \item the flow structure is qualitatively discussed based on the observation of the instantaneous vortical structures within the flow, visualised with an isosurface of constant Q-criterion extracted from the QDNS,
    \item the dynamics of the subdomain is quantitatively analysed through an energy map, as defined in \S\ref{sec:energy}, giving the distribution of the fluctuation energy in the $\omega$-$m$ domain,
    \item the structure of SPOD modes from the highest-energy parts of this map is discussed,
    \item a resolvent analysis of the subdomain is carried out, and the results are compared to the SPOD analysis.   {In particular, the energy maps are compared to maps of $\mu_0^2$ and $c_0$ to identify whether or not the high-energy structures result from a linear convective instability.}
\end{enumerate}

%
%
%
%
%
%

\subsection{Attached boundary layer}
\label{sec:BL}
Boundary layer instabilities most likely play an important role in the transition process. For that reason, this section focuses on the attached boundary layer (\textit{i.e.} the domain downstream of the interaction is discarded, focusing only on $X\in[0,0.16]$ or ${X}/{L}\in[0,0.63]$). This corresponds to the first of the three regions of interest defined in \S\ref{sec:setup}. Boundary layer profiles from the mean flow are in agreement with self similar solutions and can be found in appendix \ref{sec:profiles}.

\begin{figure}
    \centering
    \includegraphics[trim= 5cm 0.2cm 5cm 0.15cm,width=0.85\linewidth,clip=true]{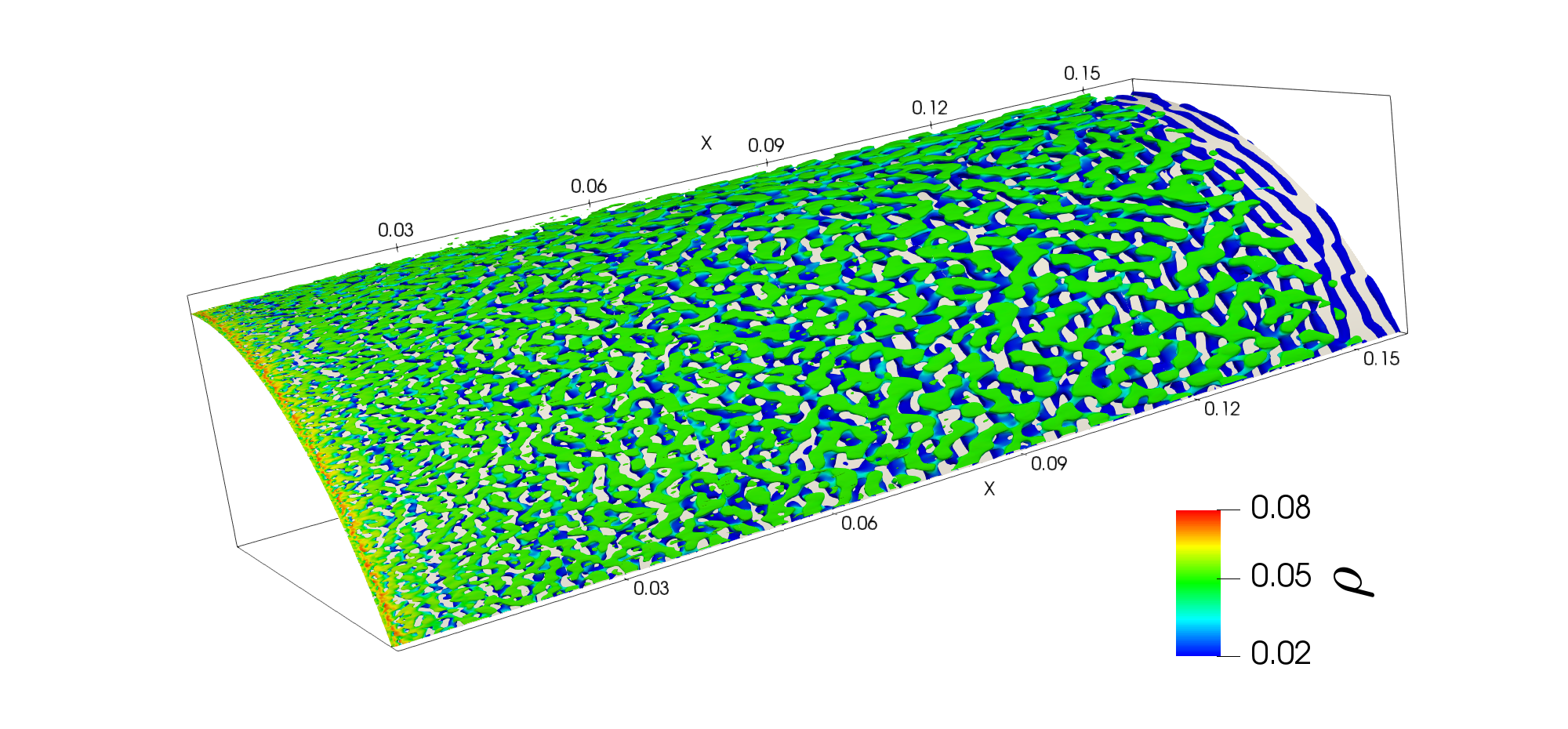}
    \caption{Isosurface of Q criterion ($Q=2\times10^{-6}{U^2}/{\delta^2}$) coloured by density for the attached boundary layer from an instantaneous snapshot of the QDNS.}
    \label{fig:QCRITBL}
\end{figure}

First of all, a surface of constant Q-criterion extracted from the QDNS is presented in figure~\ref{fig:QCRITBL}. 
The green cross-shaped patterns in the upper part of the boundary layer indicate that oblique first mode structures (both clockwise and counter-clockwise) are present. 
To a lesser extent, elongated azimuthal structures are also visible in the lower part of the boundary layer, which suggests the presence of second mode disturbances.

\begin{figure}
  \centering
  \begin{tabular}[b]{c}
    \includegraphics[trim= 0.2cm 0cm 0.2cm 0cm,width=0.45\linewidth,clip=true]{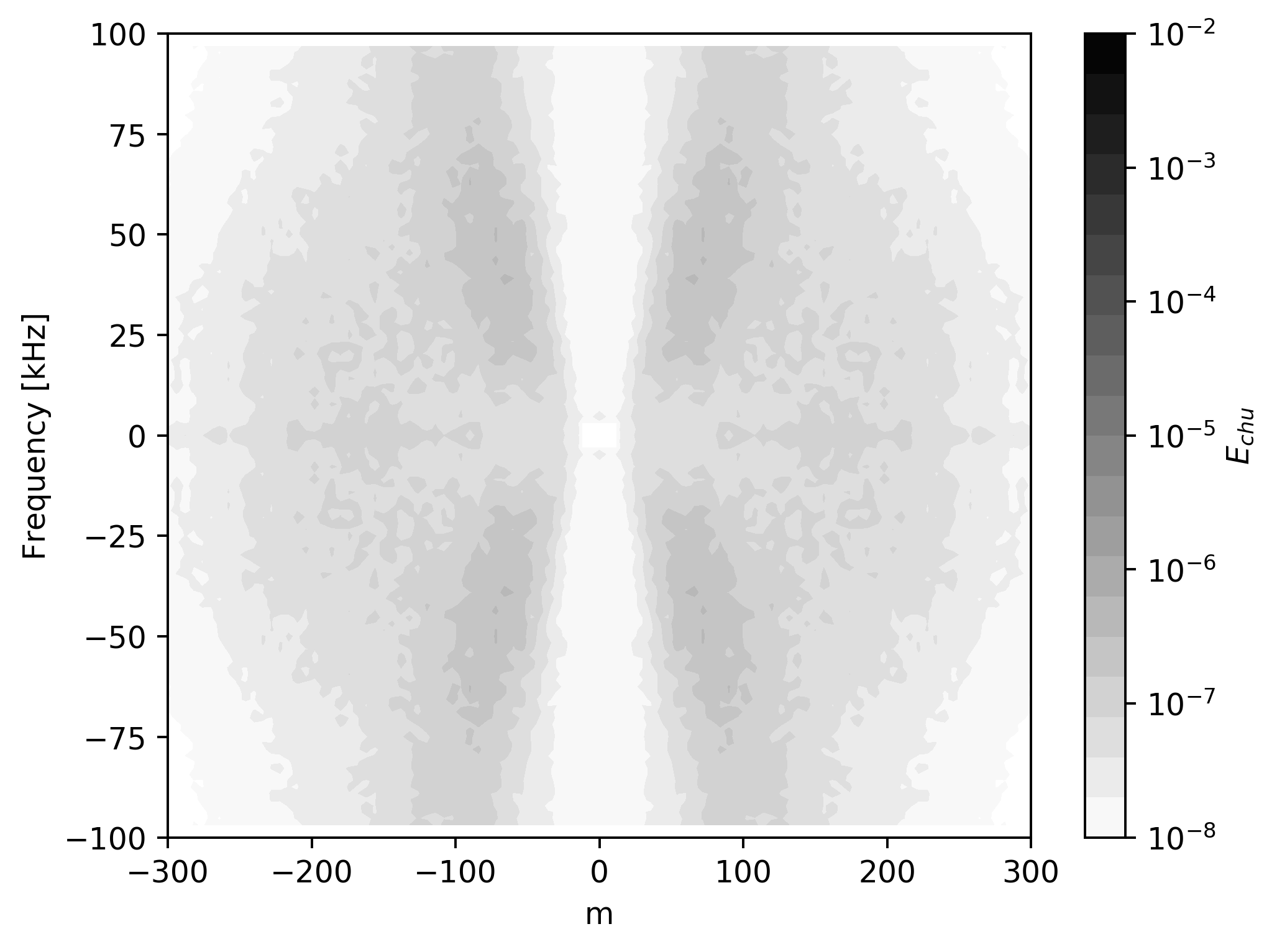}\\
    \small (a)
  \end{tabular} \qquad
  \begin{tabular}[b]{c}
  \includegraphics[trim= 0.2cm 0cm 0.2cm 0cm,width=0.45\linewidth,clip=true]{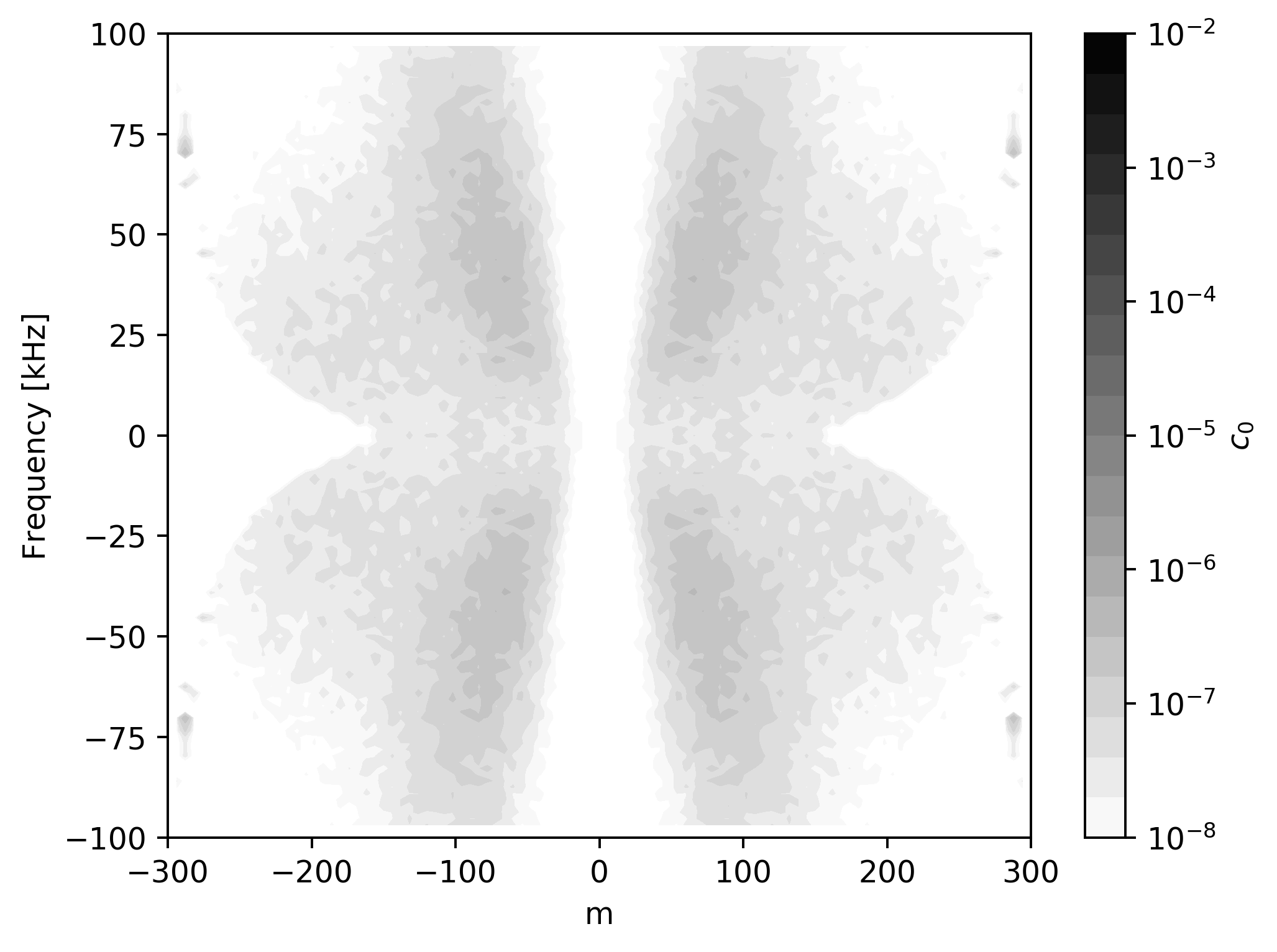}\\
    \small (b)
  \end{tabular}
  \caption{  {Map of (a) the distribution of the fluctuation energy from the QDNS (b) the $c_0$  coefficient for the dominant linear mechanism} against frequency and azimuthal wavenumber for the attached boundary layer region.}
  \label{fig:QDNSEnergyBL}
\end{figure}

The fluctuation energy distribution in the $(m,\omega)$ domain presented in figure \ref{fig:QDNSEnergyBL} confirms the importance of the first oblique modes.
Energy linked to these structures is contained in the four diagonal branches of the diagram. They represent the majority of the fluctuation energy from the boundary layer.

  {Oblique modes appear on a wide range of frequencies (from $20kHz$ to up to $100kHz$) and wavenumbers (from $20$ up to $125$) with peak amplification around $m=\pm60$ and $f=\pm40kHz$.}
The broadband nature of boundary layer instabilities shows that it was relevant to inject white noise rather than a specific forcing designed to focus on a given instability mode. It unveils the complexity of this flow, where a wide range of structures may develop, interact, and compete.

\begin{figure}
  \centering
  \begin{tabular}[b]{c}
    \includegraphics[trim=1cm 0.1cm 1cm 0.1cm,width=0.45\linewidth,clip=true]{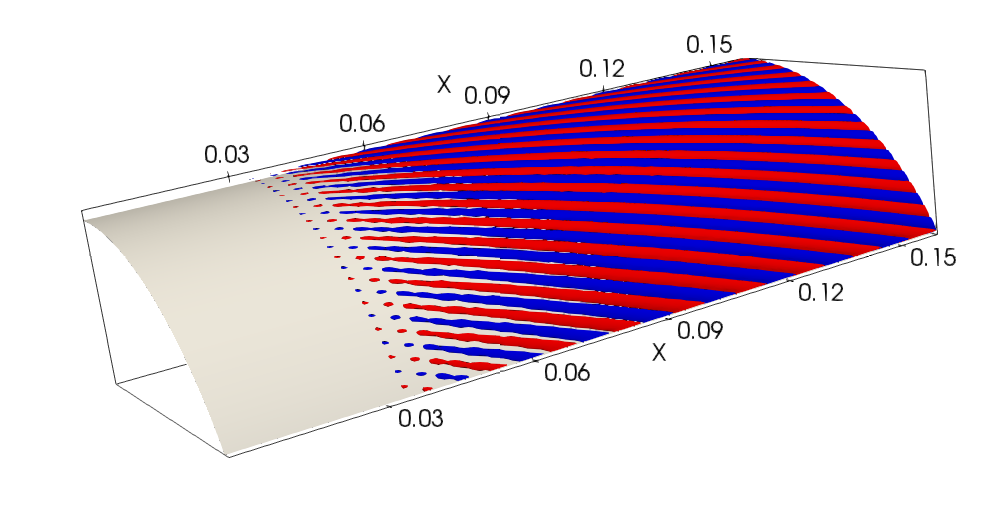}\\
    \small (a)
  \end{tabular} \qquad
  \begin{tabular}[b]{c}
    \includegraphics[trim=1cm 0.1cm 1cm 0.1cm,width=0.45\linewidth,clip=true]{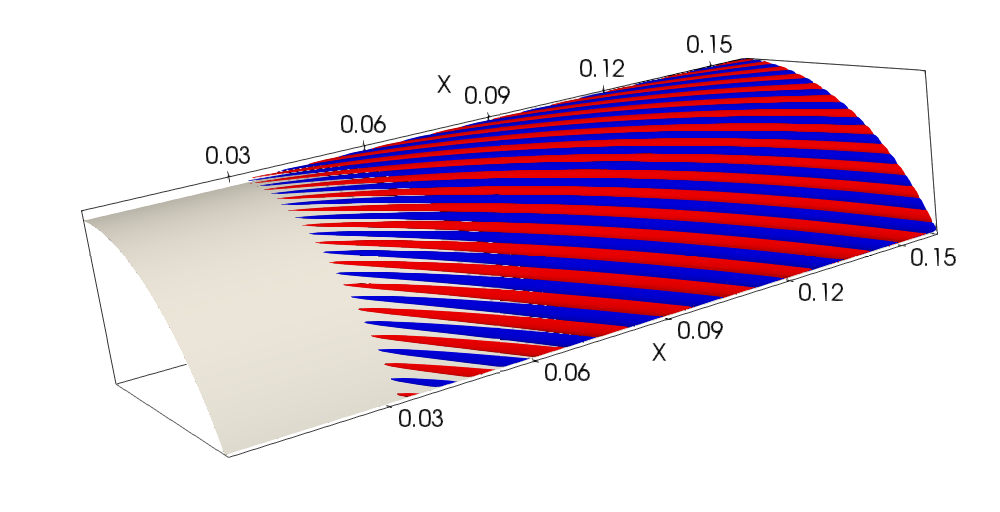}\\
    \small (b)
  \end{tabular}
  \caption{3D reconstruction (isosurface of equal positive and negative density fluctuations) of  (a) the leading SPOD mode   {($r_0>92\%$)}, (b) the optimal response, for the attached boundary layer at $m=72$ and $f=51kHz$, showing oblique first mode structures.}
  \label{fig:RESPOD_OBL}
\end{figure}

It is then interesting to look at the structure of the SPOD modes link to four of the most energetic $(m,\omega)$-points associated with the four diagonal branches.  As discussed in \S\ref{sec:energy}, two points correspond to the same clockwise SPOD mode ($m$ and $\omega$ of same sign) and the two others to the same anti-clockwise SPOD modes, which are a mirror symmetry of the former. Thus, only the structure of the clockwise leading SPOD mode is shown in figure \ref{fig:RESPOD_OBL}(a), revealing that it is indeed an oblique mode.
This leading SPOD mode accounts for more than   {$92\%$} of the energy with respect to other lesser-ranked SPOD modes at the same frequency/wavenumber. This strong dominance of the first SPOD mode occurs for the whole energetic oblique branches from figure \ref{fig:QDNSEnergyBL}(a). Thus, other modes will neither be presented nor discussed.

\begin{figure}
  \centering
  \begin{tabular}[b]{c}
    \includegraphics[trim= 0.2cm 0cm 0.2cm 0cm,width=0.45\linewidth,clip=true]{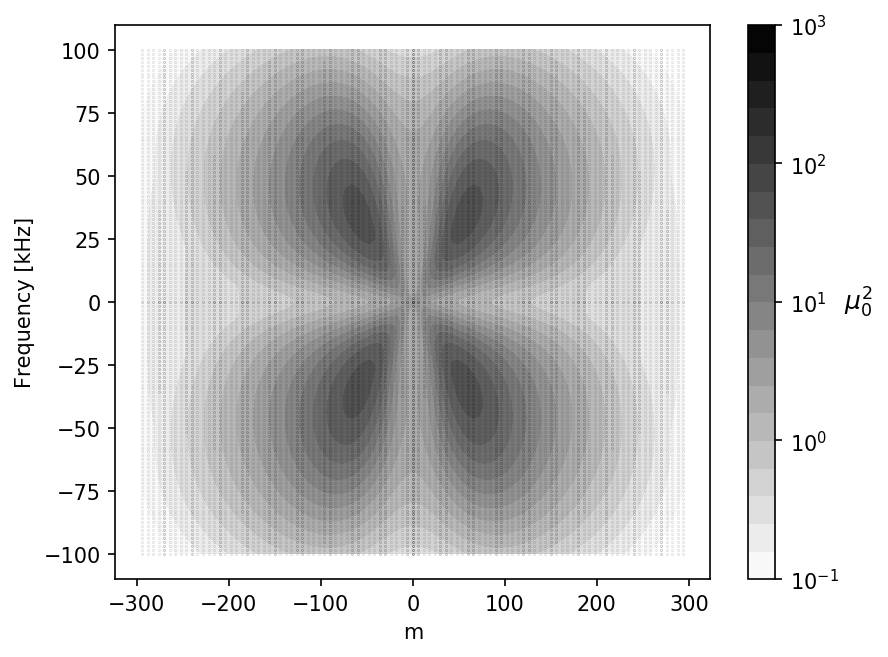}\\
    \small (a)
  \end{tabular} \qquad
  \begin{tabular}[b]{c}
  \includegraphics[trim= 0.2cm 0cm 0.2cm 0cm,width=0.45\linewidth,clip=true]{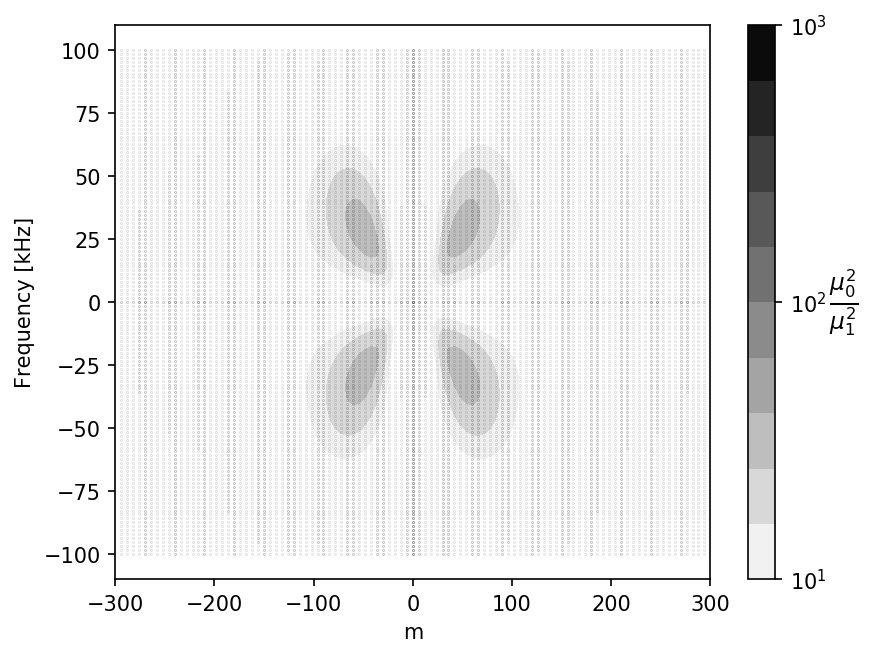}\\
    \small (b)
  \end{tabular}
  \caption{Map of (a) the gain ($\mu_0^2$) from the resolvent analysis (b) the separation between the two first eigenvalues of equation \ref{equ:eigprob} (${\mu_0^2}/{\mu_1^2}$) against frequency and azimuthal wavenumber for the boundary layer region.}
  \label{fig:gainSepBL}
      \label{fig:GainSepBL}
\end{figure}

As oblique modes are linearly amplified convective instabilities, the resolvent analysis should yield high optimal gain and a strong separation of singular values for the corresponding range of wavenumbers and frequencies (see \S\ref{sec:resolv}).
The separation ratio of the first two largest eigenvalues is presented in figure \ref{fig:gainSepBL}(b), the largest eigenvalue is at least one order of magnitude larger than the second one in the zone of interest for the study of oblique modes. As expected, this zone also displays the highest linear amplification (see figure \ref{fig:gainSepBL}(a)) such that the energy distribution of figure \ref{fig:QDNSEnergyBL}(a) is very close to the maps of figure \ref{fig:gainSepBL}. 
The frequencies and wavenumbers of highest amplification match those of energetic structures that develop in the QDNS. The optimal response at the same wavenumber and frequency than the SPOD mode of figure \ref{fig:RESPOD_OBL}(a) is shown in figure \ref{fig:RESPOD_OBL}(b), and their structures seem identical.   {This similarity may be quantified by computing an `alignment coefficient', defined as the modulus of the scalar product of the two normalised modes $|\langle \boldsymbol{\Psi} , \boldsymbol{\psi} \rangle|$ (the modes are normalised such that $\langle \boldsymbol{\Psi} , \boldsymbol{\Psi} \rangle= \langle \boldsymbol{\psi} , \boldsymbol{\psi} \rangle =1$). If this value is 1, the modes are aligned and thus represent exactly the same structure, while a null alignment coefficient means that the modes are orthogonal and have nothing in common. This is commonly used to asses the correspondence of SPOD and resolvent response modes \citep{towne2018spectral}.
The modes presented in figure \ref{fig:RESPOD_OBL} yield  $|\langle \boldsymbol{\Psi} ,  \boldsymbol{\psi} \rangle|=0.98$,}
  {which confirms that the observed oblique modes relate to a linear non-normal (convective-type) amplification mechanism, excited by the inlet white noise. }

\begin{figure}
    \centering
    \includegraphics[trim=0cm 1cm 0cm 0.25cm,width=\SizeDDD\linewidth,clip=true]{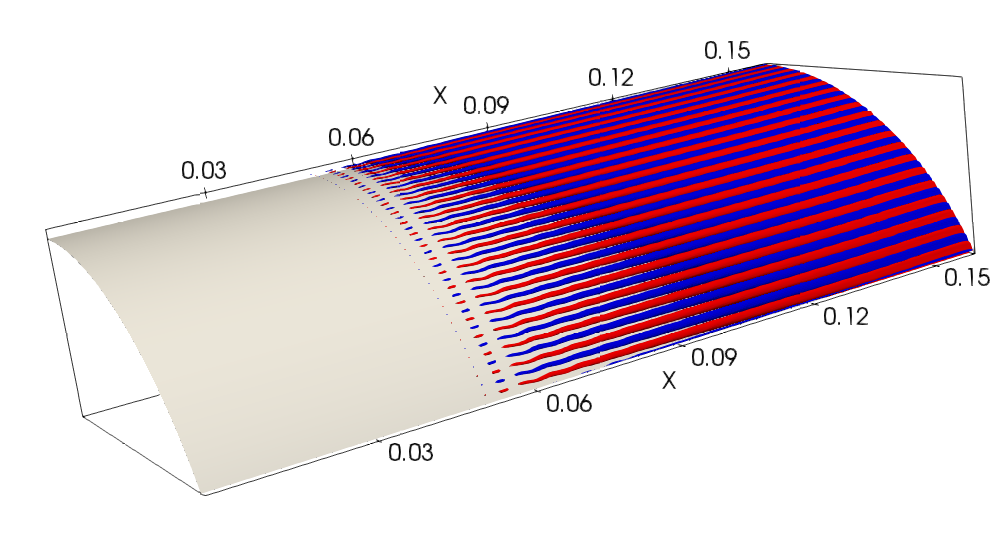}
    \caption{3D reconstruction (isosurface of equal positive and negative density fluctuations) of the leading SPOD mode   {($r_0>83\%$)} for the attached boundary layer at $m=120$ and $f=1.5kHz$, showing quasi-stationary streamwise streaks.}
    \label{fig:PODCLSTREAK}
\end{figure}

Other structures than oblique modes appear in the boundary layer: a wide zone of less-energetic fluctuations is visible in figure \ref{fig:QDNSEnergyBL} close to the $\omega=0$ axis. 
The SPOD analysis reveals that it corresponds to elongated streamwise structures, that will be called streaks (see figure \ref{fig:PODCLSTREAK}).
Note that these streaks are barely visible in the Q criterion isosurface   {as it does not allow for the visualisation of velocity deficit}.

As there is no corresponding zone of amplification in the gain map (figure \ref{fig:gainSepBL}(a)), these structures do not stem from a strong linear amplification mechanism excited by the inlet white noise   and may come from either a weak linear mechanism strongly excited by the injected noise (receptivity), or a non-linear interaction.
As explained in section \ref{sec:resolv}, the former hypothesis may easily be discarded by computing the coefficient $c_0$ which is presented figure \ref{fig:QDNSEnergyBL}(b). For the $(\omega,m)$ couple linked to oblique modes, the $c_0$ map is very close to the $E_{chu}$ map, however, for those linked  to streaks, this coefficient is several orders of magnitude lower than the energy present in the QDNS.
This result was actually expected since the apparition of these streaks relates to a well-known non-linear mechanism. Since the DNS investigations for a Mach 1.6 flat plate boundary layer by \citet{thumm1991numerische,fasel1991direct,fasel1993direct}, it is known that a pair of oblique modes of opposite spanwise wavenumbers (\textit{e.g.} the clockwise rotating mode $(\omega,m)$ and the counterclockwise rotating one $(\omega,-m)$)
can interact non-linearly to generate streamwise stationary streaks $(0, 2m)$. Given the symmetry of the modes presented in section \ref{sec:energy}, modes of opposite frequency $(\omega,m)$ and $(-\omega,m)$ have the same interaction toward $(0, 2m)$.
This explains the diffused energetic zones between the opposed frequency oblique branch of the energy map (figure \ref{fig:QDNSEnergyBL}(a)), which result from the non-linear interaction of these branches. This interaction is linked to an oblique breakdown mechanism that has been widely studied, mostly for flat-plate boundary layer, for example by   {\citet{fasel1993direct}}, \citet{chang1994oblique}, \citet{sandham1995direct},  or more recently \citet{franko_lele_2013,leleAdverse}. 
  {Here, this non-linear interaction is causing the birth of low-energy streaks in the boundary layer which will play an important role in the transition process, as shown in the next section. }

\begin{figure}
  \centering
  \begin{tabular}[b]{c}
   \includegraphics[width=0.45\linewidth]{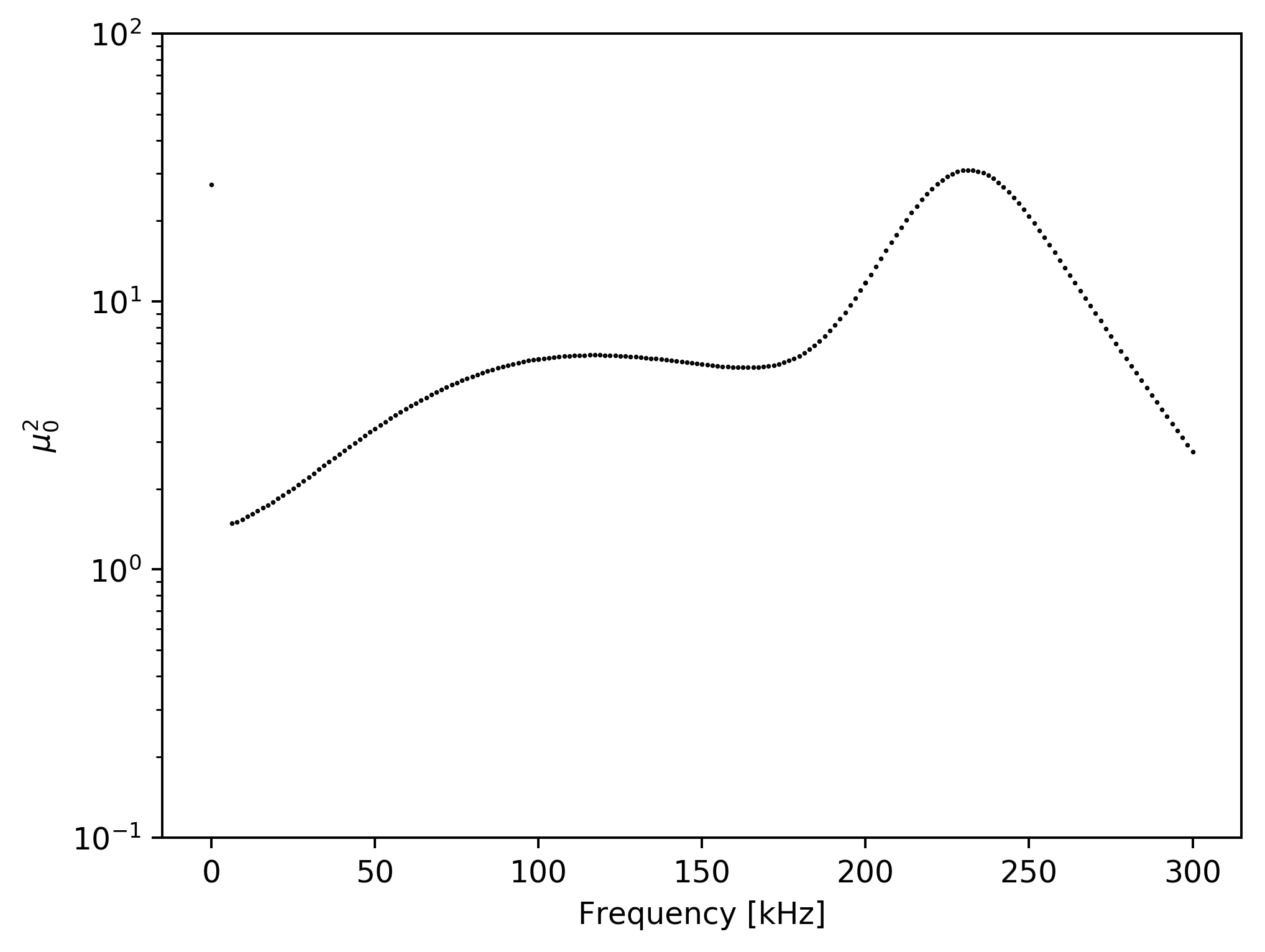}\\
    \small (a)
  \end{tabular} \qquad
  \begin{tabular}[b]{c}
  \includegraphics[width=0.45\linewidth]{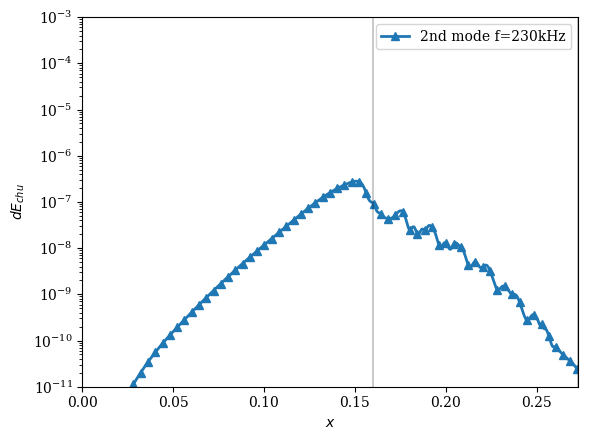}\\
    \small (b)
  \end{tabular}
  \caption{(a) resolvent gain against frequency for $m=0$ showing the second mode peak  (b) streamwise distribution of the energy predicted by the resolvent analysis for the second mode. The grey line represents the limit between the boundary layer and the mixing layer region. The amplitude of the linear prediction is arbitrary and only the longitudinal evolution should be considered.}
     \label{fig:DampingSM}
\end{figure}

\begin{figure}
    \centering
    \includegraphics[width=0.85\linewidth]{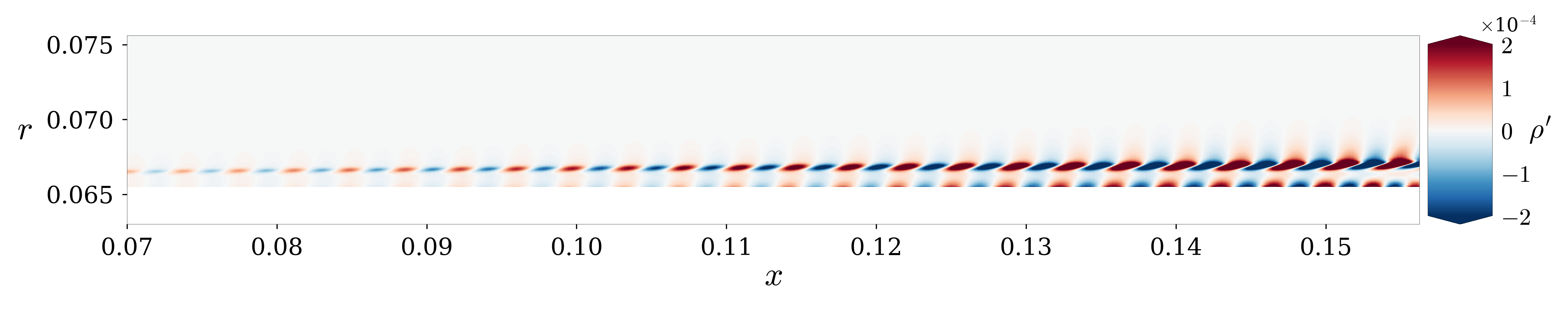}
    \caption{Optimal response (density fluctuation) for the attached boundary layer at $m=0$ and $f=230kHz$, showing second Mack mode structures.}
    \label{fig:resolvCL}
\end{figure}

Lastly, axisymmetrical ($m=0$) second mode disturbances are also predicted by the resolvent analysis (  {see figure \ref{fig:DampingSM}(a) which presents the resolvent gain against frequency for axisymetrical structures $m=0$}), with a local peak amplification at $f=230kHz$ and a maximal associated gain less than half that of the most amplified oblique first mode (see figure \ref{fig:gainSepBL}(a) for the gain of the most amplified first mode). The optimal response associated with this maximal gain value is presented in figure \ref{fig:resolvCL} and displays the typical structure of second mode disturbances \citep{laurence2016experimental,bugeat20193d}. The low-intensity, high-frequency peak from the Power Spectral Densities (PSD) presented in figure \ref{fig:PSD_capt} is a sign of those weaker instabilities. The peak visible on the green curve (${X}/{L}=0.588$, near the end of the boundary layer region) around $f=230kHz$ confirms that the most amplified frequencies in the QDNS are matching the resolvent prediction. 

  {Figure \ref{fig:DampingSM}(b) presents the streamwise distribution of the energy of second mode instabilities as predicted by the resolvent analysis. The quantity $dE_{chu}$ is computed by integrating the local Chu energy contribution of the optimal response mode along the gridlines in the wall-normal direction (the gridlines are not exactly perpendicular to the wall in the cylinder-flare junction zone due to the mesh construction) for each streamwise location. This is similar to what has been done by \citet{article}, or more recently by \citet{bugeat2017stabilite} and \citet{bugeat20193d}. It is important to note here that the energy level presented in this figure relates to an eigenmode, which is defined up to a multiplicative constant by construction.
Figure \ref{fig:DampingSM}(b) shows that the second mode instability is strongly damped as soon as it enters the separated region, such that at the end of the mixing layer, and thus before the transition, it has reached negligible levels. This explains why past the separation point, the PSDs from figure \ref{fig:PSD_capt} do not display any high-frequency peak. Therefore, the second mode is not considered to play a role in the transition scenario since it is virtually absent from the flow at the transition location.
Note that \citet{marxen2010disturbance} showed that lower frequency second mode instabilities might be further amplified in the separated region. 
However, in the present case, even if some of the lower frequency second mode instability are less damped, none of them are further amplified downstream of the interaction and thus all of them become several orders of magnitude less energetic than other modes.
This difference with the results of \citet{marxen2010disturbance} is probably due to the fundamental topology difference between the separated region as the recirculation region studied here is massive.
Because of the damping of second mode instabilities, the computational and storage cost linked to a higher sampling frequency for the QDNS snapshots (that would allow the extraction of SPOD mode for second mode instabilities) is unnecessary and it was chosen as previously explained to use a sampling frequency of $200kHz$.}

  {The resolvent results in the boundary layer are consistent with the LST results in the literature. Even if the second mode can be locally more amplified than the first mode for a Mach 5 boundary layer, the present results should be compared with integrated values such as the N factor, which is often higher for the oblique mode due to the larger instability domain of the first mode \citep{adams1993numerical}: the first mode gets amplified earlier than the second mode, resulting in a higher amplitude despite a possible locally weaker amplification.}

To conclude, oblique first mode structures are dominant in the boundary layer. They are due to a   {convective-type} non-normal linear mechanism and are perfectly predicted by the resolvent analysis.
They interact non-linearly to create low-energy quasi-steady streaks.
Second mode instabilities are also present in the QDNS and predicted by the resolvent analysis, but of much lower energy.

%
%
%
%
%
%
%
\subsection{Mixing layer}
\label{sec:ML}
As it encounters the separation shock, the boundary layer is subject to an adverse pressure gradient and separates, creating a recirculation bubble. A mixing layer appears between the high-speed flow outside of the recirculation region and the reversed flow inside of it, changing the topology of the flow and marking the entry in the second region defined in \S\ref{sec:setup}.
This section focuses on the portion of the flow that is downstream of the separation shock but upstream of the reattachment point, \textit{i.e.} $X\in[0.16,0.28]$ or ${X}/{L}\in[0.63,1.11]$. As such, as explained in \S \ref{sec:resolv}, the resolvent analysis spatially constrains the response to this region, but not the forcing, in order to account for the amplification of structures that have developed in the boundary layer.

\begin{figure}
    \centering
    \includegraphics[trim= 4cm 0.15cm 4cm 0.15cm,width=0.85\linewidth,clip=true]{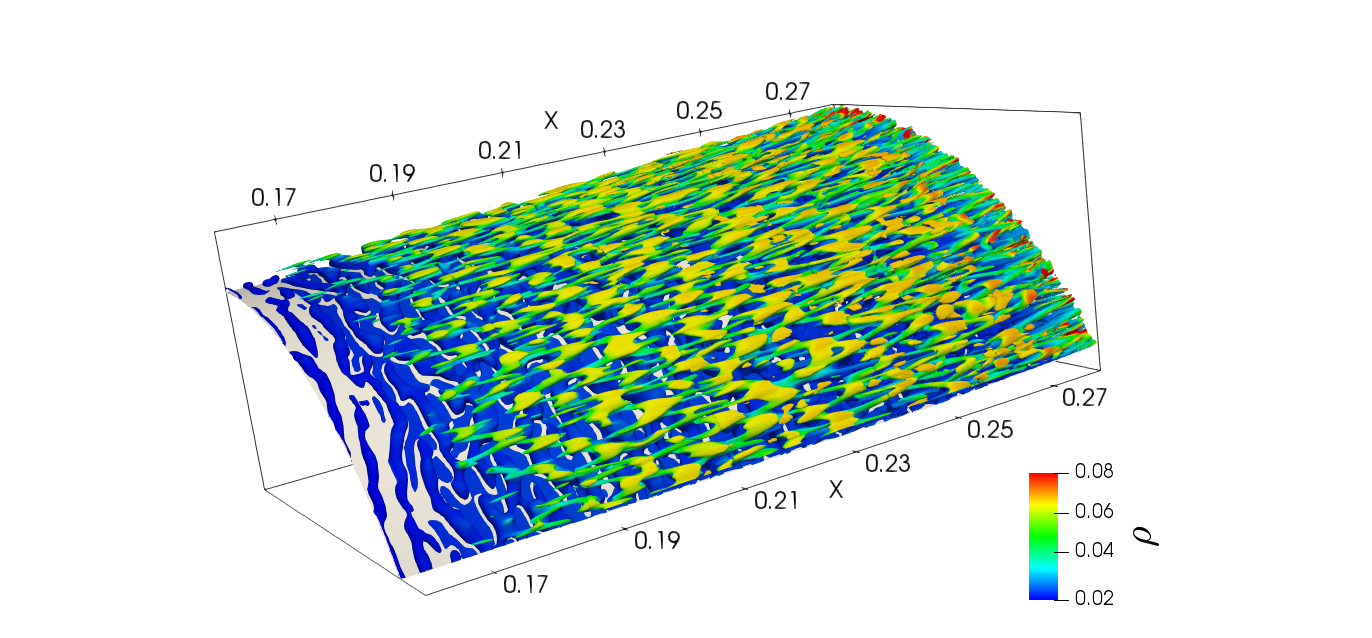}
    \caption{Isosurface of Q criterion ($Q=9\times10^{-6} {U^2}/{\delta^2}$) coloured by density for the mixing layer from an instantaneous snapshot of the QDNS.}
    \label{fig:QCRITML}
\end{figure}

Figure \ref{fig:QCRITML} presents an isosurface of Q criterion coloured by density for the mixing layer. Oblique mode structures are still visible, but they have a larger wavelength than those of the boundary layer. Quasi-axisymmetrical structures can also be seen near the wall and are either linked to the bubble dynamics (see \S \ref{sec:bulle}) or to convected second mode structures. 

There are fewer small structures at the beginning of the mixing layer than at the end of the boundary layer, which may be seen more clearly in figure \ref{fig:QCRIT_INTRO}. This indicates that the separation region damps high-frequency instabilities coming from the boundary layer. 

\begin{figure}
  \centering
  \begin{tabular}[b]{c}
    \includegraphics[trim= 0.2cm 0cm 0.2cm 0cm,width=0.45\linewidth,clip=true]{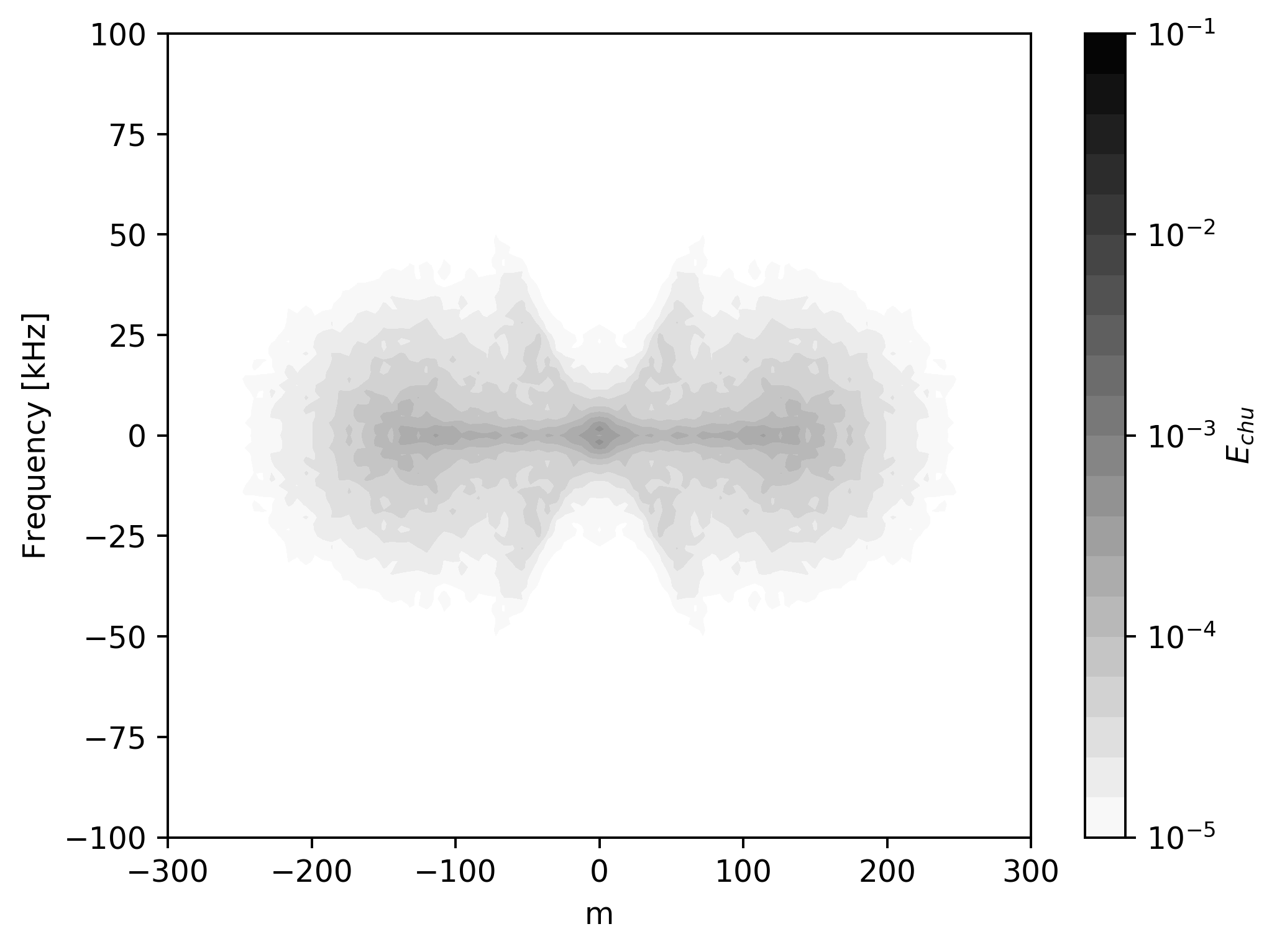}\\
    \small (a)
  \end{tabular} \qquad
  \begin{tabular}[b]{c}
  \includegraphics[trim= 0.2cm 0cm 0.2cm 0cm,width=0.45\linewidth,clip=true]{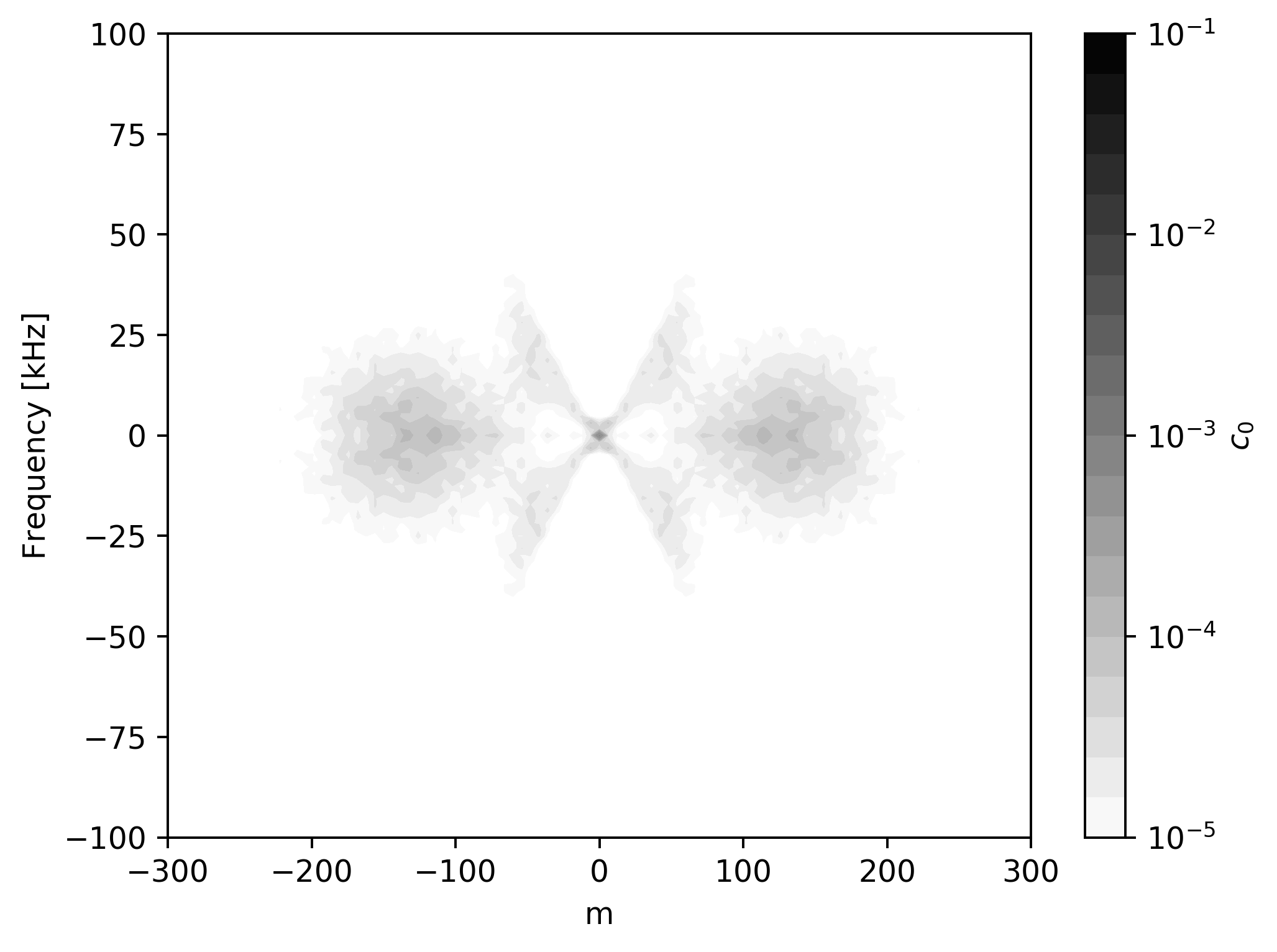}\\
    \small (b)
  \end{tabular}
  \caption{  {Map of (a) the distribution of the fluctuation energy from the QDNS (b) the $c_0$  coefficient for the dominant linear mechanism} against frequency and azimuthal wavenumber for the mixing layer region.}
     \label{fig:QDNSEnergyML}
\end{figure}

Figure \ref{fig:QDNSEnergyML} (a) presents the fluctuation energy distribution in the mixing layer. Consistently with the Q-criterion results, oblique modes are still present in the flow but they are of lower frequency/wavenumber than in the boundary layer region. The four oblique branches are indeed still visible but shorter: only structures below $f=40$kHz/$m=100$ have been amplified, which confirms the qualitative observation that the separation region filters a large part of high frequency fluctuations. Moreover, the oblique modes are not dominant anymore, due to the appearance of high-energy streaks around $f=0$Hz for wavenumbers up to $|m|\approx200$.

\begin{figure}
  \centering
  \begin{tabular}[b]{c}
    \includegraphics[trim= 0.2cm 0cm 0.2cm 0cm,width=0.45\linewidth,clip=true]{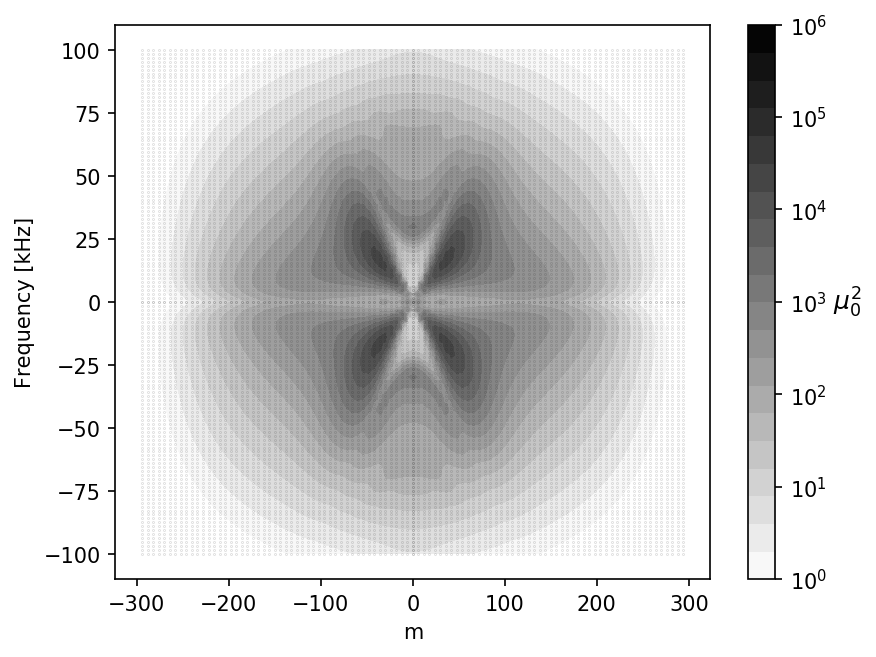}\\
    \small (a)
  \end{tabular} \qquad
  \begin{tabular}[b]{c}
  \includegraphics[trim= 0.2cm 0cm 0.2cm 0cm,width=0.45\linewidth,clip=true]{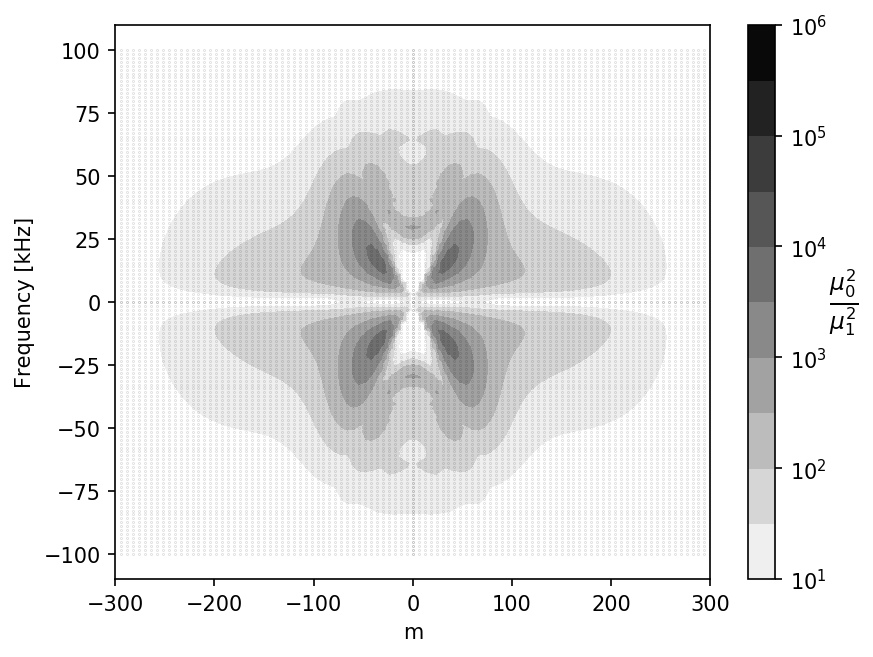}\\
    \small (b)
  \end{tabular}
  \caption{Map of (a) the gain ($\mu_0^2$) from the resolvent analysis (b) the separation between the two first eigenvalues of equation \ref{equ:eigprob} (${\mu_0^2}/{\mu_1^2}$) against frequency and azimuthal wavenumber for the mixing layer region.}
  \label{fig:gainSepML}
\end{figure}

\begin{figure}
  \centering
  \begin{tabular}[b]{c}
    \includegraphics[trim=1cm 0.1cm 1cm 0.1cm,width=0.45\linewidth,clip=true]{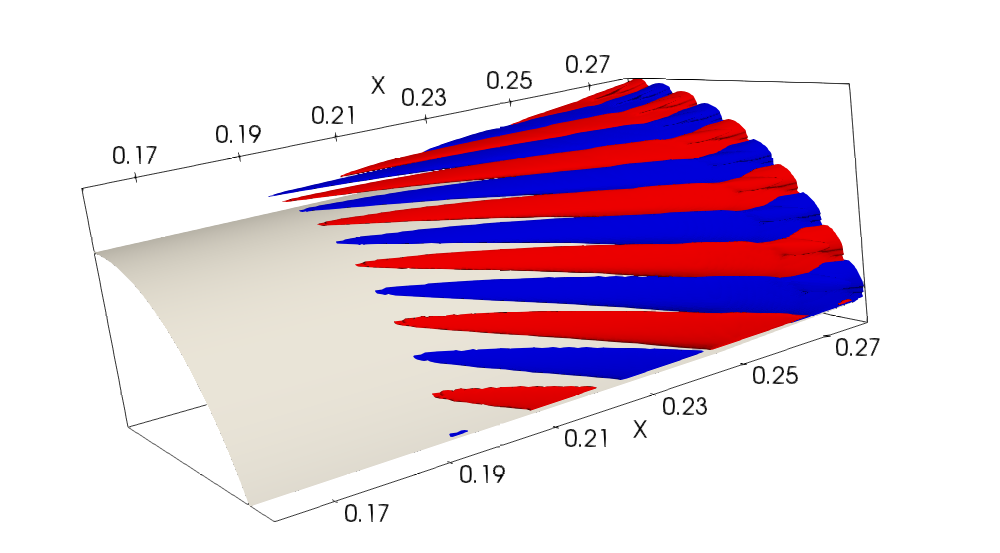}\\
    \small (a)
  \end{tabular} \qquad
  \begin{tabular}[b]{c}
    \includegraphics[trim=1cm 0.1cm 1cm 0.1cm,width=0.45\linewidth,clip=true]{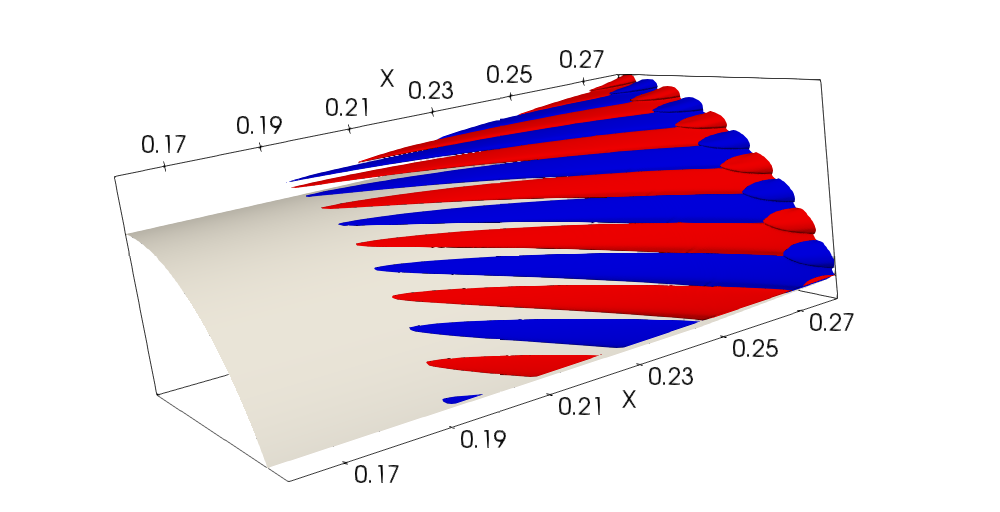}\\
    \small (b)
  \end{tabular}
  \caption{3D reconstruction (isosurface of equal positive and negative density fluctuations) of  (a) the leading SPOD mode ($r_0>87\%$), (b) the optimal response  of the mixing layer at $m=30$ and $f=15kHz$, showing oblique first mode structures.}
  \label{fig:RESPOD_OBL_ML}
\end{figure}

Then, two points need to be addressed: the filtering of high-frequency oblique modes, and the appearance of high-energy quasi-steady streaks. The former point may be straightforwardly explained by the resolvent analysis. Figure \ref{fig:gainSepML} presents the results of the analysis for the mixing layer. Overall the results are comparable to what was observed in the boundary layer: oblique modes are still dominant, but compared to the boundary layer, the mixing layer mainly amplifies structures below $f=50$kHz. Physically, this may be caused by the sudden increase of equivalent boundary layer thickness due to the separation, leading to a weaker wall-normal velocity gradient in this region. Therefore, the filtering property of the separation region is the consequence of the abrupt change of the topology of the flow, which shifts the frequency range of the linear amplification mechanisms towards lower values. Consequently, high-frequency structures coming from the boundary layer, which have transferred a part of their energy non-linearly to streamwise structures (see \S\ref{sec:BL}) are not as amplified as they were in the boundary layer. Meanwhile, lower frequency oblique structures, such as presented in figure \ref{fig:RESPOD_OBL_ML}, start to be more strongly amplified when entering the separation region. Thus, the resolvent analysis yields again consistent explanations concerning oblique modes in the mixing layer. 
  {Once again, the alignment coefficient between the most amplified oblique mode and the corresponding SPOD mode such as presented figure \ref{fig:RESPOD_OBL_ML} is high: $|\langle \boldsymbol{\Psi} , \boldsymbol{\psi} \rangle| =0.84$, which shows that the SPOD oblique modes match their resolvent counterpart.
}

\begin{figure}
  \centering
  \begin{tabular}[b]{c}
     \includegraphics[trim=0cm 0cm 0cm 0cm,width=0.45\linewidth,clip=true]{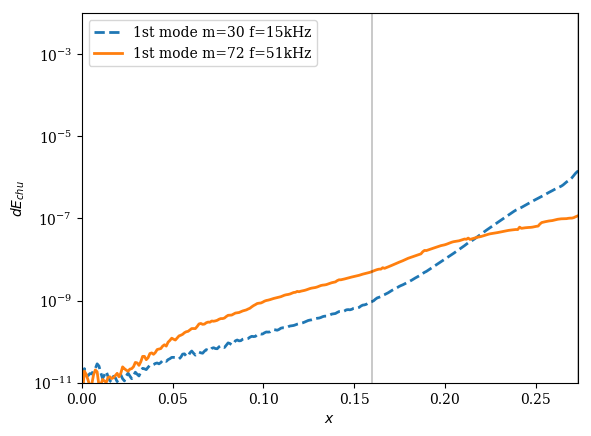}\\
    \small (a)
  \end{tabular} \qquad
  \begin{tabular}[b]{c}
    \includegraphics[trim=0cm 0cm 0cm 0cm,width=0.45\linewidth,clip=true]{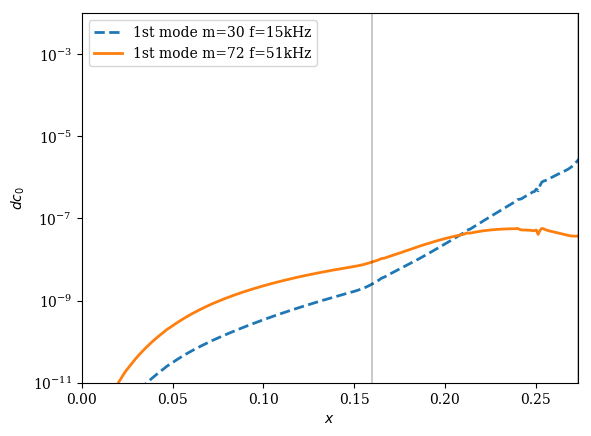}\\
    \small (b)
  \end{tabular}
  \caption{  {Streamwise distribution of the energy of two oblique modes of interest. Panel (a) shows the energy evolution for the SPOD modes extracted from the QDNS, (b) shows the energy evolution predicted by the resolvent analysis, the grey line represents the limit between the boundary layer and the mixing layer region. $dE_{chu}$ and $dc_0$ can be quantitatively compared.}}
  \label{fig:EvoluttionML}
\end{figure}

  {
Figure \ref{fig:EvoluttionML} presents the streamwise distribution of the energy of two oblique modes of interest both for (a) the SPOD mode in the QDNS and (b) the linear prediction by the resolvent analysis. In the same way as what was presented in figure \ref{fig:DampingSM}, the quantities $dE_{chu}$ is computed by integrating the local Chu energy contribution along the gridlines in the wall-normal direction for each streamwise location. Regarding the resolvent mode, the streamwise energy distribution  $dc_0$ is computed in a similar way from the optimal response normalised based on $c_0$ (i.e. receptivity is taken into account) such that the curves from \ref{fig:DampingSM}(a) can be quantitatively compared to \ref{fig:DampingSM}(b).
}{  Figure \ref{fig:EvoluttionML}(b) confirms the frequency filtering in the separated region: the linear amplification of low-frequency oblique modes (blue dashed line) becomes significantly stronger than that of high-frequency modes (plain orange line) in the separated region. This result is consistent with the observation from \citet{marxen2010disturbance}. 
However, unlike \citet{marxen2010disturbance}, there is a good agreement between the predicted linear energy growth of oblique modes in the separated region (figure \ref{fig:EvoluttionML}(b)) and the actual growth in the QDNS (figure \ref{fig:EvoluttionML}(a)), showing the advantage of global resolvent analysis against LST. 
This is due to the ability of the present linear stability study to account for both non-parallel effects and component-type non-normalities, which are the two main limitations of the study of \citet{marxen2010disturbance} (the non-parallel effect most probably  being the main cause of error for the oblique modes in the separated region).
Finally, a similar amplification of the 2D first mode in the separated region to that discussed by \citet{marxen2010disturbance} was also observed in the present study. It is not presented here as it is several orders of magnitude less energetic than the oblique modes.
}

  {
Regarding the energetic streaks observed in figure \ref{fig:QDNSEnergyML}(a) they become one of the most energetic structures in the mixing layer. This cannot be explained by the resolvent gain alone, which displays low values for quasi-steady structures (figure \ref{fig:gainSepML}(a)). This means that existing linear mechanisms that may generate streaks are very weak. }

\begin{figure}
  \centering
  \begin{tabular}[b]{c}
     \includegraphics[trim=1cm 0.1cm 1cm 0.1cm,width=0.45\linewidth,clip=true]{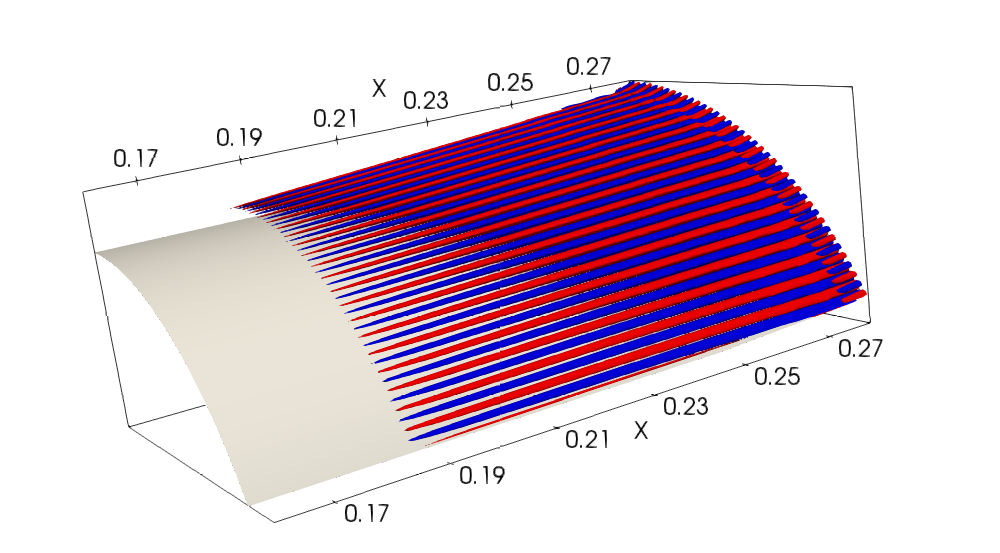}\\
    \small (a)
  \end{tabular} \qquad
  \begin{tabular}[b]{c}
    \includegraphics[trim=1cm 0.1cm 1cm 0.1cm,width=0.45\linewidth,clip=true]{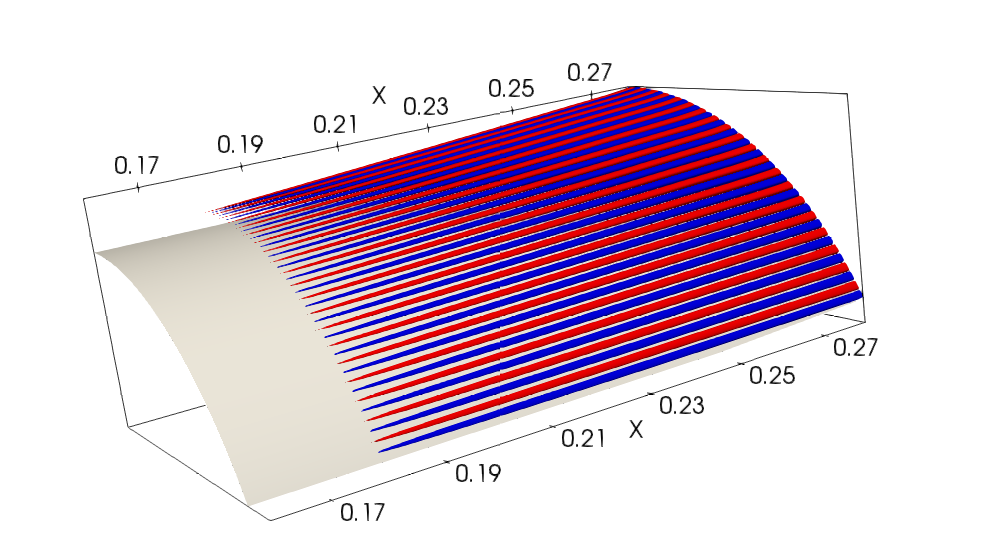}\\
    \small (b)
  \end{tabular}
  \caption{3D reconstruction (isosurface of equal positive and negative density fluctuations) of  (a) the leading SPOD mode ($r_0>83\%$), (b) the optimal response  of the mixing layer at $m=120$ and $f=1.5kHz$, showing similar quasi-stationary streamwise streaks.}
  \label{fig:PODMLSTREAK}
\end{figure}

Yet, the leading SPOD modes corresponding to the streaks, which dominate the flow for low frequencies, have a structure similar to the linear optimal responses at the same frequency/wavenumber. An example is presented in figure \ref{fig:PODMLSTREAK} for $m=120$ and $f=1.5kHz$.   {The two modes are globally similar, except for some small structures in the SPOD mode around $X=0.26$, which are absent from the resolvent mode. This may be due to intermittent turbulent spots (the intermittency function is no longer zero in this region, see figure \ref{fig:intemrittency})) or to the presence of a nearby shock. The alignment coefficient is $|\langle \boldsymbol{\Psi}, \boldsymbol{\psi} \rangle|=0.66$, which is lower than what was observed in the previous region of the flow.} 
It is nonetheless still rather high and may indicate that the streaks are due to the weak linear amplification mechanisms mentioned above, that may lead to high-energy structure through receptivity processes. 
  {This may be investigated by computing the $c_0$  coefficient distribution in the $(\omega,m)$-domain (see \S\ref{sec:resolv}), and comparing to the energy distribution of the fluctuations. Note that $c_0$ and $E_{chu}$ are homogeneous quantities that can be quantitatively compared (see \S\ref{sec:resolv}).}

\begin{figure}
    \centering
    \includegraphics[trim=3cm 3cm 3cm 0.15cm,width=0.9\linewidth,clip=true]{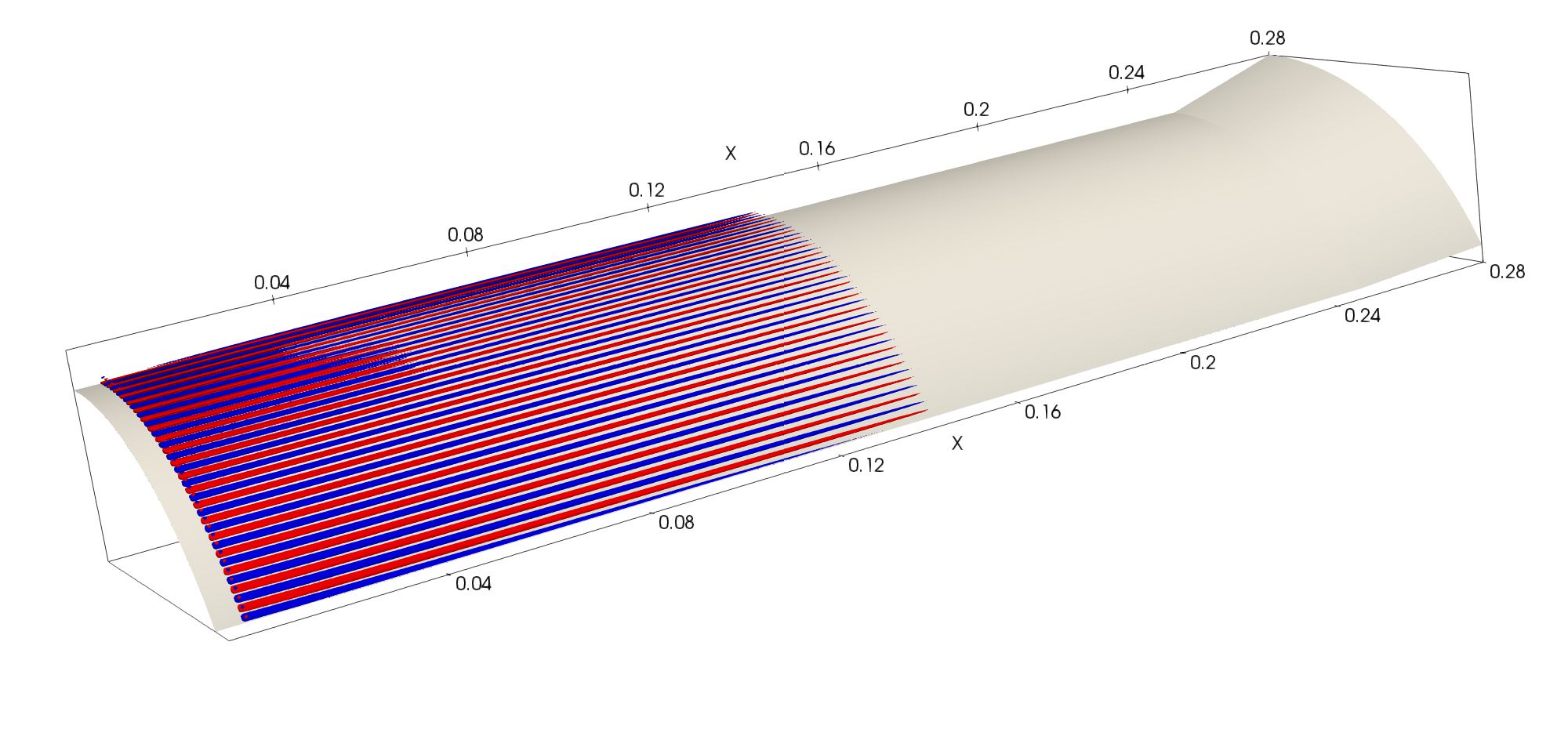}
    \caption{3D reconstruction of an optimal forcing (isosurface of equal positive and negative density   {forcing}) at $f=1.5kHz$ and $m=120$ corresponding to linear amplification of streaks developing due to the non-linear forcing in the mixing layer. Boundary layer region is shown as the forcing is mainly located upstream of the mixing layer. }
    \label{fig:STREAKSFORCML}
\end{figure}

Figure \ref{fig:QDNSEnergyML} (b) presents the resulting projection map. Compared to the gain map (figure \ref{fig:gainSepML}(a)), a high-energy region appears   {for the streaks up to $|m|=200$ and low frequency. 
It is interesting to note that because of these discrepancies, contrary to the gain map, the $c_0$ coefficient map is very similar to the fluctuation energy map form the QDNS.} 
{  
This proves that the quasi-steady streaks are mainly the result of a weak linear mechanism, which is strongly excited either by the injected noise or by the non-linear forcing. As a result, due to this selective nature of the excitation, quasi-steady streaks becomes as energetic as oblique modes, despite their much lower amplification gain.
The particular role of the inlet noise in this receptivity process may be ruled out by computing the coefficient $c_r$ (see \S\ref{sec:resolv}), which was found several orders of magnitude smaller than $c_0$ and $E_{chu}$. Therefore, these structures are driven by the nonlinear forcing rather than directly created by the inlet noise. In particular, the nonlinear structures created upstream in the boundary layer play an important role. This may be underpinned
by computing the optimal forcing (figure \ref{fig:STREAKSFORCML}), which is located in the upstream part of the boundary layer. Additionally, even though they are not homogeneous to a forcing, it is interesting to observe that the elongated structures created in the boundary layer (figure \ref{fig:PODCLSTREAK}) turn out to be very close to the optimal forcing exciting streaks in the mixing layer (figure \ref{fig:STREAKSFORCML}).
Finally, the $c_0$-map reveals that above $|m|=200$, the streaks are almost not excited anymore, which is consistent with the QDNS results where structures with $|m|>200$ display very low levels of energy.}

  {A linear streak growth  triggered by the non-linear interaction of oblique modes was already observed for supersonic boundary layer by \citet{laible2016continuously}, who concluded that the non-linear interaction of oblique modes acted as an "actuator" that forces component-type non-normal growth of the streaks, in the same way as it was described by \citet{schmid1992new} for incompressible channel flow.
This mechanism is known as one of the fastest ways to transition in attached boundary layers according to the studies of \citet{franko_lele_2013}.
For the present configuration, the separation induces an even stronger non-normal growth of the streaks than in the boundary layer, making this scenario even more relevant.}

To conclude, in the mixing layer, high-frequency oblique modes from the boundary layer have transferred their energy to streaks via the non-linear interaction described before and are, at best, only weakly amplified due to the effective thickening of the boundary layer linked with separation. Consequently, their relative intensity becomes very low. Some oblique modes of lower frequency continue to be linearly amplified and are thus present in the flow. But the most important finding is that the structures created by the non-linear interaction in the boundary layer are actually close to the optimal forcing that generates streaks in the mixing layer. Therefore, the flow in this region is dominated by quasi-steady streaks for wavenumbers up to $|m|\approx200$. Non-linear interactions of oblique modes in the mixing layer may also contribute to the appearance of these streaks, although probably to a lesser extent.

%
%
%
%
%
%
%

\subsection{Reattachment}
\label{sec:RE}
At some point on the flare, the mixing layer compresses, the flow reattaches, and the separation bubble no longer exists. That marks the entry in the third region of interest as defined in \S\ref{sec:setup}, the reattachment region. This is where the heat fluxes are usually the highest, particularly in transitional cases. This region also contains the most energetic fluctuations.
The reattachment region is studied in a similar way as the two previous regions by focusing on the region downstream of the reattachment (\textit{i.e.} $X\in[0.28,0.35]$ or ${X}/{L}\in[1.11,1.39]$).
Once again, for the resolvent analysis, the energy norm for the response only accounts for the reattachment region, but the forcing is not constrained.

\begin{figure}
    \centering
    \includegraphics[trim= 14cm 0.1cm 10cm 0.1cm,width=0.80\linewidth,clip=true]{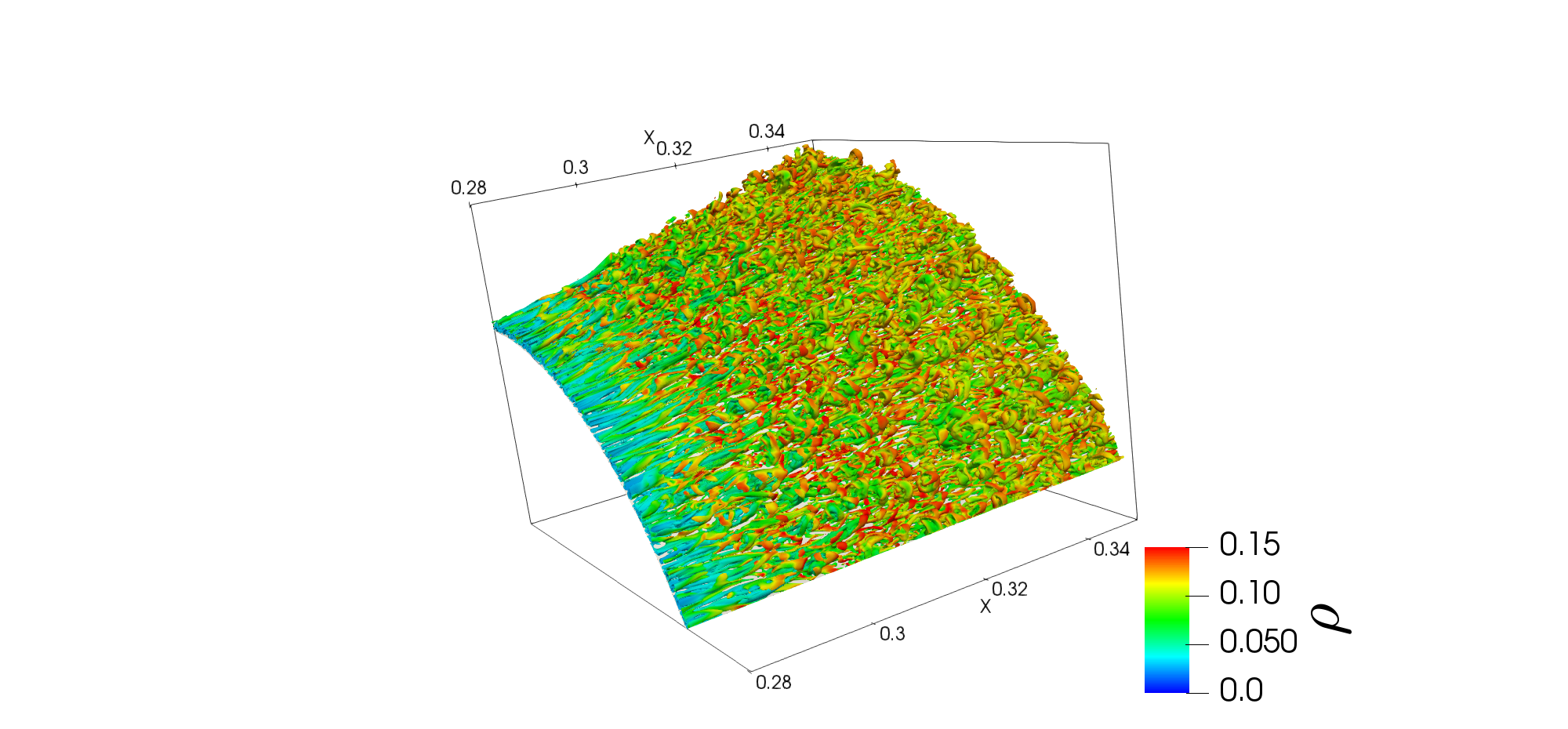}
    \caption{Isosurface of Q criterion ($Q=9\times10^{-3}{U^2}/{\delta^2}$) coloured by density for the reattachment on the flare from an instantaneous snapshot of the QDNS.}
    \label{fig:QCRITRE}
\end{figure}

Figure \ref{fig:QCRITRE} presents an isosurface of Q criterion for the reattachment region. It reveals elongated streamwise structures at the beginning of the domain, which then breakdown creating smaller structures like hairpin vortices, a sign of transition towards a turbulent flow. 
The fact that the breakdown happens at reattachment is one of the main reasons for the peak of heat flux. As observed by \citet{mayer2011direct} for boundary layer oblique breakdown, the point where the periodicity of the flow is lost (\textit{i.e.} were the streaks breakdown) is the point where the skin-friction, and thus the heat-flux, is maximal. 

\begin{figure}
  \centering
  \begin{tabular}[b]{c}
    \includegraphics[trim= 0.2cm 0cm 0.2cm 0cm,width=0.45\linewidth,clip=true]{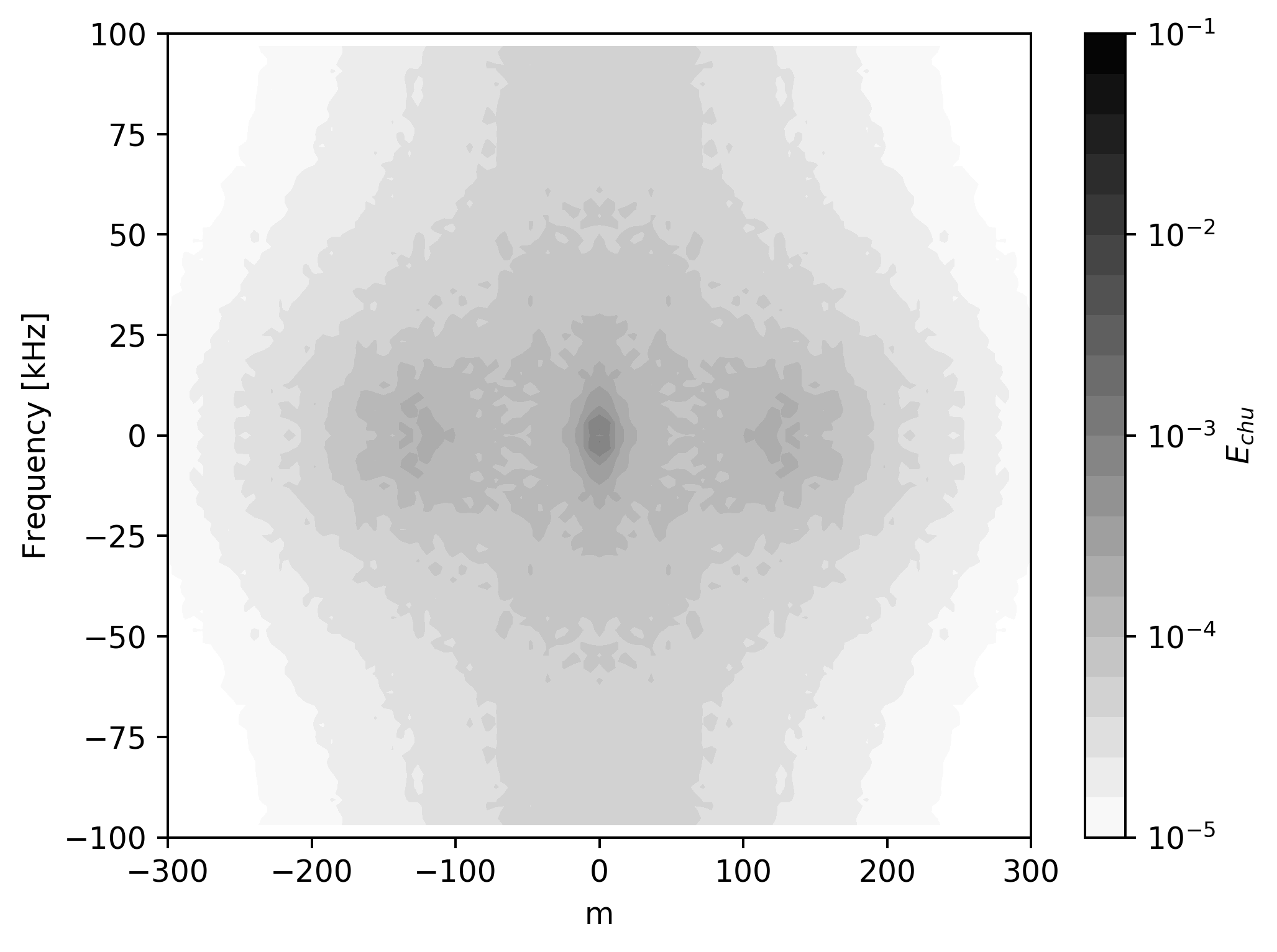}\\
    \small (a)
  \end{tabular} \qquad
  \begin{tabular}[b]{c}
  \includegraphics[trim= 0.2cm 0cm 0.2cm 0cm,width=0.45\linewidth,clip=true]{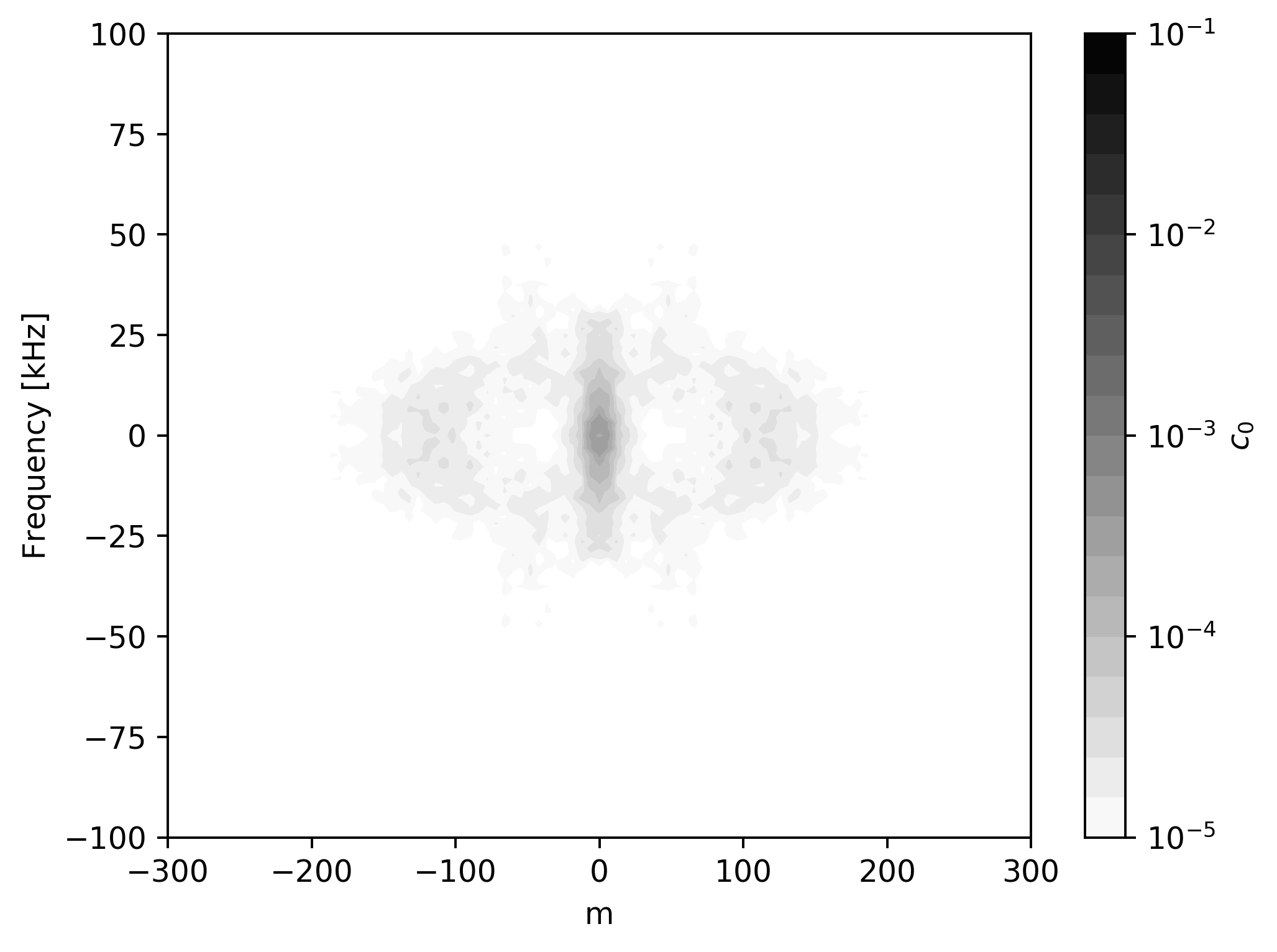}\\
    \small (b)
  \end{tabular}
  \caption{  {Map of (a) the distribution of the fluctuation energy from the QDNS (b) the $c_0$  coefficient for the dominant linear mechanism} against frequency and azimuthal wavenumber for the reattachment region.}
  \label{fig:QDNSEnergyRE}
\end{figure}

The energy distribution presented in figure \ref{fig:QDNSEnergyRE} (a) confirms that the flow is transitioning to turbulence as the energy is spread on a wide range of frequencies and wavenumbers, which is typically due to the breakdown of coherent structures into many smaller scale structures. This breakdown leads to a spread of energy from low to high frequencies and for $|m|\approx 300$ or lower.
  {The energy map also shows that there are less significant levels of energy in coherent structures like streaks and oblique modes.} Moreover, the energy levels involved are several orders of magnitude higher than those of the boundary layer  (see \S\ref{sec:BL}, figure \ref{fig:QDNSEnergyBL})
shows how intense the dynamics is in the reattachment region.

\begin{figure}
  \centering
  \begin{tabular}[b]{c}
    \includegraphics[trim= 0.2cm 0cm 0.2cm 0cm,width=0.45\linewidth,clip=true]{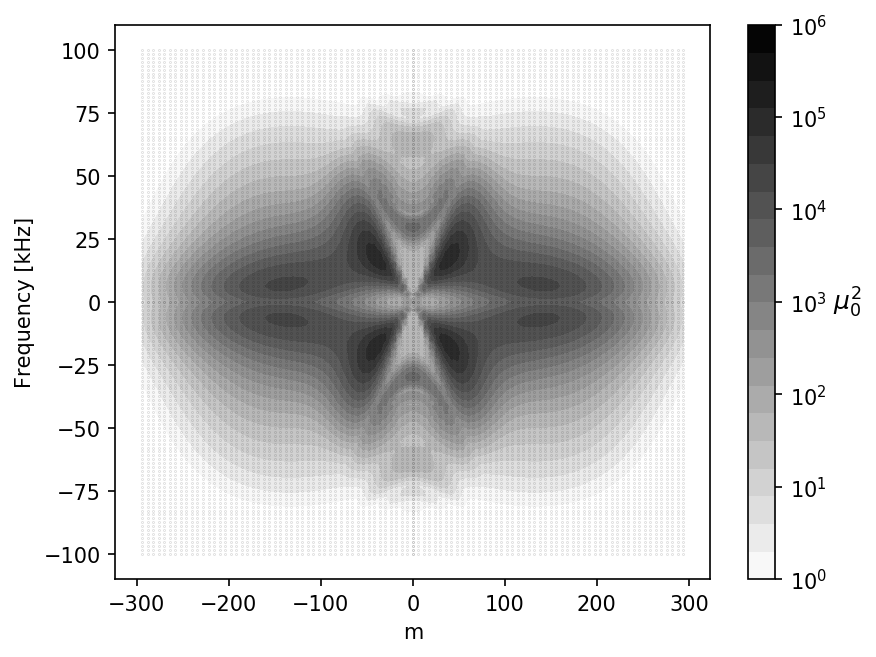}\\
    \small (a)
  \end{tabular} \qquad
  \begin{tabular}[b]{c}
  \includegraphics[trim= 0.2cm 0cm 0.2cm 0cm,width=0.45\linewidth,clip=true]{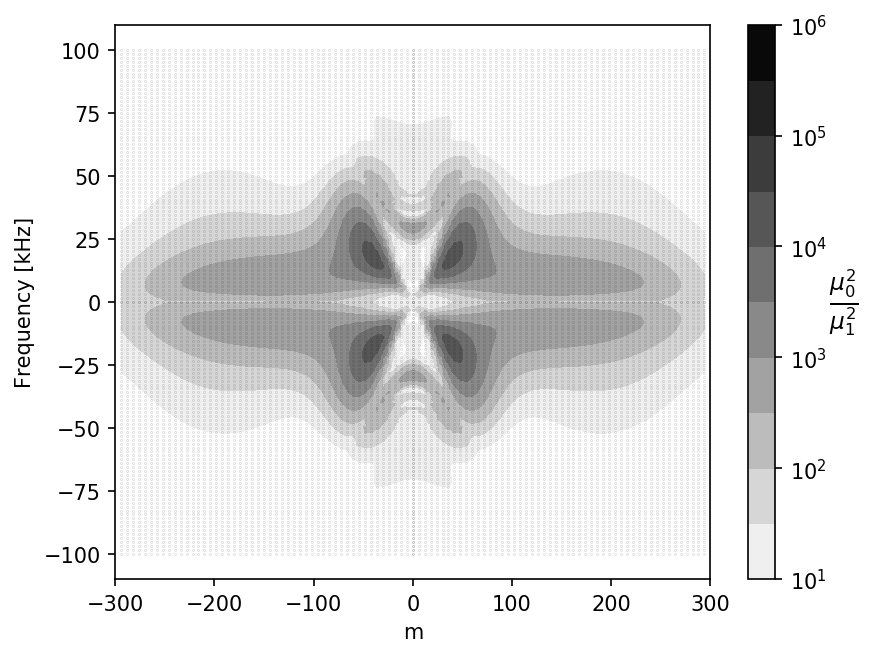}\\
    \small (b)
  \end{tabular}
  \caption{Map of (a) the gain ($\mu_0^2$) from the resolvent analysis (b) the separation between the two first eigenvalues of equation \ref{equ:eigprob} (${\mu_0^2}/{\mu_1^2}$) against frequency and azimuthal wavenumber for the reattachment region.}
  \label{fig:gainSepRE}
\end{figure}

Let us now compare these results to the resolvent analysis. Figure \ref{fig:gainSepRE} presents the resolvent results for the reattachment region: oblique modes display the highest gain values. However, the energy map from the QDNS (figure \ref{fig:QDNSEnergyRE}(a)) shows that they are far from being dominant in the reattachment region   {as there is no clearly defined energetic region for these structures.} Beside oblique modes, a new secondary zone of amplification appears at frequencies and wavenumbers corresponding to already existing streaks. The local maxima of this zone agree well with the energy map of figure \ref{fig:QDNSEnergyRE}(a) (the maximal linear amplification of streaks occurs around $|m|=120$ ).   {However, the energy map presents energy spread on a wider range of frequencies and wavenumbers than the gain map.}

\begin{figure}
    \centering
    \includegraphics[trim=0.15cm 2.5cm 0.15cm 2cm,width=0.9\linewidth,clip=true]{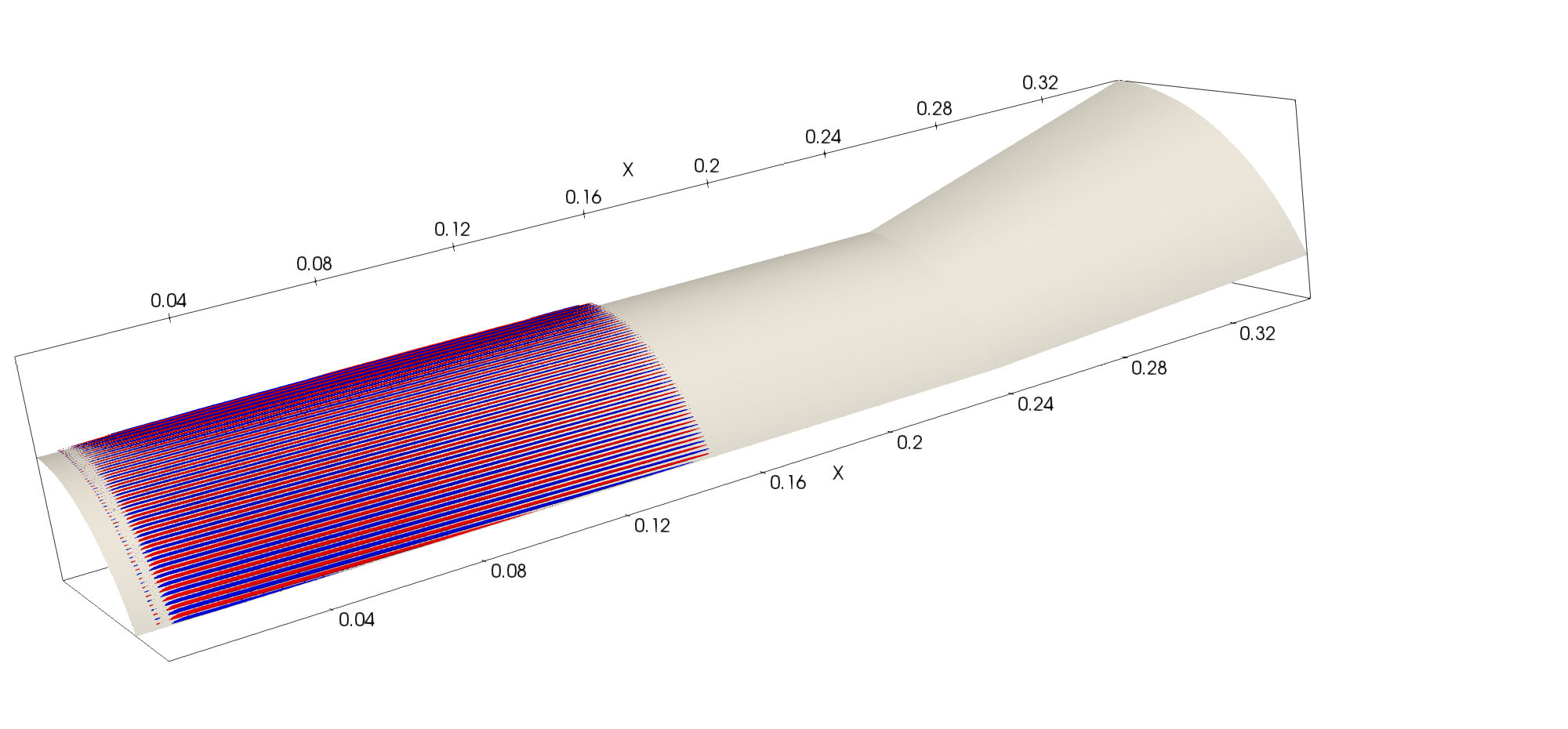}
    \caption{3D reconstruction (isosurface of equal positive and negative density   {forcing}) of the optimal forcing linked to streak amplification for the reattachment at $m=174$ and $f=7kHz$.}
    \label{fig:FORCSTREAK}
\end{figure}

\begin{figure}
  \centering
  \begin{tabular}[b]{c}
    \includegraphics[trim=8cm 0cm 8cm 0cm,width=0.45\linewidth,clip=true]{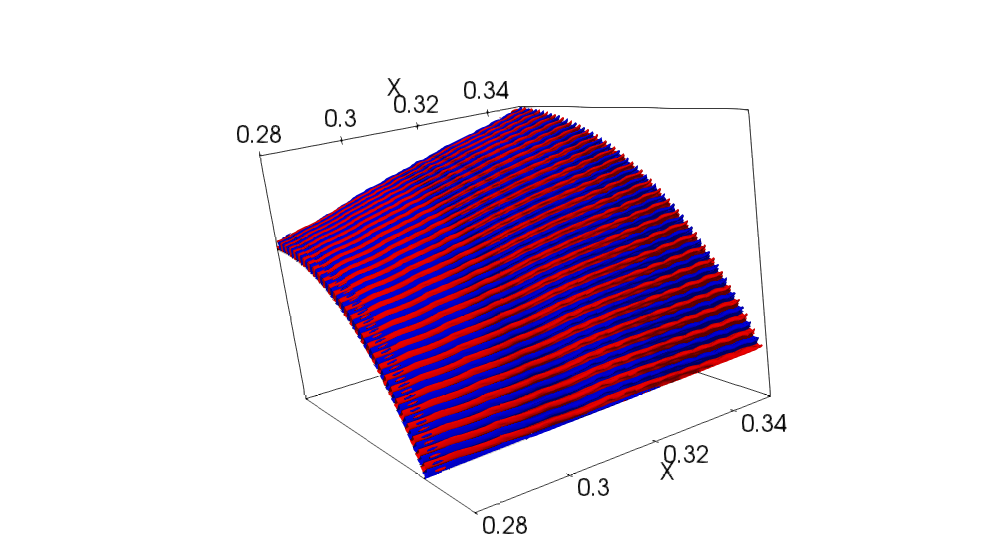}\\
    \small (a)
  \end{tabular} \qquad
  \begin{tabular}[b]{c}
    \includegraphics[trim=8cm 0cm 8cm 0cm,width=0.45\linewidth,clip=true]{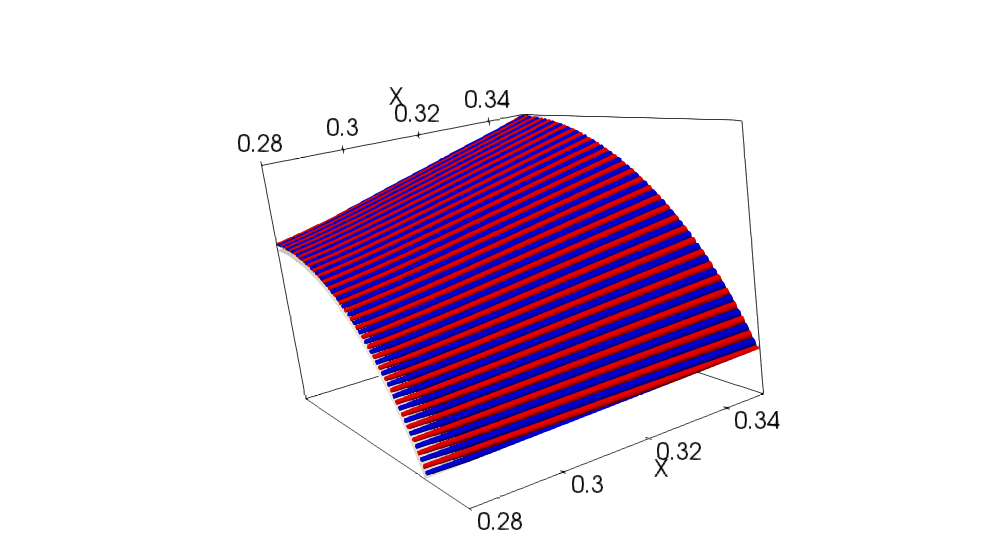}\\
    \small (b)
  \end{tabular}
  \caption{3D reconstruction (isosurface of equal positive and negative density fluctuations) of  (a) the leading SPOD mode   {($r_0>86\%$)}, (b)  the optimal response of the reattachment region at $m=174$ and $f=7kHz$, showing elongated streaks.}
  \label{fig:RESPOD_STRE}
\end{figure}

\begin{figure}
    \centering
    \includegraphics[width=0.55\linewidth]{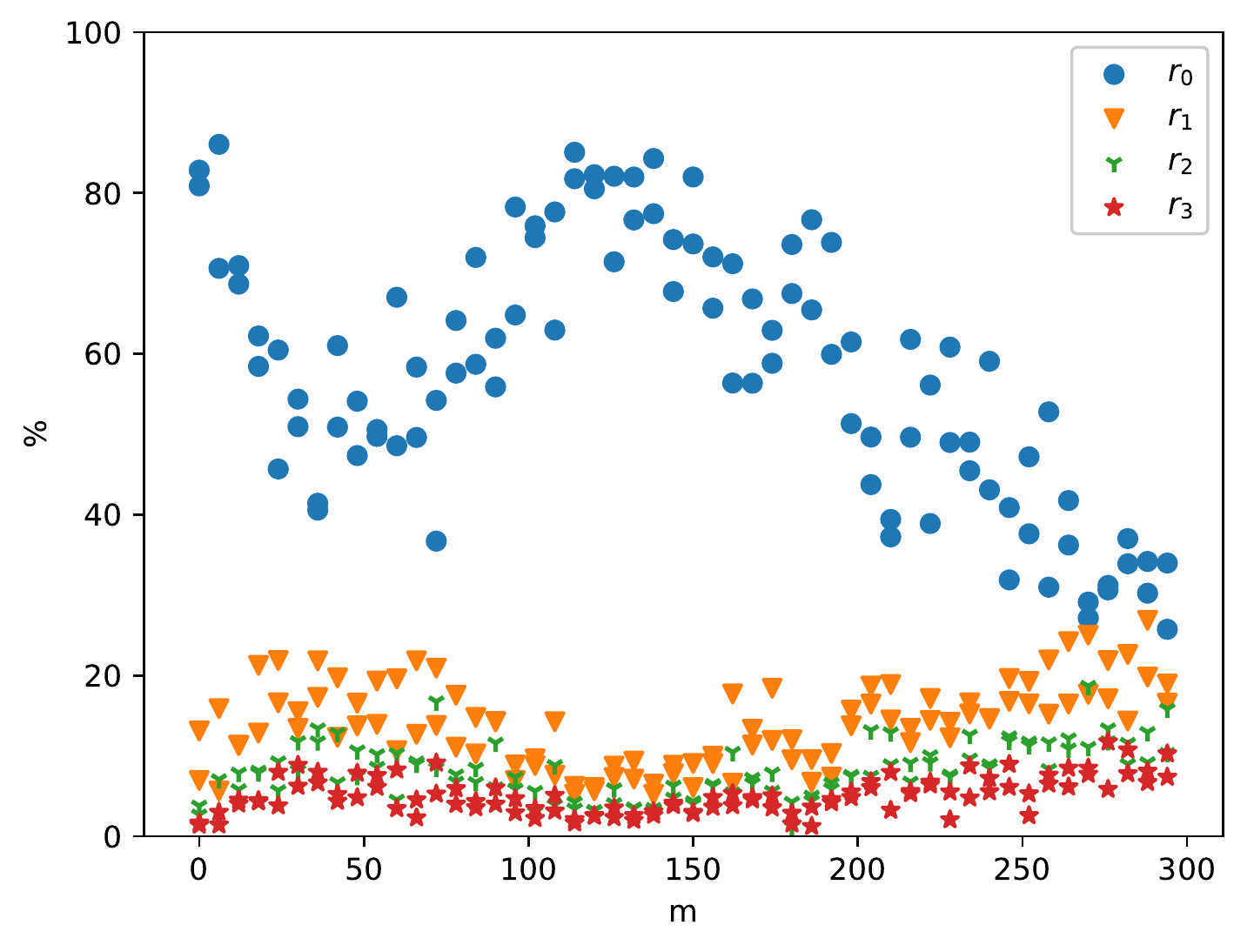}
    \caption{Percentage of energy contained in the first 4 SPOD modes for the reattachment at $f=1.5kHz$ depending on azimuthal wavenumber.}
    \label{fig:EnergyPODRE}
\end{figure}

These discrepancies may be investigated following the same approach as in \S\ref{sec:ML} by   {computing a map of the $c_0$  coefficient} in the $(m,\omega)$ domain (see \S\ref{sec:resolv}). 
Figure \ref{fig:QDNSEnergyRE}(b) presents the results of this analysis. As in the mixing layer, the forcing term plays a significant role in the selection of linearly amplified structures. It completely shifts the amplification map from an oblique-mode-dominated configuration to a streaks-dominated one.   {This is similar to the situation of the mixing layer: the strong linear mechanism for oblique modes is very weakly excited while streaks are nearly-optimally forced by higher-energy structures. Indeed,  as shown in figure \ref{fig:FORCSTREAK}, the optimal forcing associated with streaks is once again located far upstream, and is reminiscent of the structures that developed through nonlinear interaction of oblique modes in the boundary layer.}
  {Thus, the dynamics of the boundary layer plays a critical role in the transition process, even though the boundary layer instabilities are around three orders of magnitude less energetic.} The corresponding optimal response of a streak mode is compared to SPOD results for the QDNS in figure \ref{fig:RESPOD_STRE}. Again, there is a good agreement between the predicted linearly amplified structures and those that develop in the QDNS.
  {Another interesting point is that, contrary to the mixing layer region, the $c_0$ map for the reattachement region (figure \ref{fig:QDNSEnergyRE} (b)) is quite different from the energy map from the QDNS (figure \ref{fig:QDNSEnergyRE} (a)) and is lacking a lot of energy that is spread on a wide range of wavenumber and frequency.}
{  
This is a sign that the coherent structures are breaking down. This breakdown implies a transfer of energy from the streaks to a multitude of other spatio-temporal scales associated with turbulence. 
  {This is confirmed by figure \ref{fig:EnergyPODRE}, which shows that the low-frequency dynamics at moderate $|m|$ values is dominated by one single SPOD mode associated with streaks. But as $|m|$ increases above approximately 200, there is almost no separation between the leading SPOD mode and others, which reveals the spatio-temporally uncorrelated (turbulent) nature of the flow. In such conditions, as explained by \citet{towne2018spectral}, the resolvent analysis is expected to differ from the actual dynamics, since only a limited number of resolvent modes cannot characterise the dynamics anymore. This is confirmed by the alignment coefficient $|\langle \boldsymbol{\Psi} , \boldsymbol{\psi} \rangle|=0.37$ (for $m=174$, $f=7kHz$) which is very low. An explanation for this low value can be found in figure \ref{fig:RESPOD_STRE}. Even if the modes look very similar in the beginning of the region, the SPOD mode begin to meander as soon as we reach the turbulent zone. While the alignment coefficient at the beginning of the domain would be high as the linearly amplified structure is very similar to the one present in the QNDS, its value is plummeting in the downstream part of the region due to the breakdown. 
The same logic applies for all the energy spread on a wide frequency-wavenumber range in this region, as the energy is spread by the breakdown to turbulence and the flow is no longer dominated by a single dominant mechanism, the leading resolvent mode is unable to describe it correctly, causing the discrepancies between figure \ref{fig:QDNSEnergyRE} (a) and (b).}
}
  {
Even with these discrepancies, it is still interesting to notice that the linear amplification of streaks is increasingly stronger in the mixing layer and at the beginning of the reattachment region than in the boundary layer, due to an increasingly stronger linear mechanism. As previously discussed, oblique breakdown is already known to be one of the fastest ways to create turbulence in attached boundary layers \citep{franko_lele_2013,leleAdverse,laible2016continuously}, the fact that linear mechanisms associated with streaks become stronger after the separation point makes it even more relevant for SBLI flow.}

To summarise these findings, in the reattachment region, the streaks caused by a non-linear interaction in the boundary which are linearly amplified in the mixing layer are further amplified by a linear mechanism. Then, they quickly breakdown in the way described by \citet{mayer2011direct} for the flat plate boundary layer as the tip of the streamwise structure lift up from the wall and break to turbulence.

\section{Proposed scenario for the transition process}\label{sec:scenario}
The findings of the previous section yields a transition scenario for the studied case. Even if the scenario is not new, as it is built on mechanisms that are well-known in the literature, it is the first time that it is studied in a complex configuration.   {Note that despite the generic broadband nature of the noise injected within the simulation that excites a wide range of mechanisms (see \S\ref{sec:inletpert}), the scenario might differ for a flow subject to a different type of perturbations (for example, in the case of purely acoustic perturbations). }

The scenario is presented in figure \ref{fig:roadSchema} and can be followed step by step :

\begin{figure}
    \centering
    \includegraphics[width=0.70\linewidth]{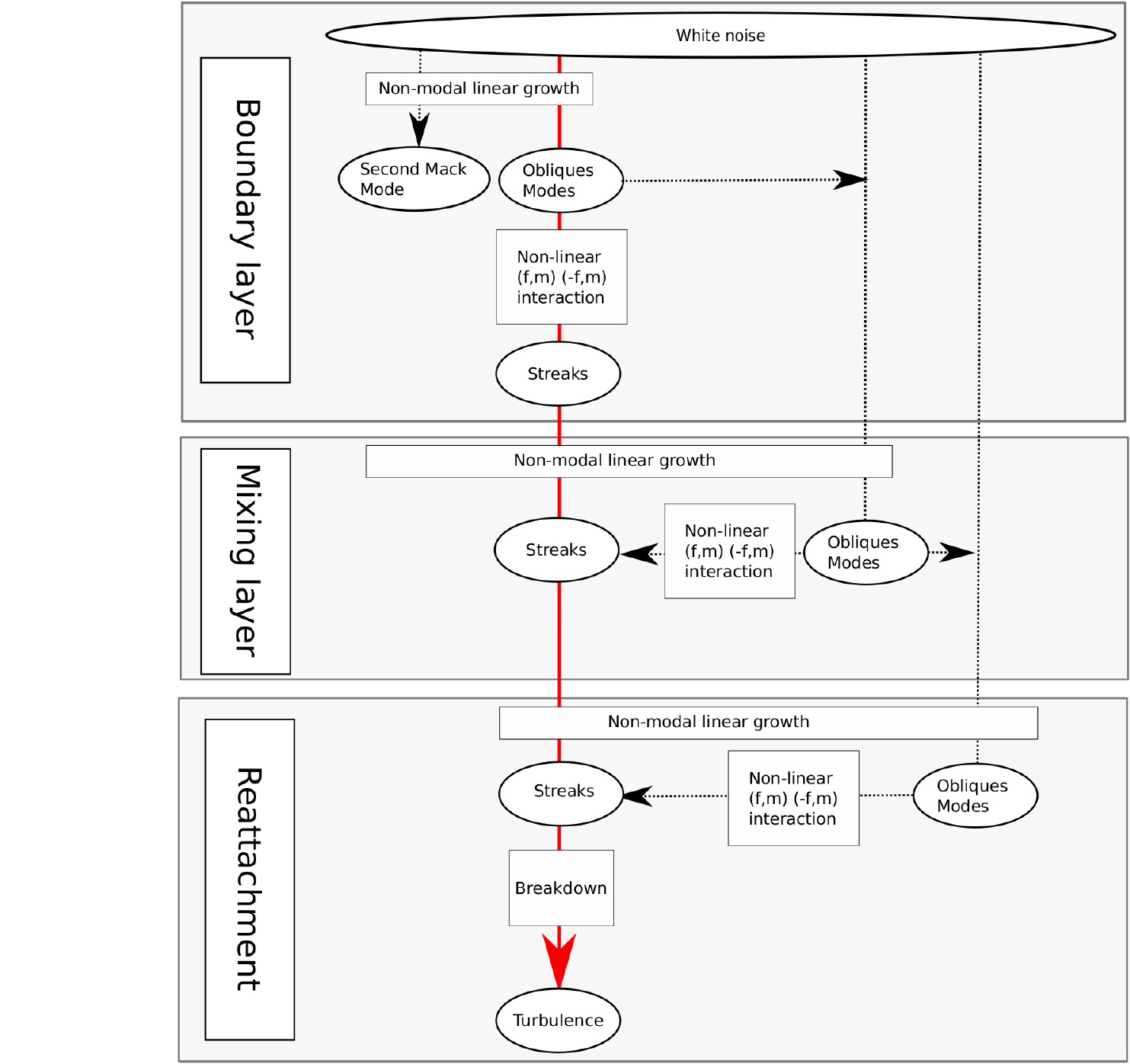}
    \caption{Proposed scenario for the transition process.}
    \label{fig:roadSchema}
\end{figure}

\begin{enumerate}
    \item Boundary layer
    \begin{enumerate}
    \item Some white noise is injected in the boundary layer, it triggers the linear growth of two well-known instabilities over a wide range of frequencies and wavenumbers:
    \begin{enumerate}
        \item Second Mack mode instabilities.
        \item Oblique first mode instabilities  {, they become dominant due to their larger instability domain.}
    \end{enumerate}
    \item The oblique modes interact non-linearly in the way proposed by \citet{thumm1991numerische}, creating quasi-steady streaks over a wide range of wavenumber.
\end{enumerate}
\item Mixing layer
\begin{enumerate}
    \item Oblique modes continue to be amplified (albeit for lower frequencies   {due to the thicker shear layer compared to the upstream boundary layer}) and to interact non-linearly to feed energy to the streaks.
    \item   {The non-linear forcing linked to} streaks created in the boundary layer trigger a weak linear amplification mechanism in the mixing layer, making streaks the dominant structure in the flow.
\end{enumerate}
\item Reattachment
\begin{enumerate}
    \item   {The non-linear forcing linked to} streaks created in the boundary layer continue to trigger a linear amplification mechanism in the reattachment region. The streaks finally break down, creating turbulent structures.
\end{enumerate}
\end{enumerate}

\begin{figure}
    \centering
    \includegraphics[width=0.70\linewidth]{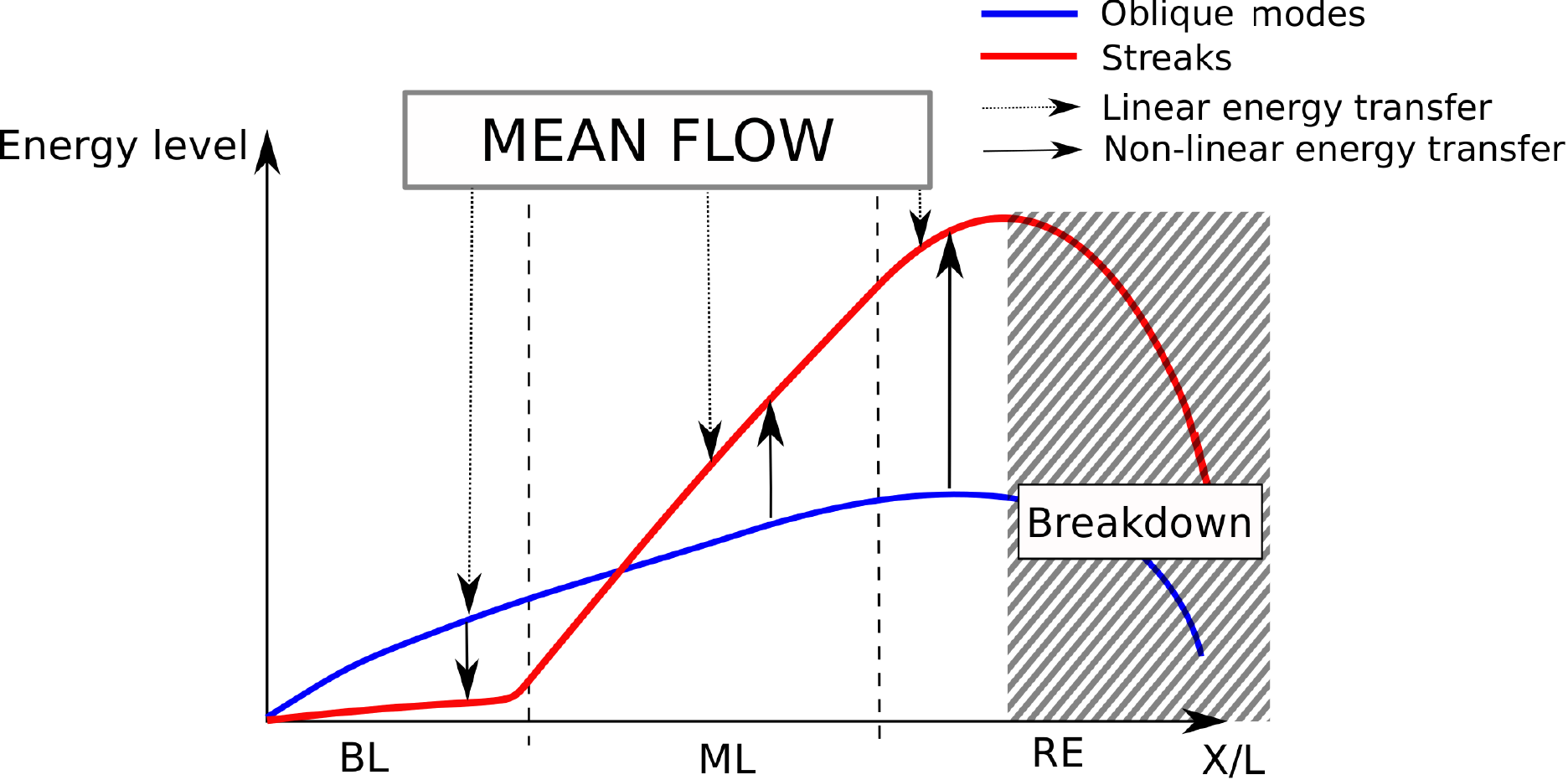}
    \caption{Energy transfer during the transition process.}
    \label{fig:roadCourbe}
\end{figure}

Energy transfers linked to that scenario are sketched in figure \ref{fig:roadCourbe}, which shows that the mean flow transfers energy to oblique modes (via a linear mechanism). These modes then transfer energy to streaks via a non-linear interaction. The mean flow also feeds energy directly to streaks but as shown in figure \ref{fig:roadSchema}, the non-linear interaction in the boundary layer is necessary to trigger this linear amplification mechanism.
An important conclusion of the study is that even if the energy transfers in the boundary layer are of very low intensity when compared to energy transfers in other regions of the flow, the transition scenario is highly dependent on the low-energy structures that develop due to boundary layer instabilities.

This work draws a clearer picture of the dynamics of the flow. Yet it did not discuss from a physical viewpoint the nature of the linear amplification mechanism of streaks, which plays a central role in the overall dynamics of the flow.
  {Even if good candidates for the linear amplification of longitudinal structures would be the centrifugal effect linked to G\"{o}rtler instabilities \citep{benaytransitionalSWBLI,GORTLER_oil_CJ_cowl_murray_hillier_williams_2013,NAVARROMARTINEZ2005225}   {or the lift-up effect such as pointed by \citet{bugeat20193d} (albeit only for boundary layer flow)},
a recent study by \citet{dwivedi2018reattachment} tends to show that they are due to baroclinic effects. }
The work of \citet{dwivedi2018reattachment} focuses on a laminar flow and does not address directly the question of transition, but nonetheless, it provides insights about the amplification mechanism of so-called 'reattachment vortices' in an hypersonic compression ramp flow. Their physical analysis, based on the study of the inviscid transport equations and particulary of the contribution of base flow gradients to the production of streamwise velocity, vorticity, and temperature perturbations, showed that the streamwise deceleration through the recirculation region caused the amplification of streamwise velocity perturbations, and that baroclinic effects were the main cause of the amplification of streamwise vorticity. Therefore, they concluded that the linear amplification of these longitudinal vortical structure was due to the baroclinic effects.

\section{Conclusion}\label{sec:concl}
An in-depth study of the transition process has been conducted for an axisymmetrical compression ramp at a Mach number of 5 and a transitional Reynolds number. A  QDNS was carried out and compared to a  global resolvent analysis. In contrast to most of the studies on the subject, the transition scenario was not chosen beforehand, and the excitation of convective instabilities in the QDNS was designed not to promote only a single instability, leading to  the amplification and interaction of instabilities on a wide range of frequencies and wavenumbers. The dominant mechanism appearing in these conditions relies on the amplification of broadband first oblique modes in the boundary layer,   beating the second mode growth because of their upstream domain of instability. These oblique modes interact non-linearly to create streaks such as already documented in many cases of supersonic and hypersonic transition \citep{fasel1993direct,laible2009numerical,mayer2011direct,franko_lele_2013, leleAdverse,fasel2015numerical}.
Then, the non-linear forcing linked to those streaks trigger a linear amplification mechanism,   {either due to centrifugal, baroclinic, or the lift-up effect}, in the mixing layer and reattachment region which lead to breakdown. Even if the breakdown is linked to linear amplification, the non-linear interaction of oblique modes was found to be essential for this transition scenario. The combination of both QDNS with SPOD and resolvent analysis has proven to be a highly efficient toolset to understand the physical mechanisms behind transition, especially when dealing with both linearly amplified instabilities and non-linear interactions.

The transition process presented is conjectured to be dominant for comparable high supersonic/low hypersonic flow. However, in the actual context of hypersonic vehicles development, higher flight Mach numbers are not uncommon. Similar studies could be conducted for Mach numbers of 6, 7, or even higher. The tools presented here could allow studying transition scenarios where the second mode instabilities play an increasing role. This may reveal possible interactions between second mode and oblique modes or streaks, and for even higher Mach numbers, a transition dominated by second mode secondary instabilities. 

The global dynamics of the flow along an axisymmetric compression ramp for a Mach number of 5 in the transitional regime is also not fully explained yet. The present work highlighted a low-frequency low-wavenumber motion, most probably linked to an unstable global mode of the recirculation bubble, which would be worth investigating in a future study.\\

This work was supported by the French Alternative Energies and Atomic Energy Commission (CEA) under the grant CEA 4600334751.   The fully resolved DNS was performed on the OCCIGEN supercomputer at CINES under the GENCI allocation A0072A11041.

\bibliographystyle{jfm}
\bibliography{BIBLIO.bib}

\begin{thebibliography}{66}
\expandafter\ifx\csname natexlab\endcsname\relax\def\natexlab#1{#1}\fi
\def\au#1{#1} \def\ed#1{#1} \def\yr#1{#1}\def\at#1{#1}\def\jt#1{\textit{#1}}
  \def\bt#1{#1}\def\bvol#1{\textbf{#1}} \def\vol#1{#1} \def\pg#1{#1}
  \def\publ#1{#1}\def\arxiv#1{#1}\def\org#1{#1}\def\st#1{\textit{#1}}

\bibitem[Adams \& Kleiser(1993)]{adams1993numerical}
{\sc \au{Adams, N} \& \au{Kleiser, L}} \yr{1993} Numerical simulation of
  fundamental breakdown of a laminar boundary-layer at mach 4.5.  \bt{In {\em
  5th International Aerospace Planes and Hypersonics Technologies
  Conference\/}},  \pg{p. 5027}.

\bibitem[Adams(2000)]{DNS_adams_2000}
{\sc \au{Adams, Nikolaus~A.}} \yr{2000}  \at{Direct simulation of the turbulent
  boundary layer along a compression ramp at {M} = 3 and ${R}e_\theta$ = 1685}.
   \jt{Journal of Fluid Mechanics}  \bvol{420},  \pg{47–83}.

\bibitem[Amestoy {\em et~al.\/}(2001)Amestoy, Duff, L'Excellent \&
  Koster]{amestoy2001fully}
{\sc \au{Amestoy, Patrick~R}, \au{Duff, Iain~S}, \au{L'Excellent, Jean-Yves} \&
  \au{Koster, Jacko}} \yr{2001}  \at{A fully asynchronous multifrontal solver
  using distributed dynamic scheduling}.  \jt{SIAM Journal on Matrix Analysis
  and Applications}  \bvol{23}~(1),  \pg{15--41}.

\bibitem[Arnal(1989)]{arnal1989laminar}
{\sc \au{Arnal, Daniel}} \yr{1989}  \at{Laminar-turbulent transition problems
  in supersonic and hypersonic flows}.  \jt{AGARD, Special Course on
  Aerothermodynamics of Hypersonic Vehicles 45 p(SEE N 89-29306 24-02)} .

\bibitem[Arnal \& Juillen(1977)]{arnal}
{\sc \au{Arnal, Daniel} \& \au{Juillen, J.-C.}} \yr{1977}  \at{Etude de
  l'intermittence dans une r{\'{e}}gion de transition de la couche limite}.
  \jt{La Recherche Aérospatiale}  \pg{pp. 147--166}.

\bibitem[Benay {\em et~al.\/}(2006)Benay, Chanetz, Mangin, Vandomme \&
  Perraud]{benaytransitionalSWBLI}
{\sc \au{Benay, Richard}, \au{Chanetz, Bruno}, \au{Mangin, Bruno},
  \au{Vandomme, Ludovic} \& \au{Perraud, Jean}} \yr{2006}  \at{Shock
  wave/transitional boundary-layer interactions in hypersonic flow}.  \jt{AIAA
  Journal}  \bvol{44}~(6),  \pg{1243--1254}.

\bibitem[Beneddine(2017)]{bened2017}
{\sc \au{Beneddine, Samir}} \yr{2017}  \at{Characterization of unsteady flow
  behavior by linear stability analysis}. PhD thesis, ONERA.

\bibitem[Beneddine {\em et~al.\/}(2016)Beneddine, Sipp, Arnault, Dandois \&
  Lesshafft]{beneddine2016conditions}
{\sc \au{Beneddine, Samir}, \au{Sipp, Denis}, \au{Arnault, Anthony},
  \au{Dandois, Julien} \& \au{Lesshafft, Lutz}} \yr{2016}  \at{Conditions for
  validity of mean flow stability analysis}.  \jt{Journal of Fluid Mechanics}
  \bvol{798},  \pg{485--504}.

\bibitem[Berkooz {\em et~al.\/}(1993)Berkooz, Holmes \&
  Lumley]{berkooz1993proper}
{\sc \au{Berkooz, Gal}, \au{Holmes, Philip} \& \au{Lumley, John~L}} \yr{1993}
  \at{The proper orthogonal decomposition in the analysis of turbulent flows}.
  \jt{Annual review of fluid mechanics}  \bvol{25}~(1),  \pg{539--575}.

\bibitem[Bonne {\em et~al.\/}(2019)Bonne, Brion, Garnier, Bur, Molton, Sipp \&
  Jacquin]{bonne2019analysis}
{\sc \au{Bonne, N}, \au{Brion, V}, \au{Garnier, E}, \au{Bur, R}, \au{Molton,
  P}, \au{Sipp, D} \& \au{Jacquin, L}} \yr{2019}  \at{Analysis of the
  two-dimensional dynamics of a {M}ach 1.6 shock wave/transitional boundary
  layer interaction using a {RANS} based resolvent approach}.  \jt{Journal of
  Fluid Mechanics}  \bvol{862},  \pg{1166--1202}.

\bibitem[Brandt {\em et~al.\/}(2011)Brandt, Sipp, Pralits \&
  Marquet]{brandt2011effect}
{\sc \au{Brandt, Luca}, \au{Sipp, Denis}, \au{Pralits, Jan~O} \& \au{Marquet,
  Olivier}} \yr{2011}  \at{Effect of base-flow variation in noise amplifiers:
  the flat-plate boundary layer}.  \jt{Journal of Fluid Mechanics}  \bvol{687},
   \pg{503--528}.

\bibitem[Bugeat(2017)]{bugeat2017stabilite}
{\sc \au{Bugeat, Benjamin}} \yr{2017}  \at{Stabilit{\'e} et perturbations
  optimales globales d'{\'e}coulements compressibles pari{\'e}taux}. PhD
  thesis, Paris 6.

\bibitem[Bugeat {\em et~al.\/}(2019)Bugeat, Chassaing, Robinet \&
  Sagaut]{bugeat20193d}
{\sc \au{Bugeat, B}, \au{Chassaing, J-C}, \au{Robinet, J-C} \& \au{Sagaut, P}}
  \yr{2019}  \at{3d global optimal forcing and response of the supersonic
  boundary layer}.  \jt{Journal of Computational Physics}  \pg{p. 108888}.

\bibitem[Bur \& Chanetz(2009)]{BURTransitionalSWBLI}
{\sc \au{Bur, R.} \& \au{Chanetz, B.}} \yr{2009}  \at{Experimental study on the
  {PRE-X} vehicle focusing on the transitional shock-wave/boundary-layer
  interactions}.  \jt{Aerospace Science and Technology}  \bvol{13}~(7),
  \pg{393--401}.

\bibitem[Chang \& Malik(1994)]{chang1994oblique}
{\sc \au{Chang, Chau-Lyan} \& \au{Malik, Mujeeb~R}} \yr{1994}  \at{Oblique-mode
  breakdown and secondary instability in supersonic boundary layers}.
  \jt{Journal of Fluid Mechanics}  \bvol{273},  \pg{323--360}.

\bibitem[Chu(1965)]{chu1965energy}
{\sc \au{Chu, Boa-Teh}} \yr{1965}  \at{On the energy transfer to small
  disturbances in fluid flow (part i)}.  \jt{Acta Mechanica}  \bvol{1}~(3),
  \pg{215--234}.

\bibitem[Clemens \&
  Narayanaswamy(2014)]{Clemens_Review_upstream_downstream_mecanism}
{\sc \au{Clemens, Noel~T.} \& \au{Narayanaswamy, Venkateswaran}} \yr{2014}
  \at{Low-frequency unsteadiness of shock wave/turbulent boundary layer
  interactions}.  \jt{Annual Review of Fluid Mechanics}  \bvol{46}~(1),
  \pg{469--492},  \arxiv{arXiv:
  https://doi.org/10.1146/annurev-fluid-010313-141346}.

\bibitem[Dwivedi {\em et~al.\/}(2019)Dwivedi, Sidharth, Nichols, Candler \&
  Jovanovi{\'c}]{dwivedi2018reattachment}
{\sc \au{Dwivedi, Anubhav}, \au{Sidharth, GS}, \au{Nichols, Joseph~W},
  \au{Candler, Graham~V} \& \au{Jovanovi{\'c}, Mihailo~R}} \yr{2019}
  \at{Reattachment streaks in hypersonic compression ramp flow: an
  input--output analysis}.  \jt{Journal of Fluid Mechanics}  \bvol{880},
  \pg{113--135}.

\bibitem[Fasel \& Thumm(1991)]{fasel1991direct}
{\sc \au{Fasel, H} \& \au{Thumm, A}} \yr{1991}  \at{Direct numerical simulation
  of three-dimensional breakdown in supersonic boundary layer transition}.
  \jt{Bull. Am. Phys. Soc}  \bvol{36},  \pg{2701}.

\bibitem[Fasel {\em et~al.\/}(2015)Fasel, Sivasubramanian \&
  Laible]{fasel2015numerical}
{\sc \au{Fasel, Hermann~F}, \au{Sivasubramanian, Jayahar} \& \au{Laible,
  Andreas}} \yr{2015}  \at{Numerical investigation of transition in a flared
  cone boundary layer at mach 6}.  \jt{Procedia IUTAM}  \bvol{14},
  \pg{26--35}.

\bibitem[Fasel {\em et~al.\/}(1993)Fasel, Thumm \& Bestek]{fasel1993direct}
{\sc \au{Fasel, Hermann~F}, \au{Thumm, A} \& \au{Bestek, H}} \yr{1993} Direct
  numerical simulation of transition in supersonic boundary layers: oblique
  breakdown.  \bt{In {\em Fluids Engineering Conference\/}},  \pg{pp. 77--92}.
  Publ by ASME.

\bibitem[Franko \& Lele(2014)]{leleAdverse}
{\sc \au{Franko, Kenneth~J.} \& \au{Lele, Sanjiva}} \yr{2014}  \at{Effect of
  adverse pressure gradient on high speed boundary layer transition}.
  \jt{Physics of Fluids}  \bvol{26}~(2),  \pg{024106},  \arxiv{arXiv:
  https://doi.org/10.1063/1.4864337}.

\bibitem[Franko \& Lele(2013)]{franko_lele_2013}
{\sc \au{Franko, Kenneth~J.} \& \au{Lele, Sanjiva~K.}} \yr{2013}  \at{Breakdown
  mechanisms and heat transfer overshoot in hypersonic zero pressure gradient
  boundary layers}.  \jt{Journal of Fluid Mechanics}  \bvol{730},
  \pg{491–532}.

\bibitem[Gallaire {\em et~al.\/}(2007)Gallaire, Marquillie \&
  Ehrenstein]{gallaire2007three}
{\sc \au{Gallaire, Fran{\c{c}}ois}, \au{Marquillie, Matthieu} \&
  \au{Ehrenstein, Uwe}} \yr{2007}  \at{Three-dimensional transverse
  instabilities in detached boundary layers}.  \jt{Journal of Fluid Mechanics}
  \bvol{571},  \pg{221--233}.

\bibitem[Garnier {\em et~al.\/}(2009)Garnier, Adams \&
  Sagaut]{garnier2009large}
{\sc \au{Garnier, Eric}, \au{Adams, Nikolaus} \& \au{Sagaut, Pierre}} \yr{2009}
  {\em Large eddy simulation for compressible flows\/}.  \publ{Springer Science
  \& Business Media}.

\bibitem[Garnier {\em et~al.\/}(2002)Garnier, Sagaut \&
  Deville]{garnier2002large}
{\sc \au{Garnier, Eric}, \au{Sagaut, Pierre} \& \au{Deville, Michel}} \yr{2002}
   \at{Large eddy simulation of shock/boundary-layer interaction}.  \jt{AIAA
  journal}  \bvol{40}~(10),  \pg{1935--1944}.

\bibitem[George \& Sujith(2011)]{JOSEPHGEORGE20115280}
{\sc \au{George, K.~Joseph} \& \au{Sujith, R.I.}} \yr{2011}  \at{On {C}hu's
  disturbance energy}.  \jt{Journal of Sound and Vibration}  \bvol{330}~(22),
  \pg{5280 -- 5291}.

\bibitem[Georgiadis {\em et~al.\/}(2010)Georgiadis, Rizzetta \&
  Fureby]{georgiadis2010large}
{\sc \au{Georgiadis, Nicholas~J}, \au{Rizzetta, Donald~P} \& \au{Fureby,
  Christer}} \yr{2010}  \at{Large-eddy simulation: current capabilities,
  recommended practices, and future research}.  \jt{AIAA journal}
  \bvol{48}~(8),  \pg{1772--1784}.

\bibitem[G{\"{o}}rtler(1940)]{Gortler_original}
{\sc \au{G{\"{o}}rtler, H.}} \yr{1940}  \at{Instabilit{\"{a}}t laminarer
  grenzschichten an konkaven w{\"{a}}nden gegen{\"{u}}ber gewissen
  dreidimensionalen st{\"{o}}rungen}.  \jt{ZAMM - Journal of Applied
  Mathematics and Mechanics / Zeitschrift für Angewandte Mathematik und
  Mechanik}  \bvol{21}~(4),  \pg{250--252}.

\bibitem[Gudmundsson \& Colonius(2011)]{gudmundsson2011instability}
{\sc \au{Gudmundsson, K} \& \au{Colonius, Tim}} \yr{2011}  \at{Instability wave
  models for the near-field fluctuations of turbulent jets}.  \jt{Journal of
  Fluid Mechanics}  \bvol{689},  \pg{97--128}.

\bibitem[Hader \& Fasel(2018)]{hader2018towards}
{\sc \au{Hader, Christoph} \& \au{Fasel, Hermann~F}} \yr{2018}  \at{Towards
  simulating natural transition in hypersonic boundary layers via random inflow
  disturbances}.  \jt{Journal of Fluid Mechanics}  \bvol{847}.

\bibitem[Hanifi {\em et~al.\/}(1996)Hanifi, Schmid \&
  Henningson]{hanifi1996transient}
{\sc \au{Hanifi, Ardeshir}, \au{Schmid, Peter~J} \& \au{Henningson, Dan~S}}
  \yr{1996}  \at{Transient growth in compressible boundary layer flow}.
  \jt{Physics of Fluids}  \bvol{8}~(3),  \pg{826--837}.

\bibitem[Hildebrand {\em et~al.\/}(2018)Hildebrand, Dwivedi, Nichols,
  Jovanovi{\'c} \& Candler]{hildebrand2018simulation}
{\sc \au{Hildebrand, Nathaniel}, \au{Dwivedi, Anubhav}, \au{Nichols, Joseph~W},
  \au{Jovanovi{\'c}, Mihailo~R} \& \au{Candler, Graham~V}} \yr{2018}
  \at{Simulation and stability analysis of oblique shock-wave/boundary-layer
  interactions at {M}ach 5.92}.  \jt{Physical Review Fluids}  \bvol{3}~(1),
  \pg{013906}.

\bibitem[Laible {\em et~al.\/}(2009)Laible, Mayer \&
  Fasel]{laible2009numerical}
{\sc \au{Laible, Andreas}, \au{Mayer, Christian} \& \au{Fasel, Hermann}}
  \yr{2009} Numerical investigation of transition for a cone at mach 3.5:
  oblique breakdown.  \bt{In {\em 39th AIAA Fluid Dynamics Conference\/}},
  \pg{p. 3557}.

\bibitem[Laible \& Fasel(2016)]{laible2016continuously}
{\sc \au{Laible, Andreas~C} \& \au{Fasel, Hermann~F}} \yr{2016}
  \at{Continuously forced transient growth in oblique breakdown for supersonic
  boundary layers}.  \jt{Journal of Fluid Mechanics}  \bvol{804},
  \pg{323--350}.

\bibitem[Laurence {\em et~al.\/}(2016)Laurence, Wagner \&
  Hannemann]{laurence2016experimental}
{\sc \au{Laurence, SJ}, \au{Wagner, A} \& \au{Hannemann, K}} \yr{2016}
  \at{Experimental study of second-mode instability growth and breakdown in a
  hypersonic boundary layer using high-speed schlieren visualization}.
  \jt{Journal of Fluid Mechanics}  \bvol{797},  \pg{471--503}.

\bibitem[Lehoucq {\em et~al.\/}(1998)Lehoucq, Sorensen \&
  Yang]{lehoucq1998arpack}
{\sc \au{Lehoucq, Richard~B}, \au{Sorensen, Danny~C} \& \au{Yang, Chao}}
  \yr{1998} {\em ARPACK users' guide: solution of large-scale eigenvalue
  problems with implicitly restarted Arnoldi methods\/}, ,  \vol{vol.~6}.
  \publ{Siam}.

\bibitem[Lumley(1970)]{lumley1970stochastic}
{\sc \au{Lumley, JL}} \yr{1970}  \at{Stochastic tools in turbulence}.  \jt{New
  York: Academic} .

\bibitem[Mack(1975)]{MACK}
{\sc \au{Mack, Leslie~M.}} \yr{1975}  \at{Linear stability theory and the
  problem of supersonic boundary- layer transition}.  \jt{AIAA Journal}
  \bvol{13}~(3),  \pg{278--289}.

\bibitem[Marquet {\em et~al.\/}(2009)Marquet, Lombardi, Chomaz, Sipp \&
  Jacquin]{marquet2009direct}
{\sc \au{Marquet, Olivier}, \au{Lombardi, Matteo}, \au{Chomaz, Jean-Marc},
  \au{Sipp, Denis} \& \au{Jacquin, Laurent}} \yr{2009}  \at{Direct and adjoint
  global modes of a recirculation bubble: lift-up and convective
  non-normalities}.  \jt{Journal of Fluid Mechanics}  \bvol{622},  \pg{1--21}.

\bibitem[Marxen {\em et~al.\/}(2010)Marxen, Iaccarino \&
  Shaqfeh]{marxen2010disturbance}
{\sc \au{Marxen, Olaf}, \au{Iaccarino, Gianluca} \& \au{Shaqfeh, Eric~SG}}
  \yr{2010}  \at{Disturbance evolution in a mach 4.8 boundary layer with
  two-dimensional roughness-induced separation and shock}.  \jt{Journal of
  Fluid Mechanics}  \bvol{648},  \pg{435--469}.

\bibitem[Marxen \& Rist(2010)]{marxen2010mean}
{\sc \au{Marxen, Olaf} \& \au{Rist, Ulrich}} \yr{2010}  \at{Mean flow
  deformation in a laminar separation bubble: separation and stability
  characteristics}.  \jt{Journal of Fluid Mechanics}  \bvol{660},  \pg{37--54}.

\bibitem[Mary \& Sagaut(2001)]{AUSMONERA}
{\sc \au{Mary, Ivan} \& \au{Sagaut, Pierre}} \yr{2001}  \at{Large eddy
  simulation of flow around an airfoil near stall}.  \jt{AIAA Journal}
  \bvol{40}.

\bibitem[Mayer {\em et~al.\/}(2011)Mayer, Von~Terzi \& Fasel]{mayer2011direct}
{\sc \au{Mayer, Christian~SJ}, \au{Von~Terzi, Dominic~A} \& \au{Fasel,
  Hermann~F}} \yr{2011}  \at{Direct numerical simulation of complete transition
  to turbulence via oblique breakdown at mach 3}.  \jt{Journal of Fluid
  Mechanics}  \bvol{674},  \pg{5--42}.

\bibitem[Murray {\em et~al.\/}(2013)Murray, Hillier \&
  Williams]{GORTLER_oil_CJ_cowl_murray_hillier_williams_2013}
{\sc \au{Murray, N.}, \au{Hillier, R.} \& \au{Williams, S.}} \yr{2013}
  \at{Experimental investigation of axisymmetric hypersonic
  shock-wave/turbulent-boundary-layer interactions}.  \jt{Journal of Fluid
  Mechanics}  \bvol{714},  \pg{152–189}.

\bibitem[Navarro-Martinez \& Tutty(2005)]{NAVARROMARTINEZ2005225}
{\sc \au{Navarro-Martinez, S.} \& \au{Tutty, O.R.}} \yr{2005}  \at{Numerical
  simulation of {G}{\"{o}}rtler vortices in hypersonic compression ramps}.
  \jt{Computers \& Fluids}  \bvol{34}~(2),  \pg{225--247}.

\bibitem[Paladini {\em et~al.\/}(2019)Paladini, Beneddine, Dandois, Sipp \&
  Robinet]{paladini2019transonic}
{\sc \au{Paladini, Edorado}, \au{Beneddine, Samir}, \au{Dandois, Julien},
  \au{Sipp, Denis} \& \au{Robinet, Jean-Christophe}} \yr{2019}  \at{Transonic
  buffet instability: From two-dimensional airfoils to three-dimensional swept
  wings}.  \jt{Physical Review Fluids}  \bvol{4}~(10),  \pg{103906}.

\bibitem[P{\'e}ron {\em et~al.\/}(2017)P{\'e}ron, Renaud, Mary, Benoit \&
  Terracol]{peron2017immersed}
{\sc \au{P{\'e}ron, St{\'e}phanie}, \au{Renaud, Thomas}, \au{Mary, Ivan},
  \au{Benoit, Christophe} \& \au{Terracol, Marc}} \yr{2017} An immersed
  boundary method for preliminary design aerodynamic studies of complex
  configurations.  \bt{In {\em 23rd AIAA Computational Fluid Dynamics
  Conference\/}},  \pg{p. 3623}.

\bibitem[Priebe \& Mart{\'\i}n(2012)]{priebe2012low}
{\sc \au{Priebe, Stephan} \& \au{Mart{\'\i}n, M~Pino}} \yr{2012}
  \at{Low-frequency unsteadiness in shock wave--turbulent boundary layer
  interaction}.  \jt{Journal of Fluid Mechanics}  \bvol{699},  \pg{1--49}.

\bibitem[Renard \& Deck(2016)]{renard2016theoretical}
{\sc \au{Renard, Nicolas} \& \au{Deck, S{\'e}bastien}} \yr{2016}  \at{A
  theoretical decomposition of mean skin friction generation into physical
  phenomena across the boundary layer}.  \jt{Journal of Fluid Mechanics}
  \bvol{790},  \pg{339--367}.

\bibitem[Robinet(2007)]{robinet2007bifurcations}
{\sc \au{Robinet, J-Ch}} \yr{2007}  \at{Bifurcations in
  shock-wave/laminar-boundary-layer interaction: global instability approach}.
  \jt{Journal of Fluid Mechanics}  \bvol{579},  \pg{85--112}.

\bibitem[Sandham {\em et~al.\/}(1995)Sandham, Adams \&
  Kleiser]{sandham1995direct}
{\sc \au{Sandham, ND}, \au{Adams, Nikolaus~A} \& \au{Kleiser, Leonhard}}
  \yr{1995}  \at{Direct simulation of breakdown to turbulence following oblique
  instability waves in a supersonic boundary layer}.  \jt{Applied scientific
  research}  \bvol{54}~(3),  \pg{223--234}.

\bibitem[Sandham {\em et~al.\/}(2014)Sandham, Sch{\"u}lein, Wagner, Willems \&
  Steelant]{sandham2014transitional}
{\sc \au{Sandham, ND}, \au{Sch{\"u}lein, E}, \au{Wagner, Andrew}, \au{Willems,
  S} \& \au{Steelant, Johan}} \yr{2014}  \at{Transitional
  shock-wave/boundary-layer interactions in hypersonic flow}.  \jt{Journal of
  Fluid Mechanics}  \bvol{752},  \pg{349--382}.

\bibitem[Schmid \& Henningson(1992)]{schmid1992new}
{\sc \au{Schmid, PJ} \& \au{Henningson, DS}} \yr{1992}  \at{A new mechanism for
  rapid transition involving a pair of oblique waves}.  \jt{Physics of Fluids
  A: Fluid Dynamics}  \bvol{4}~(9),  \pg{1986--1989}.

\bibitem[Schmid(2007)]{schmid2007nonmodal}
{\sc \au{Schmid, Peter~J}} \yr{2007}  \at{Nonmodal stability theory}.
  \jt{Annu. Rev. Fluid Mech.}  \bvol{39},  \pg{129--162}.

\bibitem[Schmid {\em et~al.\/}(2017)Schmid, de~Pando \&
  Peake]{schmid2017stability}
{\sc \au{Schmid, Peter~J}, \au{de~Pando, Miguel~Fosas} \& \au{Peake, Nigel}}
  \yr{2017}  \at{Stability analysis for n-periodic arrays of fluid systems}.
  \jt{Physical Review Fluids}  \bvol{2}~(11),  \pg{113902}.

\bibitem[Schneider(2008)]{schneider2008development}
{\sc \au{Schneider, Steven~P}} \yr{2008}  \at{Development of hypersonic quiet
  tunnels}.  \jt{Journal of Spacecraft and Rockets}  \bvol{45}~(4),
  \pg{641--664}.

\bibitem[Sidharth {\em et~al.\/}(2018)Sidharth, Dwivedi, Candler \&
  Nichols]{sidharth2018onset}
{\sc \au{Sidharth, GS}, \au{Dwivedi, Anubhav}, \au{Candler, Graham~V} \&
  \au{Nichols, Joseph~W}} \yr{2018}  \at{Onset of three-dimensionality in
  supersonic flow over a slender double wedge}.  \jt{Physical Review Fluids}
  \bvol{3}~(9),  \pg{093901}.

\bibitem[Sipp \& Marquet(2013)]{article}
{\sc \au{Sipp, Denis} \& \au{Marquet, Olivier}} \yr{2013}  \at{Characterization
  of noise amplifiers with global singular modes: The case of the leading-edge
  flat-plate boundary layer}.  \jt{Theoretical and Computational Fluid
  Dynamics}  \bvol{27},  \pg{617--635}.

\bibitem[Spalart(2000)]{spalart2000strategies}
{\sc \au{Spalart, Philippe~R}} \yr{2000}  \at{Strategies for turbulence
  modelling and simulations}.  \jt{International Journal of Heat and Fluid
  Flow}  \bvol{21}~(3),  \pg{252--263}.

\bibitem[Teramoto(2005)]{teramoto2005large}
{\sc \au{Teramoto, Susumu}} \yr{2005}  \at{Large-eddy simulation of
  transitional boundary layer with impinging shock wave}.  \jt{AIAA journal}
  \bvol{43}~(11),  \pg{2354--2363}.

\bibitem[Thumm(1991)]{thumm1991numerische}
{\sc \au{Thumm, Andreas}} \yr{1991}  \at{Numerische untersuchungen zum
  laminar-turbulenten str{\"o}mungsumschlag in transsonischen
  grenzschichtstr{\"o}mungen}. PhD thesis, University of Stuttgart.

\bibitem[Timme(2018)]{timme2018global}
{\sc \au{Timme, Sebastian}} \yr{2018}  \at{Global instability of wing shock
  buffet}.  \jt{arXiv preprint arXiv:1806.07299} .

\bibitem[Towne {\em et~al.\/}(2018)Towne, Schmidt \&
  Colonius]{towne2018spectral}
{\sc \au{Towne, Aaron}, \au{Schmidt, Oliver~T} \& \au{Colonius, Tim}} \yr{2018}
   \at{Spectral proper orthogonal decomposition and its relationship to dynamic
  mode decomposition and resolvent analysis}.  \jt{Journal of Fluid Mechanics}
  \bvol{847},  \pg{821--867}.

\bibitem[Wu \& Martin(2007)]{wu2007direct}
{\sc \au{Wu, Minwei} \& \au{Martin, M~Pino}} \yr{2007}  \at{Direct numerical
  simulation of supersonic turbulent boundary layer over a compression ramp}.
  \jt{AIAA journal}  \bvol{45}~(4),  \pg{879--889}.

\bibitem[Zhuang {\em et~al.\/}(2018)Zhuang, Tan, Li, Sheng \&
  Zhang]{zhuang2018gortler}
{\sc \au{Zhuang, Yi}, \au{Tan, Hui-jun}, \au{Li, Xin}, \au{Sheng, Fa-jia} \&
  \au{Zhang, Yu-chao}} \yr{2018}  \at{G{\"o}rtler-like vortices in an impinging
  shock wave/turbulent boundary layer interaction flow}.  \jt{Physics of
  Fluids}  \bvol{30}~(6),  \pg{061702}.

\end{thebibliography}
\appendix
\section{Inner product matrices}\label{inner}
 {This appendix provides the two matrices $\mathsfbi{Q_e}$ and $\mathsfbi{Q}$ for the inner products presented in sections \S \ref{sec:SPOD}, \S \ref{sec:resolv} and for a conservative state vector (such as described in \S \ref{sec:resolv}).\\
The matrix $Q_e$ correspond to the inner product linked to Chu's energy \citep{chu1965energy} which reads, in its continuous form :
\begin{equation}
    E_{Chu}=\frac{1}{2}\int_V \bar{\rho} \left|\boldsymbol{u'}\right|^2+ \frac{a^2}{\bar{\rho}\gamma}(\rho')^2+ \frac{\bar{\rho}C_v}{\bar{T}}(T')^2 d\Omega
    \label{equ:chu}
\end{equation}
The discrete form used in this work (see equation \ref{eq:EnergyBalance}) is derived from \citep{bugeat20193d} and relies on the following inner product matrix :
\begin{equation}
    a_1 = \frac{\overline{\rho}}{C_v \overline{T}}
\end{equation}
\begin{equation}
    a_2 = \frac{\frac{|\overline{\mathbf{u}}|^2}{2} - \overline{e}}{\overline{\rho}}
\end{equation}
\begin{equation}
    Q_e = 
\Omega \left[
\begin{array}{ccccc}
 \frac{|\mathbf{\overline{u}}|^2 + R \overline{T}}{\overline{\rho}} + a_1 a_2^2 & \frac{-\overline{u_x} (1+a_1 a_2)}{\overline{\rho}}  & \frac{-\overline{u_r} (1+a_1 a_2)}{\overline{\rho}} & \frac{-\overline{u_\theta} (1+a_1 a_2)}{\overline{\rho}}  & \frac{a_1 a_2}{\overline{\rho}}   \\
   \frac{-\overline{u_x} (1+a_1 a_2)}{\overline{\rho}}  & \frac{\rho+\overline{u_x^2}a_1}{\overline{\rho^2}} & \frac{\overline{u_x u_r}a_1}{\overline{\rho^2}} & \frac{\overline{u_x u_\theta}a_1}{\overline{\rho^2}}  & - \frac{\overline{u_x}a_1}{\overline{\rho^2}} \\
  \frac{-\overline{u_r} (1+a_1 a_2)}{\overline{\rho}} & \frac{\overline{u_x u_r}a_1}{\overline{\rho^2}} & \frac{\rho+\overline{u_r^2}a_1}{\overline{\rho^2}} & \frac{\overline{u_r u_\theta}a_1}{\overline{\rho^2}}  &  - \frac{\overline{u_r}a_1}{\overline{\rho^2}} \\
  \frac{-\overline{u_\theta} (1+a_1 a_2)}{\overline{\rho}} & \frac{\overline{u_x u_\theta}a_1}{\overline{\rho^2}} & \frac{\overline{u_r u_\theta}a_1}{\overline{\rho^2}} & \frac{\rho+\overline{u_\theta^2}a_1}{\overline{\rho^2}}  &  -\frac{\overline{u_\theta}a_1}{\overline{\rho^2}} \\
   \frac{a_1 a_2}{\overline{\rho}} & -\frac{\overline{u_x} a_1}{\overline{\rho}} & -\frac{\overline{u_r} a_1}{\overline{\rho}} & -\frac{\overline{u_\theta} a_1}{\overline{\rho}} &  \frac{a_1}{\overline{\rho^2}} \\
\end{array}  \right]
\end{equation}
With $\Omega$ the local cell volume, $\overline{.}$ the temporal average and $e$ the internal energy.\\
Q is the inner product matrix linked with the standard $\mathcal{L}_2$ norm :
\begin{equation}
    Q = 
\left[
\begin{array}{ccccc}
  \Omega & 0  & 0 & 0  & 0   \\
   0 & \Omega & 0 & 0  & 0 \\
  0 & 0 & \Omega & 0 &  0 \\
  0 & 0 & 0 &\Omega  &  0 \\
   0 & 0 & 0 & 0 &  \Omega \\
\end{array}  \right]
\end{equation}
}

\section{White noise injection}\label{noise}
\label{sec:noise}
 
This section is dedicated to the technical description of the white noise injection in the DNS. As described in the section \S\ref{sec:inletpert} white noise is injected in the simulation, this is done by perturbing the density field, the resulting forcing term apply on all the conservative variables. The injection is realised 4 cells downstream ($i=4$) of the inlet boundary condition in order not to interfere with it.
The form of this injection is the following  :
\begin{equation}
    \rho'[j,k]=\rho[j,k] (1+ 0.015 r_{n}[j,k])
\end{equation}
With $r_{n}$ a random number normalised such that the root mean square on the whole injection plane is 1 :
\begin{equation}
    r_{n}[j,k] = \frac{r_{r}[j,k]}{\sqrt{\overline{r_{r}^2}}}
\end{equation}
$r_{r}$ being a numpy generated (numpy.random.random) random number from a continuous uniform distribution between $-0.5$ and $0.5$ which is seeded from the linux random generator "/dev/urandom" of the cluster, $\overline{.}$ beeing for this special case a spatial average and $j,k$ ranging the indices of the cell of the injection plan (\textit{i.e.} $j\in[0,60]$ for the wall normal direction and $k\in[0,600]$ for the azimuthal direction).
As the time step used in the computation is far less than the convection time through one cell in the streamwise direction, the spatial scheme would be unable to transport a white noise that is updated every iteration. In order to address this issue, it was chosen not to update the noise every iteration but to keep it constant for 15 iterations between each update. This ensures that the scheme is able to discretise the noise while the spectral content is still rich enough in the high frequencies for the present study.

\section{QDNS grid convergence}\label{grid}
\label{sec:grid}
 
\begin{figure}
  \centering
  \begin{tabular}[b]{c}
    \includegraphics[trim= 0cm 0cm 0cm 0cm,width=0.442\linewidth,clip=true]{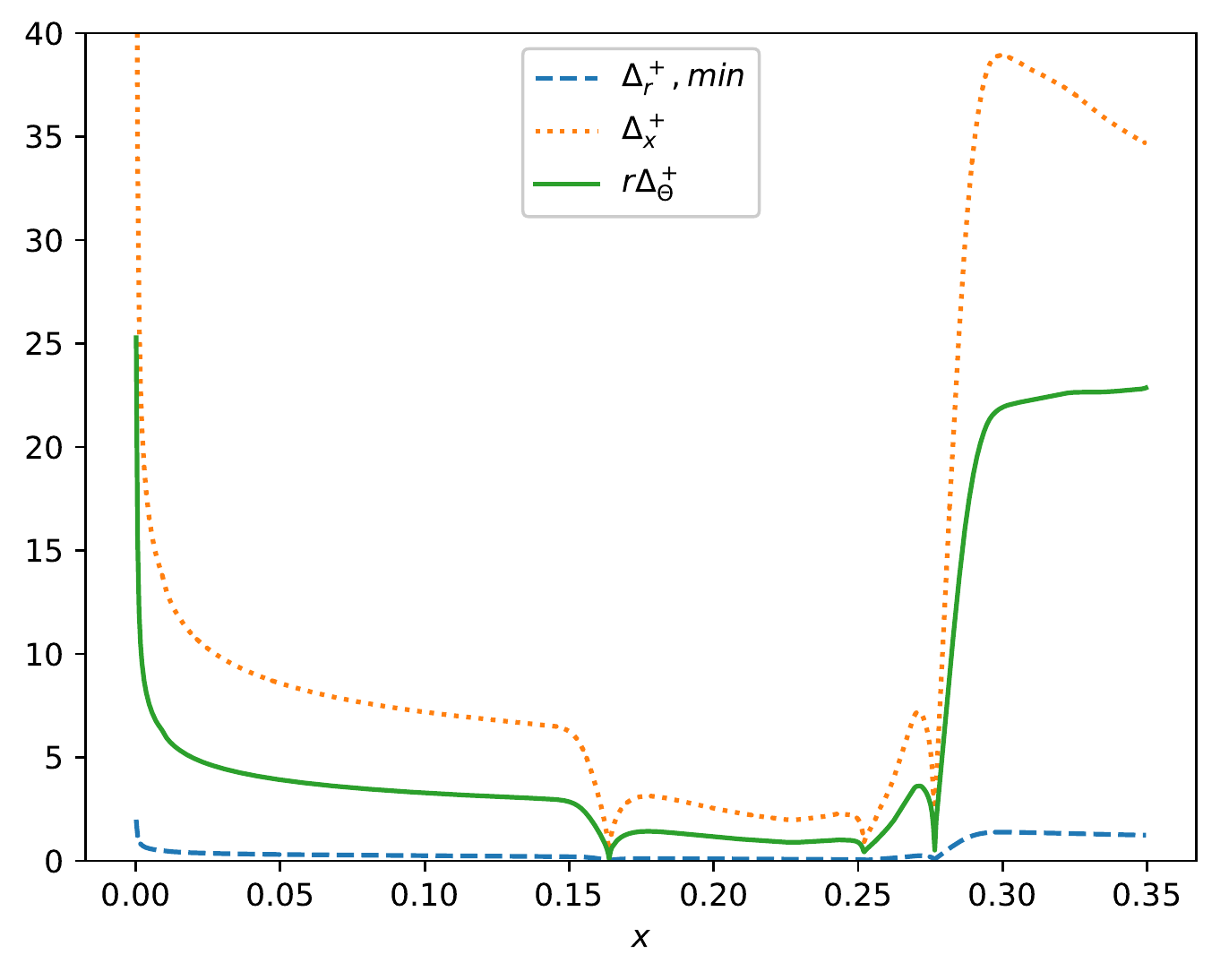}\\
    \small (a)
  \end{tabular} \qquad
  \begin{tabular}[b]{c}
  \includegraphics[trim= 0cm 0cm 0cm 0cm,width=0.458\linewidth,clip=true]{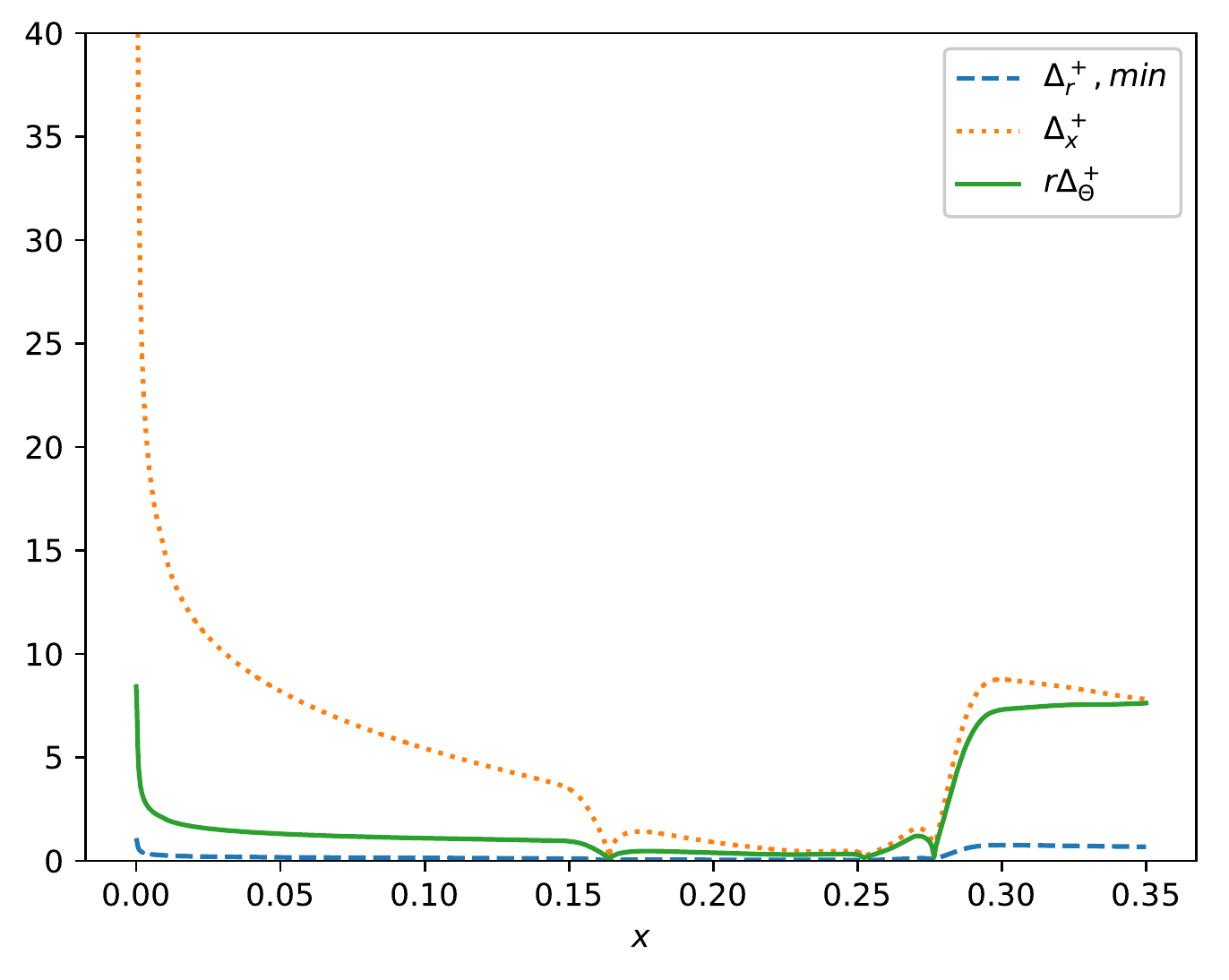}\\
    \small (b)
  \end{tabular}
   \caption{Sizing of the mesh in local wall unit for (a) the grid used for the QDNS presented in the article, (b) the grid corresponding to a stricter DNS used for validation purpose.}
    \label{fig:sizing}
\end{figure}

To assess the validity of the QDNS, a fully resolved DNS was also computed. The DNS have roughly 10 times more grid points (overall) which brings the total number of cells to around 1.5 billion. A quantitative comparison of the results obtained with the two meshes is presented in this section. Except for the grid and the timestep, which are presented in table \ref{table:numericalAnnex}, all the numerical parameters are kept the same 
The local sizing of the two grids are presented in figure \ref{fig:sizing}, showing that the sizing of the fine grid corresponds to the DNS requirement. 
\begin{table}
 \begin{center}
  \begin{tabular}{ccc}
      \textbf{Grid size}  & QDNS & DNS \\
      $n_x$  & 1409 & 3378 \\
      $n_r$   & 204 &  249\\
      $n_\theta$ & 600 & 1800 \\
      $n_{pts}$ & $172\times10^6$ & $1.514\times10^9$\\
      \hline
      $\Delta t$ & $10^{-8}s$ & $5\times 10^{-9}s$
  \end{tabular}
  \caption{Grid and timestep for the QDNS and DNS.}{\label{table:numericalAnnex}}
 \end{center}
\end{table}

\begin{figure}
  \centering
  \begin{tabular}[b]{c}
    \includegraphics[trim= 0cm 0cm 0cm 0cm,width=0.42\linewidth,clip=true]{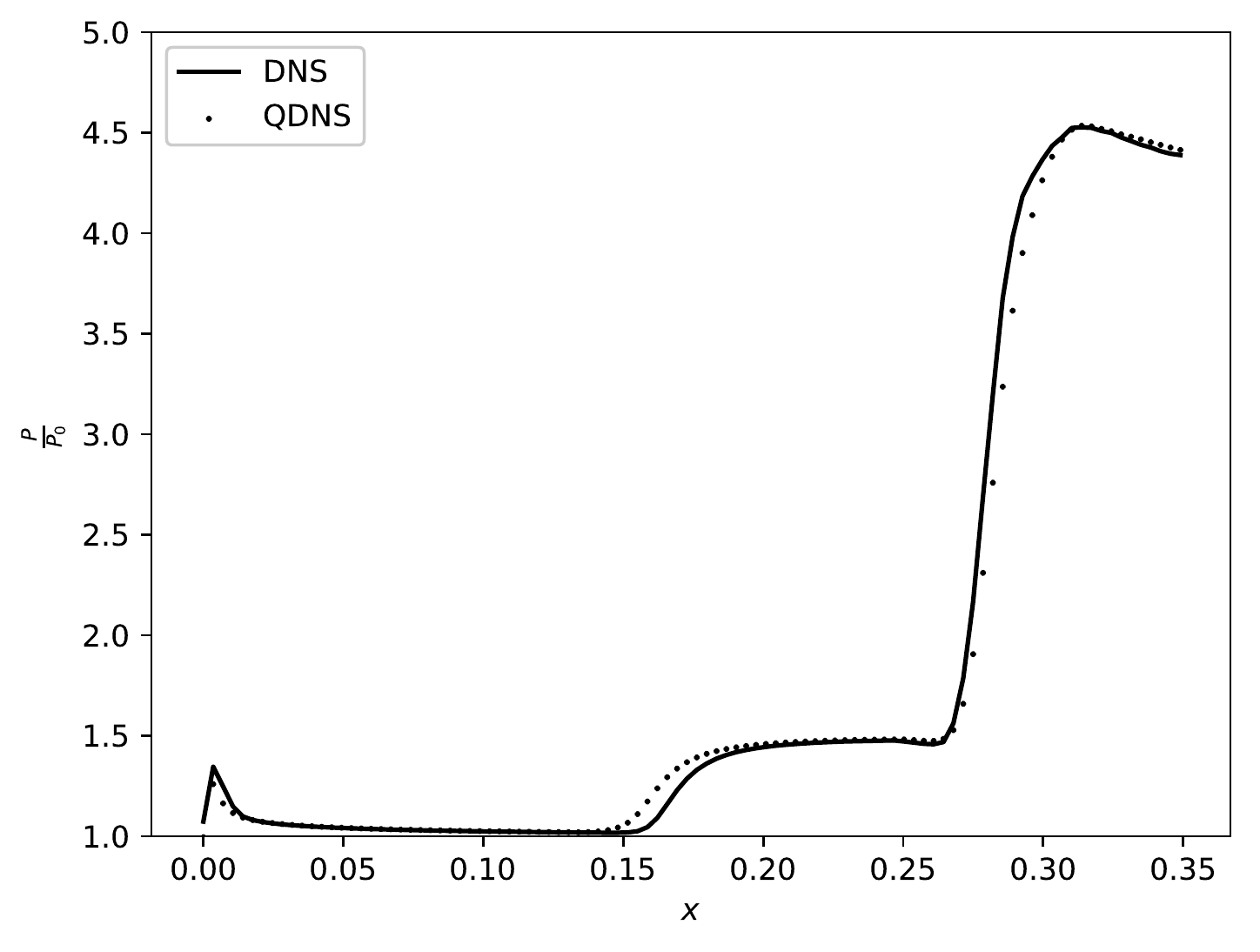}\\
    \small (a)
  \end{tabular} \qquad
  \begin{tabular}[b]{c}
    \includegraphics[trim= 0cm 0cm 0cm 0cm,width=0.42\linewidth,clip=true]{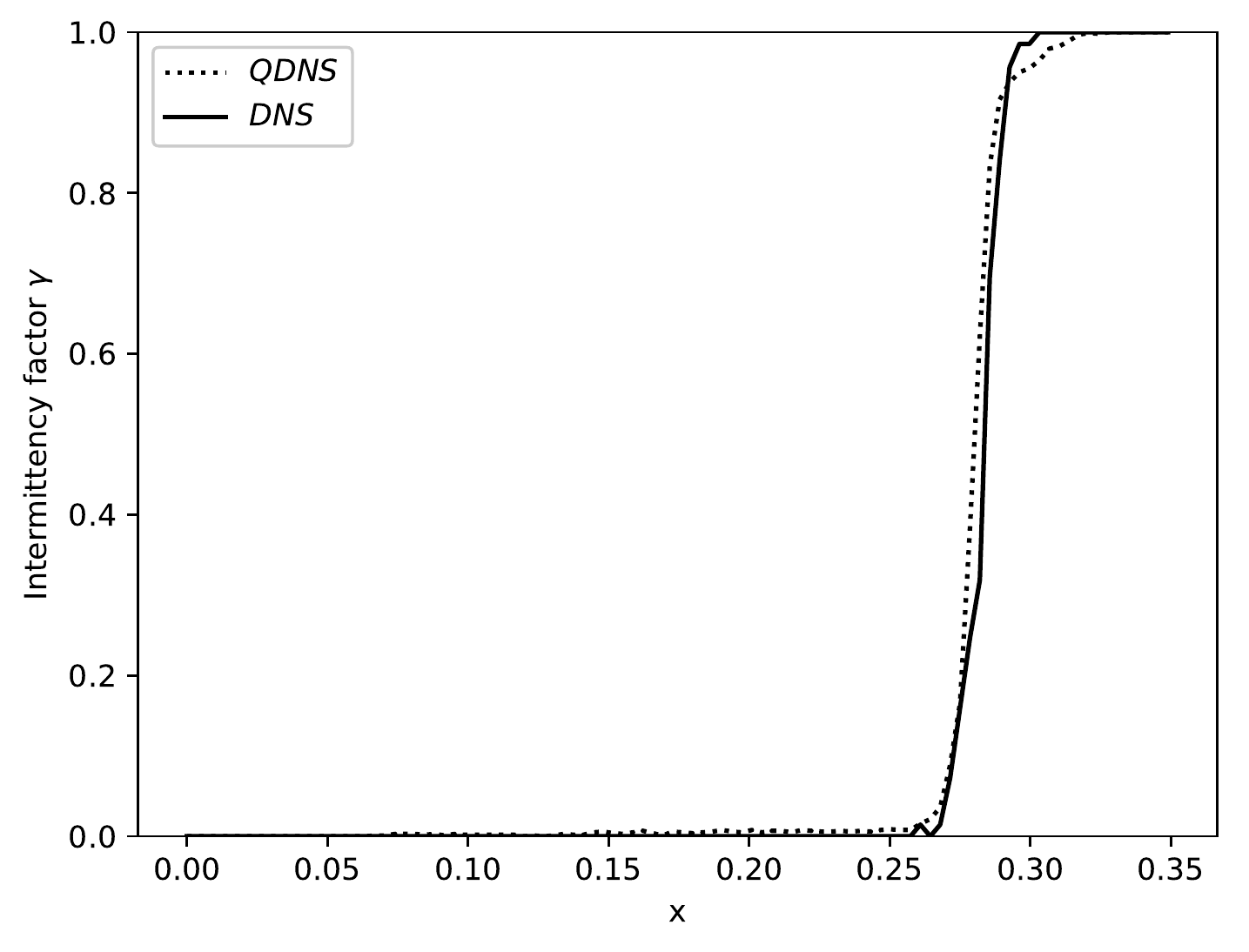}\\
    \small (b)
  \end{tabular}
   \caption{Comparison of the results of the DNS and QDNS for (a) the pressure distribution along the geometry and (b) the intermittency factor.}
    \label{fig:ValidTopoInterm}
\end{figure}

To give an idea of the computational resources used, the QDNS was run on 420 xeon cores while the DNS was run on 3840 xeon cores. The total computational cost of the QDNS presented study (convergence of multiple mean-flows, extraction of snapshots) is estimated around 500.000 CPU hours, the convergence of the mean flow of the resolved DNS alone is around the same cost.
For obvious computational cost reasons, the DNS was not run for the same amount of time than the QDNS and it was thus impossible to collect data to compute the SPOD modes.
However, a quantitative comparison is still possible on the topology of the flow.
Figure \ref{fig:ValidTopoInterm} presents the pressure distribution along the geometry and the intermittency factor for both the QDNS and DNS. It shows that except for a small overestimation of the bubble size, the QDNS is fully able to predict the right flow topology and transition point, which is an important result given the high sensitivity of the flow topology to the transition location (such as discussed \S \ref{sec:setup}).

In addition to that, and in order to be sure that all the linear instabilities are correctly described and that their higher harmonics can be adequately resolved for the nonlinear development, table \ref{table:numericalAnnexWave} presents the number of grid points per wavelength of the most amplified waves for both oblique first mode and second mode waves, showing that the spatial discretisation is able to resolve at least one higher harmonics of even the smallest energetic waves.
\\
\\

\textcolor{red}{
\begin{table}
 \begin{center}
  \begin{tabular}{ccccc}
      \textbf{Instability}  & $\omega$ [kHz] & m & $N_x$ & $N_\theta$\\
      Oblique first mode  & 51 & 72 & 65 & 50 \\
      Second mode   & 230 & 0 & 20 & undefined\\
  \end{tabular}
  \caption{Number of points per wavelength  upstream of separation for the most amplified boundary layer instabilities for the QDNS grid.}{\label{table:numericalAnnexWave}}
 \end{center}
\end{table}}

\section{Boundary layer profiles}
\label{sec:profiles}
\begin{figure}
  \centering
  \begin{tabular}[b]{c}
    \includegraphics[trim= 0cm 0cm 0cm 0cm,width=0.442\linewidth,clip=true]{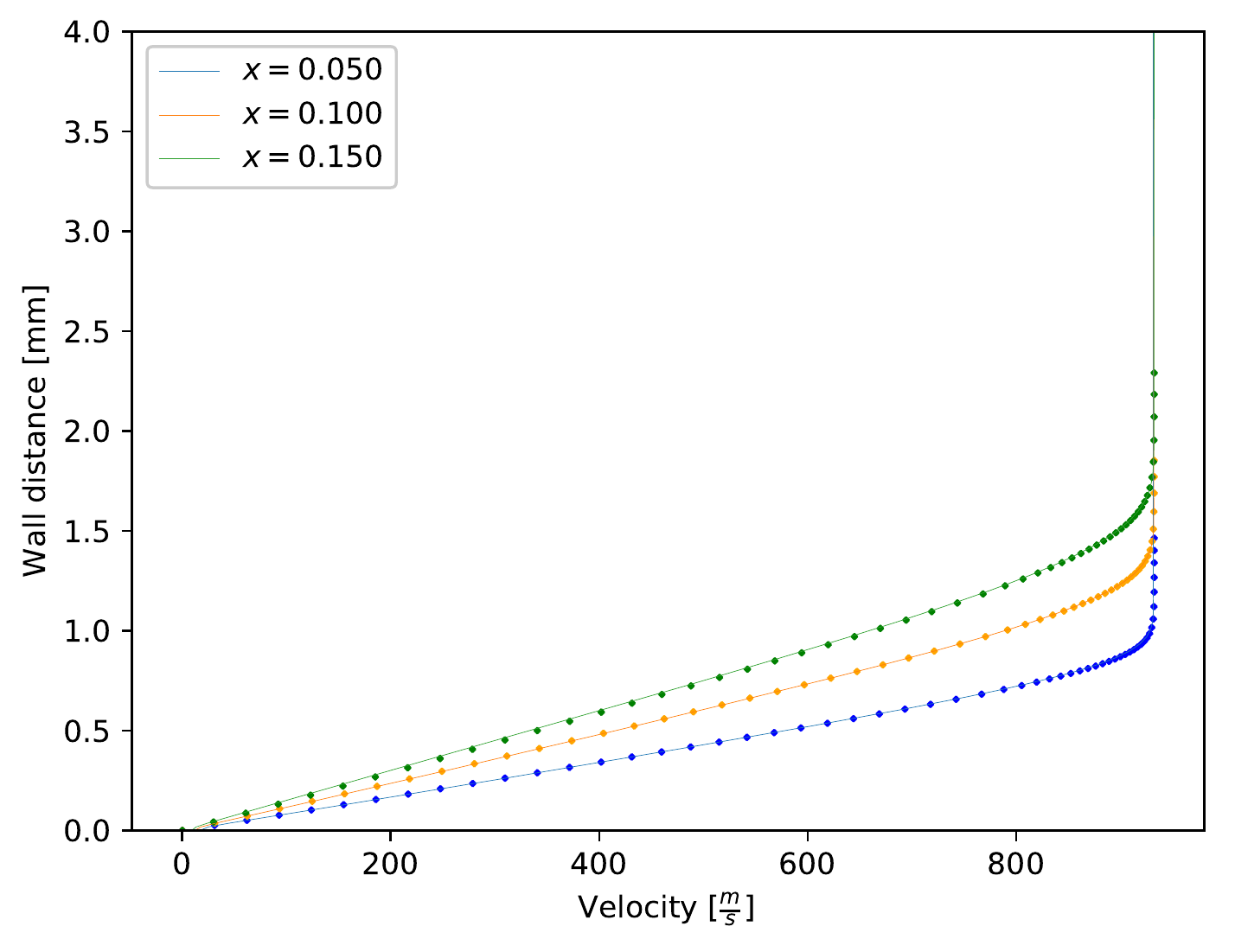}\\
    \small (a)
  \end{tabular} \qquad
  \begin{tabular}[b]{c}
  \includegraphics[trim= 0cm 0cm 0cm 0cm,width=0.442\linewidth,clip=true]{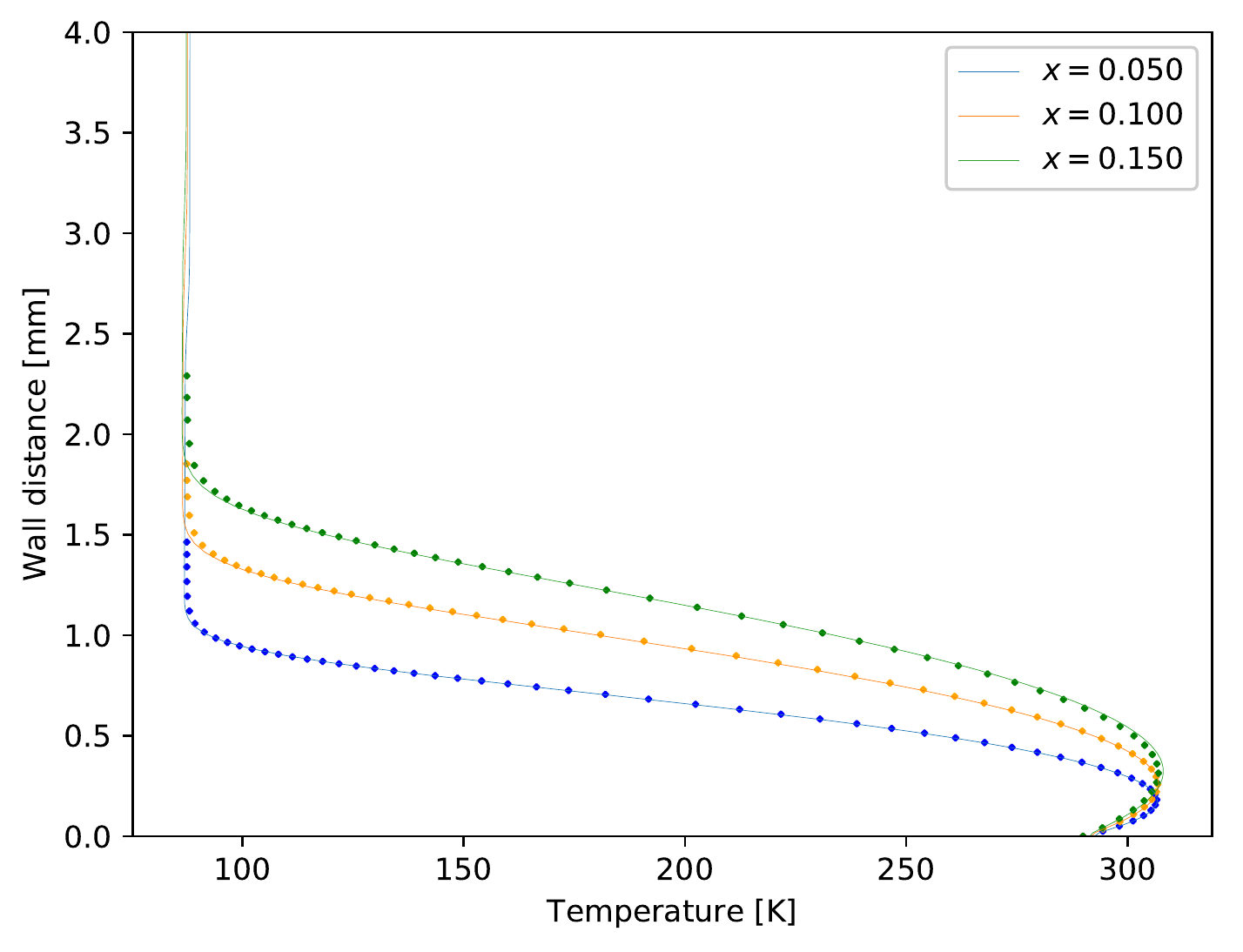}\\
    \small (b)
  \end{tabular}
  \caption{Velocity (a) and temperature (b) boundary layer profiles from the mean flow (plain lines) and self-similar solutions (dots). }
    \label{fig:profiles}
\end{figure}

\begin{figure}
  \centering
    \includegraphics[trim= 0cm 0cm 0cm 0cm,width=0.5\linewidth,clip=true]{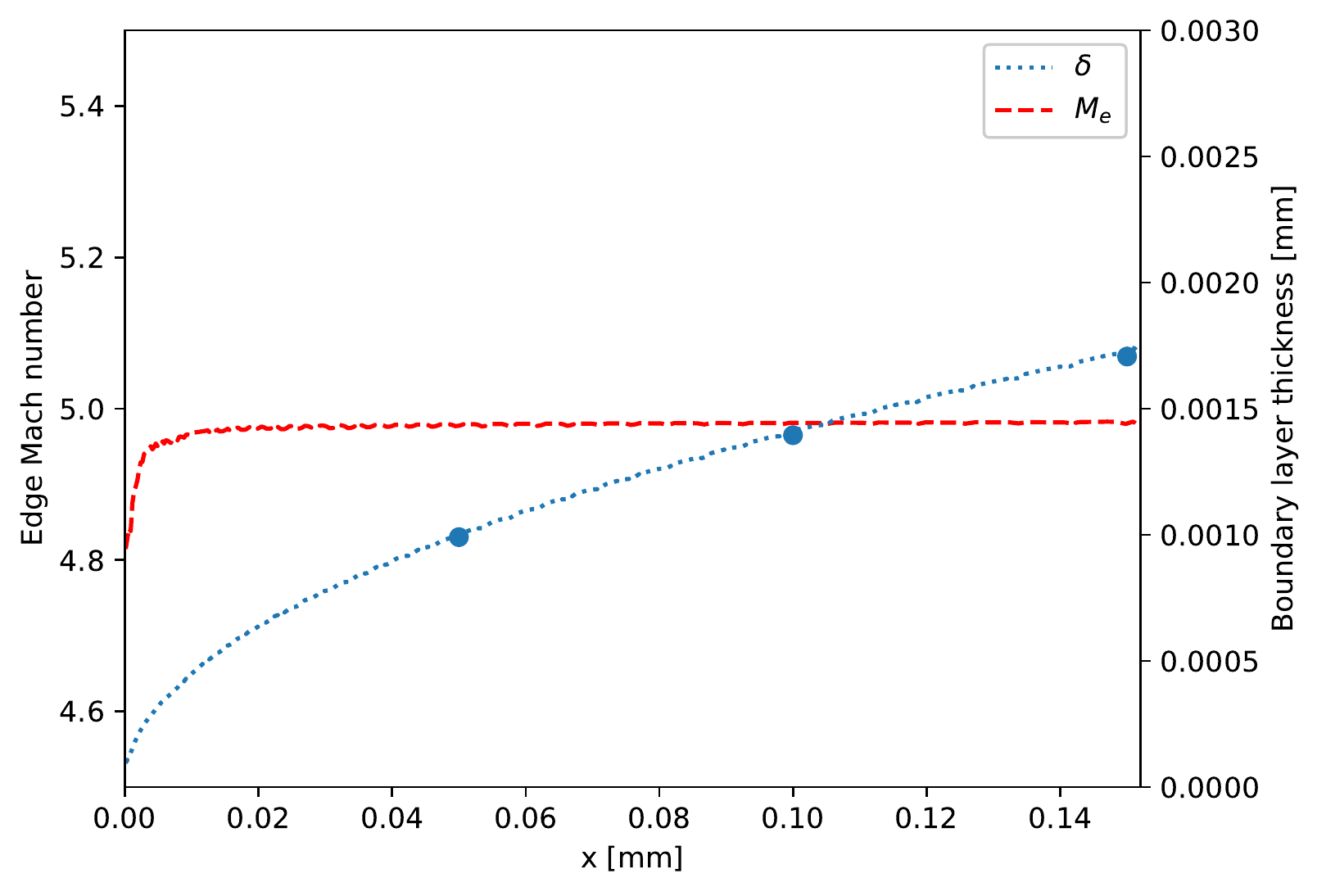}\\
  \caption{Edge Mach number and boundary layer thickness of the mean flow. The dots represents self similar verification points. }
    \label{fig:delta}
\end{figure}

Figure \ref{fig:profiles} presents the velocity and temperature boundary layer profiles from the mean flow and from a the self-similar solution computed with the ONERA boundary layer solver CLICET (see for instance \citet{renard2016theoretical}).
\\
The edge Mach number and the boundary layer thickness longitudinal evolution can be found in figure \ref{fig:delta}.  
\\

\end{document}